\documentclass[a4paper,fleqn]{cas-sc}

\usepackage[authoryear]{natbib}
\usepackage{amsfonts,amsmath,amssymb}

\usepackage{color}

\usepackage{mathtools}
\usepackage{graphicx}
\usepackage{dcolumn}
\usepackage{array}
\usepackage{lipsum}
\usepackage{bm}
\usepackage{subfigure}
\usepackage{amssymb}
\usepackage{multirow}
\usepackage{tabularx}
\usepackage{amsmath,amssymb}
\usepackage{braket}
\usepackage{mhchem}

\usepackage{centernot}

\newcommand{\br}{\bm{r}}

\newcommand{\s}{_\mathrm{{\scriptscriptstyle S}}}

\renewcommand{\vec}[1]{\mathbf{#1}}

\usepackage{tikz}

\def\tsc#1{\csdef{#1}{\textsc{\lowercase{#1}}\xspace}}
\tsc{WGM}
\tsc{QE}

\defcitealias{Fiedler_prr_2022}{Fiedler, Moldabekov et al. 2022}
\defcitealias{Boehme_PRL_2022}{B\"ohme, Moldabekov et al. 2022}

\begin{document}
\let\WriteBookmarks\relax
\def\floatpagepagefraction{1}
\def\textpagefraction{.001}

\shorttitle{From Density Response to Energy Functionals and Back}    

\shortauthors{Z. Moldabekov,  J. Vorberger, T. Dornheim}  

\title [mode = title]{From Density Response to Energy Functionals and Back: \\
An \textit{ab initio} perspective on Matter Under Extreme Conditions}  



%

\author[1]{Zhandos Moldabekov}



\ead{z.moldabekov@hzdr.de}

\affiliation[1]{organization={Center for Advanced Systems Understanding (CASUS),  Helmholtz-Zentrum Dresden-Rossendorf (HZDR)},
            city={ G\"orlitz},
            postcode={D-02826}, 
            country={Germany}}

\author[2]{Jan Vorberger}

\affiliation[2]{organization={Helmholtz-Zentrum Dresden-Rossendorf (HZDR)},
            city={Dresden},
            postcode={01328}, 
            country={Germany}}

\author[1]{Tobias Dornheim}


\begin{abstract}
Energy functionals serve as the basis for different models and methods in quantum and classical many-particle physics. Arguably, one of the most successful and widely used approaches in material science at both ambient and extreme conditions is density functional theory (DFT).
Various flavors of DFT methods are being actively used to study material properties at extreme conditions, such as in warm dense matter, dense plasmas, and nuclear physics applications. In this review, we focus on the warm dense matter regime, which occurs in the core of giant planets and stellar atmospheres, and as a transient state in inertial confinement fusion experiments.
We discuss the connection between linear density response functions and free energy functionals
as well as the utility of the linear response formalism for the construction of advanced functionals.
As a new result, we derive \textit{the stiffness theorem} linking the change in the intrinsic free energy to the density response properties of electrons.  
We review and summarize recent works that assess various exchange-correlation (XC) functionals for an inhomogeneous electron gas that is perturbed by a harmonic external field and for warm dense hydrogen using exact path integral quantum Monte Carlo data as an unassailable benchmark. This constitutes a valuable guide for selecting an appropriate XC functional for DFT calculations in the context of investigating the inhomogeneous electronic structure of warm dense matter. We stress that correctly simulating the strongly perturbed electron gas necessitates the correct UEG limit of the XC and non-interacting free-energy functionals. 
\end{abstract}

%





\maketitle

\tableofcontents
\section{Introduction} 
Matter under extreme conditions that is characterized by high densities with the Wigner–Seitz radius of the electrons $r_s$ in the vicinity of unity (density parameter) and the reduced temperature $\theta=T/T_F \sim 1$ (degeneracy parameter) is referred to as warm dense matter (WDM)~[\cite{wdm_book}] (with $r_s$ being the mean-interparticle distance in Bohr and $T_F$ is the Fermi temperature~[\cite{Ott2018}]). 
WDM is characterized by the complex interplay of quantum degeneracy, strong thermal excitations, and partial ionization.
It naturally exists in interiors of giant planets [\cite{Kraus_Nature_Astronomy_2017,Benuzzi_Mounaix_2014}] (with likely relevance also for the classification of exoplanets [\cite{Spiegel_2014,Coppari_Nature_Geo_2013}]), white dwarfs [\cite{Kritcher2020}], red dwarfs [\cite{Luetgert_pop_2022}], and in the thin outer envelope of neutron stars [\cite{Potekhin_2014}] as well as in the atmosphere of cold neutron stars [\cite{Haensel_book_2007}]. Understanding the complex physics of photon transport, thermal conductivity, and electrical conductivity in the atmospheres and interiors of stars and planets is imperative to understanding their structure and evolution. For example, in addition to estimating the distance to a star, a substantial part of the uncertainties in the measurements of the neutron star radius originates from the used atmospheric model  [\cite{Burgio_ProgPartNucPhys_2021}]. As part of laboratory astrophysics~[\cite{Takabe_Kuramitsu_2021}], WDM is actively investigated in large experimental facilities such as the European XFEL  (Germany) [\cite{Zastrau_2021}], LCLS at SLAC  (USA) [\cite{Fletcher_NaturePhotonics_2015}], the National Ignition Facility (NIF, USA) [\cite{Moses_NIF,MacDonald_pop_2023}], and at SACLA  (Japan)~[\cite{Inoue_JourSinRad_2020}].

Developing the theory of and simulation methods for WDM is also of practical importance.
In this regard, a prime example is given by inertial confinement fusion (ICF)~[\cite{Betti2016}], where recently a net positive energy gain has been reported at the NIF~[\cite{Hurricane_RevModPhys_2023,Abu_Shawareb_PRL_2024}], followed by the demonstration of hot-spot fuel gain exceeding unity in a direct drive experiment at the OMEGA laser facility in Rochester~[\cite{Williams_Nature_Phys_2024}].
These breakthroughs open up the enticing possibility of developing fusion energy into a nigh abundant source of clean energy for the future.
However, it is important to note that the path towards an industrial-scale application of ICF remains highly challenging. 
For example, during the compression and heating of an ICF target, the fuel capsule has to traverse a complex, transient WDM state~[\cite{Hu_prb_2011}].
%
%
%
%
Therefore, reliable simulation capabilities of WDM are important to understand the compression process and to achieve optimal control along the path to the ICF regime.
For an overview of the ongoing experimental, theoretical, and numerical studies of ICF applications, we refer to Refs.~ [\cite{Hurricane_RevModPhys_2023, Hu_pop_2024, Lee_pop_2023, Jacquemot_2017}].

\begin{figure}[t!]
\centering
\includegraphics[width=0.5\textwidth]{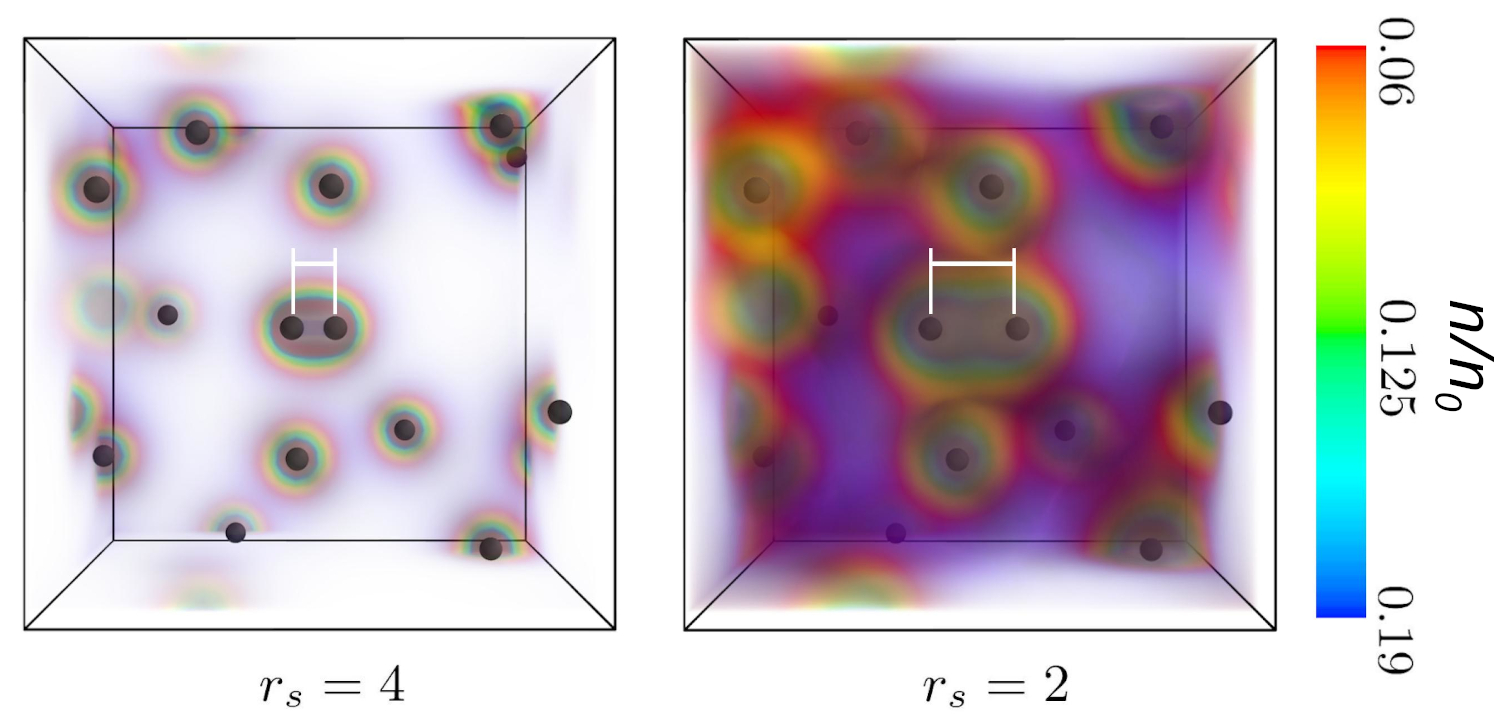}
\caption{\textit{Ab initio} path integral QMC results for the electron density in warm dense hydrogen for a snapshot of $14$ protons at $r_s=4$ (left) and $r_s=2$ (right) at $T=T_F$. In the central region of the simulation cell, two protons are positioned with a distance of $d=0.74\,$\AA$\,$  to each other (white bars); this molecular configuration is not changed for different $r_s$. The density is normalized by the mean density $n_0$. It is illustrated that upon entering the WDM regime with $r_s=2$, the segregation of electrons into bound and free becomes questionable.  
Adapted from  \cite{Moldabekov_jctc_2024}.}
\label{fig:H2_density}
\end{figure}

The theory and simulation of WDM is notoriously challenging due to the complex interplay of effects such as strong electronic correlations and partial degeneracy. In contrast to standard plasma physics, where the theory can be formulated by taking advantage of the weak electron--electron coupling, the rigorous theoretical description of WDM requires the explicit treatment of the non-ideal electrons. Compared to solid-state physics, the absence of long-range order (i.e., crystal structure) prevents the simplification of simulations by utilizing lattice symmetries.
In addition, partial degeneracy implies that neither quantum statistics nor thermal excitations can be ignored.
Due to the high densities in WDM, electronic categorization into bound and free states is often uncertain and potentially ill-defined (as illustrated in Fig. \ref{fig:H2_density}), rendering the application of chemical models questionable.
Consequently, \textit{ab initio} simulation methods are indispensable for the exploration of WDM physics and for the reliable interpretation of experimental observations~[\cite{Schoerner_PRE_2023,Dornheim_Science_2024}]. The most often used first principle approach to model WDM is based on density functional theory (DFT),
with the Kohn-Sham formalism (KS-DFT) being particularly important
[\cite{Malko2022, White_2022, Vorberger_prb_2007, wdm_book, Holst_prb_2008, White_prl_2020}].  In parallel, the development of the orbital-free formulation of DFT (OF-DFT) is actively being pursued to accelerate the simulations, which allows e.g.~to cover a larger number of particles [\cite{Lambert_cpp_2007, Dragon_2024, Fiedler_prr_2022}](see Refs.~[\cite{Wenhui_ChemRev_2023, Qiang_WIRE_2024}] for recent topical review articles). 
Moreover, a number of dedicated methods that aim to extend DFT to higher temperatures where the convergence with the number of orbitals in the usual KS-DFT becomes unfeasible have been suggested over the last years~[\cite{Zhang_pop_2016,Cytter_Rabani_prb2018,Hollebon_PRB_2022,Bethkenhagen_PRE_2023}].

A general commonality of all DFT simulation methods is that they require an approximate exchange-correlation (XC) functional as an external input; 
it decisively determines the quality of a given DFT simulation in practice. 
At ambient conditions, the accuracy of different XC functionals can be gauged by benchmarking against a massive amount of accurate experimental data on the physical and chemical properties of a host of different materials [\cite{Goerigk_PCCP_2017, Heyd2003}]. 
In contrast, performing experimental measurements with high accuracy to assess the quality of the KS-DFT method and XC functionals in the WDM regime is highly challenging; this is a direct consequence of the extreme conditions and limited time frames under which the experimental measurements are conducted.
Consequently, highly accurate and approximation-free results from quantum Monte Carlo (QMC) calculations become indispensable to test the reliability and accuracy of KS-DFT at such extreme conditions.

In practice, the construction of the XC functionals is often based on QMC results for the uniform electron gas (UEG) model~[\cite{quantum_theory,Dornheim_PhysRep_2018}]. Indeed, the most commonly used XC functionals for KS-DFT utilize ground state QMC data for the UEG~[\cite{Ceperley_Alder_PRL_1980}]. Naturally, the importance of such UEG results extends to the developments of dedicated XC-functionals that are designed for the application in the WDM regime.
This need has sparked a surge of activity in the field of QMC simulations at finite temperatures, see, e.g., Ref.~\cite{Dornheim_POP_2017} and references therein.
Here, the main obstacle is given by the so-called fermion sign problem~[\cite{troyer,dornheim_sign_problem}], which is a computational bottleneck manifesting as an exponential increase in the required compute time with respect to important parameters such as the system size.
Nevertheless, the combination of various complementary methods~[\cite{Dornheim_prl_2016,Schoof_PRL_2015,Malone_PRL_2016}] has resulted in the first accurate parametrizations of the warm dense UEG~[\cite{Groth_prl_2017,Karasiev_prl_2014}], which constitute the basis for thermal DFT simulations on the level of the local density approximation (LDA)~[\cite{Karasiev_pre_2016,kushal}] and beyond~[\cite{Sjostrom_PRB_2014,Karasiev_prl_2018,Karasiev_meta_GGA,kozlowski2023generalized}].



In practice, however, most DFT simulations of WDM have used XC functionals that were created for solid-state applications where the electrons are in their respective ground state. Therefore, it is crucial to scrutinize the performance and accuracy of these functionals at WDM conditions based on benchmarks against exact QMC results. This has been done in several recent works [\cite{Moldabekov_jcp_2021, Moldabekov_prb_2022, Moldabekov_jcp_hybrid_2023, Moldabekov_jctc_2022, Moldabekov_jctc_2023, Moldabekov_jctc_2024, Moldabekov_jpcl_2023, Moldabekov_jcp_2023_averaging, Moldabekov_prr_2023}] for the weakly and strongly perturbed warm dense electron gas, and for warm dense hydrogen, see also the recent overview by~\cite{bonitz2024principles}. The performed analyses include ground-state XC functionals with various degrees of complexity and the thermal LDA XC-functional by \cite{Groth_prl_2017}. Here, we summarize the findings from these benchmark studies and formulate recommendations regarding the utility of the considered XC functionals for WDM applications.  
 Additionally, we emphasize the importance of the UEG limit as a constraint for developing accurate XC functionals and non-interacting free energy functionals for WDM simulations and beyond. A case in point is given by the new non-empirical method for choosing the mixing coefficient in fully non-local hybrid functionals [\cite{Moldabekov_jpcl_2023}].
  
 Our analysis is built to a large degree on the fundamental connections between the static linear density response function and the XC functional. In fact, these connections constitute the essence of \textit{the stiffness theorem}, which represents the link between the change in thermodynamic potentials and density response functions~[\cite{quantum_theory}]. The stiffness theorem is well known in ground-state quantum many-body theories but rarely discussed for finite temperatures. To close this gap, we provide a detailed derivation of the relations representing the stiffness theorem for the free energy. Moreover, for the first time, we show the relation between the change in the intrinsic part of the free energy and the polarization function. This relation is of practical relevance for both KS-DFT and OF-DFT. 

An additional aim of the present work is to provide an
introduction to the main ideas of (thermal) DFT as well as to explain different approximations inherent to various XC functionals, having in mind the benefit of readers who are starting research work in the direction of matter under extreme conditions.
This is complemented by a broad overview of a number of recent developments in QMC simulations of WDM.
The developments outlined in this paper may have applicability to other systems, e.g., hydrogen at low temperatures and high pressures~[\cite{bonitz2024principles}]. Finally, we note that the DFT method is an active area of research beyond non-relativistic systems, e.g., for modeling quark matter [\cite{Blaschke_inbook, Ivanytskyi_prd_2022}], where the construction of DFT models is inspired by ideas of the orbital-free version of DFT.

The paper is organized as follows: In Sec. \ref{s:Zoo}, we introduce various types of XC functionals and non-interacting free energy functionals for KS-DFT and OF-DFT, respectively.  Then, in Sec. \ref{ss:connection}, we derive a connection between free energy functionals and various types of the density response functions appearing in DFT. This provides a basis for the derivation of the stiffness theorem connecting the change in the free energy of electrons due to an external perturbation with the static linear density response function. This derivation is provided in Sec. \ref{s:stiffness_section} for both homogeneous and inhomogeneous systems. In Sec. \ref{s:QMC}, we give a review of recent developments in the field of QMC simulations of WDM. This is followed by an in-depth summary of prior results on benchmarking XC functionals and non-interacting free energy functionals for the harmonically perturbed electron gas in Sec. \ref{s:Test_electron_gas} and for warm dense hydrogen in Sec.~\ref{s:warm_dense_hydrogen}. 
We summarize the main findings and discuss the challenges that need to be addressed for future developments in the electronic structure theory of WDM in the final Sec. \ref{s:end}.


\section{Zoo of Energy Functionals}\label{s:Zoo}

In DFT, a model for describing real matter is developed by starting from a simplified yet exactly solvable system. 
Incorporating corrections to map this model system to real matter is the second step.
Typically, these corrections are necessary due to the absence of certain interparticle correlations in the model system. This general strategy is schematically depicted in Fig. \ref{fig:diagram1}.
This workflow is similar to that used for the van der Waals equation, where the underlying solvable model system (ideal gas) is extended to describe real gases, vapors, and liquids [\cite{van_der_Waals}].   

In general, the success of a theoretical model largely depends on 
the degree to which the physical behavior of the real system is already captured by the simplified model without any additional corrections.
For most materials at ambient and extreme conditions, the ideal electron gas is a popular and appropriate choice. For example, at high enough temperatures, the ionization of atoms leads to the creation of a plasma state where the electrons are only weakly coupled with each other and with the ions. In this case, it is natural to describe the electrons and ions as two ideal subsystems of charged particles that are coupled via an electrostatic mean field, and with the correlation effects being subsequently incorporated by introducing collision frequencies between different pairs of species. A classical example of such a plasma model is the Landau-Spitzer approach for temperature relaxation [\cite{Landau1937, spitzer1962physics}]. At ambient conditions, for metals (e.g. aluminum), the UEG model provides a useful qualitative picture of the physical properties.
Among more advanced methods, 
a highly successful approach is given by 
Kohn-Sham density functional theory (KS-DFT), where the description of real materials is built upon an auxiliary model of non-interacting electrons in an effective external single-particle potential. 

\begin{figure}[t!]
\centering
\tikzset{every picture/.style={line width=0.75pt}} 
\begin{tikzpicture}[x=0.75pt,y=0.75pt,yscale=-1,xscale=1]

\draw [color={rgb, 255:red, 208; green, 2; blue, 27 }  ,draw opacity=0.8 ]  (31,149.05) .. controls (31,144.52) and (34.67,140.85) .. (39.2,140.85) -- (202.68,140.85) .. controls (207.21,140.85) and (210.88,144.52) .. (210.88,149.05) -- (210.88,173.65) .. controls (210.88,178.18) and (207.21,181.85) .. (202.68,181.85) -- (39.2,181.85) .. controls (34.67,181.85) and (31,178.18) .. (31,173.65) -- cycle ;
\draw  [color={rgb, 255:red, 208; green, 2; blue, 27 }  ,draw opacity=0.8 ] (244,149.05) .. controls (244,144.52) and (247.67,140.85) .. (252.2,140.85) -- (442.68,140.85) .. controls (447.21,140.85) and (450.88,144.52) .. (450.88,149.05) -- (450.88,173.65) .. controls (450.88,178.18) and (447.21,181.85) .. (442.68,181.85) -- (252.2,181.85) .. controls (247.67,181.85) and (244,178.18) .. (244,173.65) -- cycle ;
\draw  [color={rgb, 255:red, 208; green, 2; blue, 27 }  ,draw opacity=0.8 ] (487.88,149.05) .. controls (487.88,144.52) and (491.55,140.85) .. (496.08,140.85) -- (580.68,140.85) .. controls (585.21,140.85) and (588.88,144.52) .. (588.88,149.05) -- (588.88,173.65) .. controls (588.88,178.18) and (585.21,181.85) .. (580.68,181.85) -- (496.08,181.85) .. controls (491.55,181.85) and (487.88,178.18) .. (487.88,173.65) -- cycle ;

\draw [color={rgb, 255:red, 0; green, 0; blue, 0 }  ,draw opacity=1 ] (55,153) node [anchor=north west][inner sep=0.75pt]   [align=left] {\textbf{Exactly Solvable Model}};
\draw [color={rgb, 255:red, 0; green, 0; blue, 0 }  ,draw opacity=1 ] (271,152.05) node [anchor=north west][inner sep=0.75pt]   [align=left] {\textbf{Corrections for Non-ideality}};
\draw [color={rgb, 255:red, 0; green, 0; blue, 0 }  ,draw opacity=1 ] (218,151.45) node [anchor=north west][inner sep=0.75pt]    {$\bigoplus $};
\draw [color={rgb, 255:red, 0; green, 0; blue, 0 }  ,draw opacity=1 ] (458,155) node [anchor=north west][inner sep=0.75pt]  [font=\Large]  {$\similarrightarrow $};
\draw [color={rgb, 255:red, 0; green, 0; blue, 0 }  ,draw opacity=1 ] (502,153) node [anchor=north west][inner sep=0.75pt]   [align=left] {\textbf{Real Matter}};

\end{tikzpicture}
\caption{ \small{\textit{A general strategy of developing a theoretical model of a real matter: First, choose an exactly solvable model system relevant to a matter under consideration. Second, incorporate corrections (often approximately) missing in the model system to breach a gap between the model system and the real matter. 
}}}
\label{fig:diagram1}
\end{figure} 

Within statistical physics, all thermodynamic properties of a given system in equilibrium are readily available if an appropriate thermodynamic potential is known in terms of the corresponding natural variables. 
%
In the canonical ensemble (constant temperature $T$, number of particles $N$ and volume $V$), the thermodynamic potential is given by the free energy $F(T,V,N)$. In the following, we consider the limit $V\to \infty$ and $N\to \infty$ with $N/V={\rm const}$, and drop the dependence on $V$. 
Following the scheme depicted in Fig. \ref{fig:diagram1}, we consider the theory of electrons built upon the ideal Fermi gas model. The first and most simple correction to the Fermi gas model is added by using the potential energy in the mean-field approximation $\int n(\vec r)v(\vec r)~{\rm{d}\vec r}$, 
defined by the (local) single-particle density $n(\vec r)$ and mean field potential 
$v(\vec r)$. In the canonical ensemble, the distribution $n(\vec r)$ is constrained since $N=\int_V n(\vec r)~{\rm{d}\vec r}$. Therefore, formally, one can write $F=F([n(\vec r)],T)$ (with more reasons discussed later).
The potential $v(\vec r)$ includes the field of the ions, the mean field of electrons (as follows from the Poisson equation), and any other external potentials.  

Further corrections due to the interaction between particles originate from density fluctuations, the description of which requires information beyond the single-particle density such as the two-particle density-density correlation function. 
Let us isolate this part of the corrections by writing the intrinsic
part of the free energy of electrons as the sum of two terms:
\begin{equation}\label{eq:F}
    \mathcal{F}([n(\vec r)],T)=F([n(\vec r)],T)-\int n(\vec r)v(\vec r)~{\rm{d}\vec r}=\mathcal{F}_{\rm id}([n(\vec r)],T)+\mathcal{F}_{\rm XC}([n(\vec r)],T),
\end{equation}
where $F_{\rm id}$ is the free energy of the ideal (non-interacting) Fermi gas and  $\mathcal{F}_{\rm XC}$ is the correction describing the difference between real matter and the ideal Fermi gas. The latter is commonly referred to as the exchange-correlation (XC) free energy. 

The most simple model for $\mathcal{F}_{\rm id}([n(\vec r)],T)$ is a uniform non-interacting Fermi gas. In this case, it holds $n(\vec r)=n_0={\rm const}$ and the exact analytical solution is known and parametrized [\cite{Perrot_pra_1979}]. However, the major drawback is that it makes the task of finding $\mathcal{F}_{\rm XC}([n(\vec r)],T)$ practically unsolvable for the majority of applications where the electron--ion coupling is non-negligible. Instead, the choice of non-interacting electrons in the field $v(\vec r)$ as the auxiliary model for $\mathcal{F}_{\rm id}([n(\vec r)],T)$ has been proven to be highly successful for the description of real materials. This model system is at the heart of KS-DFT~[\cite{Jones_RMP_2015}].

\subsection{Kohn-Sham Model System and Free Energy Functionals}\label{s:KS_model}

The reason why the description based on an auxiliary non-interacting system of KS-DFT works is due to the uniqueness of the single-particle electron density, which fully determines the equilibrium state properties of a many-electron system.
This was proven by Hohenberg and Kohn for the ground state ($T=0$)~[\cite{hohenberg_inhomogeneous_1964}] and extended by Mermin to finite temperatures [\cite{Mermin_1965, Eschrig_2010}]. Part of the proof of the uniqueness of the single-particle electron density of the equilibrium state  $n(\vec r)$ is the one-to-one correspondence between $n(\vec r)$  and the external local potential $v(\vec r)$ [\cite{Aurora_inbook_2014}]. In other words, in the canonical ensemble, the free energy can be expressed exclusively as a functional of $n(\vec r)$.
Using this fact, in their seminal work, Kohn and Sham suggested considering an auxiliary non-interacting electron gas in the local effective potential $v_{\rm KS}(\vec r)$ (KS potential) that results in the density  $n(\vec r)$  minimizing the ground-state energy (free energy at $T\neq 0$). In other words, one can find the potential  $v_{\rm KS}(\vec r)$  which, when acting on a system of non-interacting fermions, generates the same single-particle electronic density $n(\vec r)$ as in the real, correlated material. Consequently, $n(\vec r)$ can be used to compute $F([n(\vec r)],T)$ and other derivative thermodynamic properties. 
In this way, the problem is reduced to the solution of a set of non-interacting single-particle Schr\"odinger type equations ~[\cite{kohn_self-consistent_1965}]:
\begin{equation}
\left[-\frac{1}{2}\nabla^2 + v_{\rm KS}(\br)\right]\phi_j(\br) = \epsilon_j \phi_j(\br), \label{eq:KS_static}
\end{equation}
where $\epsilon_j$ denotes the single-particle eigenvalues and $v\s$ the Kohn-Sham potential. 
The solution of Eq. ({\ref{eq:KS_static}}) for fermions gives one the single-electron density in terms of the orbitals $\phi_j(\br)$ as $n(\br) = \sum_{j=0}^{\rm N_b} \, f_j |\phi_j(\br)|^2$, where $f_j$ are the corresponding Fermi-Dirac occupation numbers. 
The number of bands $N_b$ must be large enough so that the total number of electrons is recovered $N=\int_V n(\vec r)~{\rm{d}\vec r}$.

The variational minimization of $F[n(\vec r)]$ (the dependence on $T$ is assumed implicitly) gives one the equation for the KS potential~[\cite{parr1994density}]:
\begin{equation}\label{eq:v_tot_ks}
    v_{\rm KS}[n](\vec r)= v_{\rm ext}[n](\vec r) + v_H[n](\vec r) + v_{\rm XC}[n](\vec r),
\end{equation}
where $v_{\rm ext}$ denotes the potential energy density due to an external potential, $v_H[n]$ the classical electrostatic (Hartree) potential energy density of electrons and ions, and $v_{\rm XC}[n]$ the XC potential of the free energy density  defined as the functional derivative of $\mathcal{F}_{\rm XC}[n(\vec r)]$:
\begin{equation}\label{eq:v_ks}
    v_{\rm XC}[n](\vec r,t)=\frac{\delta \mathcal{F}_{\rm XC}[n]}{\delta n}.
\end{equation}

The XC potential $v_{\rm XC}[n]$ (equivalently $\mathcal{F}_{\rm XC}[n]$) contains, by definition, all information that is missing from the auxiliary system of non-interacting electrons in order to reproduce the electronic density of the real system.
Therefore, $\mathcal{F}_{\rm XC}[n]$ (being provided as an input) decisively determines the level of approximation of KS-DFT in practice.
Once $\mathcal{F}_{\rm XC}[n]$ is chosen, the self-consistent solution of Eq. (\ref{eq:KS_static}) provides, in principle, all thermodynamic properties of the system (e.g., see \cite{Dornheim_pop_2023, Fiedler_prr_2022}). Following the common notation, the intrinsic free energy of the KS system excluding the XC part is denoted using a lower $S$ index [\cite{Karasiev_prb_2012}], i.e., $\mathcal{F}_{\rm id}[n(\vec r)]=\mathcal{F}_{\rm S}[n(\vec r)]$ \footnote{The notation $\mathcal{F}_{s}$ for the ideal (non-interacting) part of the intrinsic free energy is convention in the KS model system based DFT. Generally, $\mathcal{F}_{s}$ is not used in other statistical theory models, e.g., in plasma physics. Therefore, we stress that $\mathcal{F}_{s}$ indicates the ideal part of the intrinsic free energy of the KS system.}.
Many XC functionals for various applications have been developed, with many of them provided by the Libxc library of XC functionals [\cite{libxc_2012, libxc_2018}]. XC functionals are usually classified based on their degree of non-locality with respect to density dependence~[\cite{Perdew_AIP_2001}]. In the following sections, we will be discussing various classes of $\mathcal{F}_{\rm XC}[n]$ that are pertinent to WDM simulations.

The importance of the $\mathcal{F}_{\rm XC}$ contribution varies depending on temperature and density values ~[\cite{Karasiev_pre_2016, Moldabekov_jctc_2024}]. This is schematically depicted in Fig. \ref{fig:diagram2}.
In the limit of hot-dense plasmas, the thermal motion and ionization of atoms make correlation effects negligible.  In this limit, $\mathcal{F}_{\rm S}[n]$ substantially exceeds $\mathcal{F}_{\rm XC}[n]$, and simple mean-field theories provide an adequate description of plasma properties ~[\cite{kraeft2012quantum, Ramazanov_pre_2015}]. 
In stark contrast, accurate simulations in the warm dense matter regime with $\mathcal{F}_{\rm XC}[n]/\mathcal{F}_{\rm S}[n]\sim 1$ pose a notoriously difficult challenge due to the complex interplay of quantum degeneracy and XC effects~[\cite{wdm_book,Bonitz_pop_2020}]. There is no sharp boundary with respect to density or temperature that clearly delineates ideal dense plasmas from the WDM regime, and the intermediate region between these two is often referred to as hot dense matter, cf.~Fig.~\ref{fig:diagram2}.  

The advantage of the KS model is that $\mathcal{F}_{\rm S}$ is computed exactly for a given choice of the functional   $\mathcal{F}_{\rm XC}$. The main disadvantage is that the $N_b$-system of Schr\"odinger type equations has to be solved numerically. This becomes problematic at high temperatures where the Fermi-Dirac distribution is smeared significantly over a wide range of energies and, as a result, a large number of bands (orbitals) are required. For example, as illustrated by \cite{Fiedler_prr_2022}, the computational cost of the highly efficient state-of-the-art GPU implementation of Eq. (\ref{eq:KS_static}) [\cite{vaspgpu1, vaspgpu2}]  has the computation cost scaling as $\sim T^3$. This is in addition to a factor scaling as $\sim N^2$ with the number of electrons.
Therefore, the need to simulate a large number of particles at high temperatures has been motivating the development of new numerical techniques with better scaling of the computational cost e.g., see [\cite{Zhang_pop_2016, Cytter_Rabani_prb2018, White_prl_2020,Hollebon_PRB_2022, Bethkenhagen_PRE_2023}].

\begin{figure}\centering

  
\tikzset {_502il70in/.code = {\pgfsetadditionalshadetransform{ \pgftransformshift{\pgfpoint{0 bp } { 0 bp }  }  \pgftransformrotate{0 }  \pgftransformscale{2 }  }}}
\pgfdeclarehorizontalshading{_h71kl8ilp}{150bp}{rgb(0bp)=(1,0.91,0);
rgb(37.5bp)=(1,0.91,0);
rgb(62.410714285714285bp)=(0.82,0.01,0.01);
rgb(100bp)=(0.82,0.01,0.01)}
\tikzset{_8tb71pcod/.code = {\pgfsetadditionalshadetransform{\pgftransformshift{\pgfpoint{0 bp } { 0 bp }  }  \pgftransformrotate{0 }  \pgftransformscale{2 } }}}
\pgfdeclarehorizontalshading{_gifbfuz57} {150bp} {color(0bp)=(transparent!0);
color(37.5bp)=(transparent!0);
color(62.410714285714285bp)=(transparent!23);
color(100bp)=(transparent!23) } 
\pgfdeclarefading{_fvp75lshi}{\tikz \fill[shading=_gifbfuz57,_8tb71pcod] (0,0) rectangle (50bp,50bp); } 
\tikzset{every picture/.style={line width=0.75pt}} 

\begin{tikzpicture}[x=0.75pt,y=0.75pt,yscale=-1,xscale=1]

\path  [shading=_h71kl8ilp,_502il70in,path fading= _fvp75lshi ,fading transform={xshift=2}] (100,126) -- (497,126) -- (497,192) -- (100,192) -- cycle ; 
 \draw   (100,126) -- (497,126) -- (497,192) -- (100,192) -- cycle ; 

\draw    (108.99,230.47) -- (489.36,228.39)(109.01,233.47) -- (489.37,231.39) ;
\draw [shift={(496.37,229.85)}, rotate = 179.69] [color={rgb, 255:red, 0; green, 0; blue, 0 }  ][line width=0.75]    (10.93,-4.9) .. controls (6.95,-2.3) and (3.31,-0.67) .. (0,0) .. controls (3.31,0.67) and (6.95,2.3) .. (10.93,4.9)   ;
\draw [shift={(109,231.97)}, rotate = 359.69] [color={rgb, 255:red, 0; green, 0; blue, 0 }  ][line width=0.75]    (0,5.59) -- (0,-5.59)(-5.03,5.59) -- (-5.03,-5.59)   ;
\draw [line width=0.75]  [dash pattern={on 4.5pt off 4.5pt}]  (413.23,109.61) -- (195.56,110.89)(413.21,106.61) -- (195.55,107.89) ;
\draw [shift={(186.56,109.44)}, rotate = 359.66] [fill={rgb, 255:red, 0; green, 0; blue, 0 }  ][line width=0.08]  [draw opacity=0] (11.61,-5.58) -- (0,0) -- (11.61,5.58) -- (7.71,0) -- cycle    ;
\draw [shift={(413.22,108.11)}, rotate = 359.66] [color={rgb, 255:red, 0; green, 0; blue, 0 }  ][line width=0.75]    (0,6.15) -- (0,-6.15)   ;

\draw (117.35,143) node [anchor=north west][inner sep=0.75pt]  [font=\footnotesize] [align=left] {\begin{minipage}[lt]{51.87pt}\setlength\topsep{0pt}
\begin{center}
\textbf{Warm Dense}\\\textbf{Matter}
\end{center}

\end{minipage}};
\draw (268,139) node [anchor=north west][inner sep=0.75pt]  [font=\small] [align=left] {\begin{minipage}[lt]{48.14pt}\setlength\topsep{0pt}
\begin{center}
\textbf{Hot Dense}\\\textbf{Matter}
\end{center}

\end{minipage}};
\draw (405.15,139) node [anchor=north west][inner sep=0.75pt]  [font=\small] [align=left] {\begin{minipage}[lt]{53.78pt}\setlength\topsep{0pt}
\begin{center}
\textbf{Ideal Dense}\\\textbf{Plasmas}
\end{center}

\end{minipage}};
\draw [color={rgb, 255:red, 0; green, 0; blue, 0 }  ,draw opacity=1 ] (287,242.4) node [anchor=north west][inner sep=0.75pt]  [font=\normalsize]  {$\mathbf{\mathcal{F}_{\rm S} /\mathcal{F}_{\rm XC}}$};
\draw [color={rgb, 255:red, 0; green, 0; blue, 0 }  ,draw opacity=1 ] (114,202.4) node [anchor=north west][inner sep=0.75pt]  [font=\normalsize]  {$\mathbf{\mathcal{F}_{\rm XC} /\mathcal{F}_{\rm S} \sim 1}$};
\draw [color={rgb, 255:red, 0; green, 0; blue, 0 }  ,draw opacity=1 ] (402,202.4) node [anchor=north west][inner sep=0.75pt]  [font=\normalsize]  {$\mathbf{\mathcal{F}_{\rm XC} /\mathcal{F}_{\rm S} \ll 1}$};
\draw [color={rgb, 255:red, 0; green, 0; blue, 0 }  ,draw opacity=1 ] (201.33,82) node [anchor=north west][inner sep=0.75pt]  [font=\small] [align=left] {\textit{Theory and Simulation Complexity}};

\end{tikzpicture}

\caption{ \small{\textit{Theory and simulation complexity of heated dense matter vary depending on the electronic degree of correlation (XC part of free energy $\mathcal{F}_{\rm XC}$) with respect to the ideal (non-interacting) free energy $\mathcal{F}_{\rm S}$. As $\mathcal{F}_{\rm XC}/\mathcal{F}_{\rm S}$ increases, providing reliable results requires higher theoretical and computational complexity.
}}}
\label{fig:diagram2}
\end{figure} 

An alternative DFT approach that does not have these computational issues is orbital-free DFT (OF-DFT), where an explicit functional dependence of $\mathcal{F}_{\rm S}$  on $n (\vec r)$ has to be provided. 
In practice, no general functional form that is valid (or even accurate) for arbitrary model systems exists.
Instead,  $\mathcal{F}_{\rm S}[n]$ has to be approximated. This approximation is the main source of the inaccuracy in OF-DFT. Therefore, one can consider it as a computationally cheaper approximation to the KS model, which allows one to simulate up to order of $10^{5}$ atoms [\cite{Dragon_2024, DFTpy}]. Similar to the XC functionals, $\mathcal{F}_{\rm S}[n]$ can be grouped into local, semilocal, and non-local classes. On the level of the LDA, the simplest and the oldest approximation to $\mathcal{F}_{\rm S}[n]$ is the Thomas-Fermi model (e.g., see \cite{Mermin_1965, Stanton_pre_2015, Moldabekov_pop_2018}). 
A large number of semilocal level approximations for $\mathcal{F}_{\rm S}[n(\vec r)]$ were developed over the last years. Since we are interested in WDM with a certain fraction of electrons being delocalized due to partial ionization, the appropriate models for $\mathcal{F}_{\rm S}[n(\vec r)]$ for simulating WDM should be consistent with the UEG limit. A detailed discussion of such non-interacting free energy functionals is provided in a recent review paper by \cite{Mi_ChemRev_2023}. Although we focus mostly on the XC functionals in this work, we provide a few examples demonstrating the utility of the UEG limit for the construction of $\mathcal{F}_{\rm S}[n(\vec r)]$.


\subsubsection{Local and Semilocal XC Functionals}

We focus on the XC functionals for extended systems with a substantial or non-negligible part of electrons in quasi-free states.
Such systems are generated through ionization either at high temperatures or high densities. Other examples are metals and semiconductors with electrons in the valence bands under normal conditions. The XC functionals that are based on the properties of the UEG constitute an appropriate choice for such systems. A successful example from this class is the LDA based on ground state quantum Monte Carlo (QMC) results for the UEG:
\begin{equation}
\mathcal{F}_{\rm XC}[n] \approx \mathcal{F}_{\rm XC}^{\rm LDA}[n]= \int {\mathrm{d}\vec r}~n(\vec r)f^{\rm UEG}_{\rm XC}[n(\vec r)]    
\end{equation}
where $f^{\rm UEG}_{\rm XC}[n(\vec r)]$ is the XC free energy per particle of the UEG computed using the local value of the density $n(\vec r)$. 

At ambient conditions, accurate ground state QMC data for the UEG was provided by \cite{Ceperley_Alder_PRL_1980} and used to construct the analytical form for the LDA XC functional  [e.g., by \cite{Perdew_Zunger_PRB_1981} and by \cite{vwn}]. 
For finite temperatures, first QMC data for the warm dense UEG were provided by \cite{Brown_PRL_2013} based on the \emph{fixed node approximation}~[\cite{Ceperley1991}]. Subsequently, more accurate QMC results were reported by~\cite{Dornheim_prl_2016,Malone_PRL_2016,Schoof_PRL_2015}. These efforts have resulted in highly accurate parametrizations of the XC free energy of the UEG~[\cite{Karasiev_prl_2014,Groth_prl_2017,Dornheim_PhysRep_2018,Karasiev_status_2019}] that cover the entire WDM regime.


To improve LDA XC functionals, one needs to add information about non-locality, i.e., corrections due to the inhomogeneity of the density around a given point. The first step to go beyond LDA is the addition of the dependence of the XC functional on the density gradients. This is done in the class of XC functionals commonly referred to as generalized gradient approximation (GGA). In the case of the UEG, the density gradient corrections follow from the long-wavelength expansion of the density response function of the UEG, e.g., see [\cite{Stanton_pre_2015, Moldabekov_pop_2015}]. In the GGA level XC functionals, the UEG limit is combined with additional exact constraints such as Levy's uniform scaling condition [\cite{Levy_1989, Levy_pra_1991, Perdew_jcp_2014}] and the Lieb-Oxford bound [\cite{Lewin_pra_2015, Chan_pra_1999}] (for more details see, e.g, \cite{Perdew_pbcm_1991, Burke1998}). 
For solids as well as warm dense matter, arguably the most often used  GGA level functionals are PBE [\cite{PerdewPBE}] and PBEsol [\cite{Perdew_prl_2008}], which are designed using the ground state UEG data. Very recently, consistently thermal versions of the GGA have been presented by~\cite{Karasiev_prl_2018,kozlowski2023generalized}, and their application to real WDM applications is the subject of active investigation~[\cite{bonitz2024principles}].
Going beyond the GGA level, the next rung of XC functionals (meta-GGA) includes the dependence on the kinetic energy density $\tau(\br) = \sum_{j=0}^{\rm N_b}\frac{1}{2} \, f_j |\nabla \phi_j(\br)|^2$. An example of the often-used meta-GGA level functional is SCAN (strongly constrained and appropriately normed) [\cite{SCAN}], which is designed to obey various physical constraints (17 in total). Among these constraints is the linear response of the UEG in the limit of a homogeneous density [\cite{Tao_prb_2008}].  A finite-temperature version of SCAN that treats thermal effects on the level of the GGA was proposed recently by \cite{Karasiev_meta_GGA}.

\subsubsection{Non-local hybrid XC Functionals}

The GGA and meta-GGA functionals are semi-local since, at any given point, they only explicitly depend on the density distribution in the nearest vicinity of this point. Fully non-local XC functionals are expected to provide higher accuracy [\cite{PhysRevX.10.021040}]. An important class of fully non-local functionals is given by orbital-dependent hybrid XC functionals. Hybrid XC functionals are defined in the context of the so-called generalized Kohn-Sham density functional, and are of particular practical value for the simulation of material properties such as atomization and dissociation energies  [\cite{Goerigk_PCCP_2017}], lattice constants and bulk moduli [\cite{Heyd2003}], reflectivity [\cite{Goshadze_PRB_2023, Witte_prl_2017}], and conductivity [\cite{Witte_prl_2017}]. 
Arguably, the most often used hybrid XC functionals are constructed by combining exact Hartree-Fock (HF) exchange with a standard GGA level XC functional \cite{Perdew_jcp_1996, Becke}.
Indeed, the HF exchange naturally incorporates missing information about non-locality effects. Consequently, it was shown that hybrid XC functionals can lead to a more accurate description of various properties such as lattice constants [\cite{Paier}] and atomization energies  [\cite{Becke}].

Hybrid XC functionals can be derived starting from the \emph{adiabatic connection} formula for the XC free energy [\cite{Pittalis_prl_2011}]:
\begin{equation}\label{eq:adi_con}
    \mathcal{F}_{\rm XC}[n]=\int_0^{1}\frac{\mathrm{d}\lambda}{\lambda}\, \mathcal{W}_{\lambda}=\int_0^{1}{\mathrm{d}\lambda}\, \mathcal{U}_{\rm XC}^{\lambda}\,,
\end{equation}
where 
$\mathcal{W}_{\lambda}=V_{ee}^{\lambda}-V_H^{\lambda}$ is the XC potential energy of electrons with the pair potential $v_\lambda(\vec r,\vec r^{\prime})=\lambda/\left|\vec r-\vec r^{\prime}\right|$, 
$\lambda$ is the dimensionless coupling parameter controlling the interaction strength between electrons, and $V_H^{\lambda}$ is the corresponding Hartree potential energy of electrons. In Eq. (\ref{eq:adi_con}), following \cite{Becke}, we introduced 
$\mathcal{U}_{\rm XC}^{\lambda}= \mathcal{W}_{\lambda}/\lambda$ denoting the XC energy evaluated using the Coulomb potential $\lambda^{-1} v_\lambda(\vec r,\vec r^{\prime})$.
We note that the integration in Eq.~(\ref{eq:adi_con}) is to be performed with the density $n(\mathbf{r})$ being kept constant at the fully interacting value.

In practice, hybrid XC functionals are constructed by approximating the dependence of  $\mathcal{U}_{\rm XC}^{\lambda}$ on $\lambda$ in the form of a polynomial interpolation between the limits $\mathcal{U}_{\rm XC}^{\lambda=0}$ and $\mathcal{U}_{\rm XC}^{\lambda=1}$ [\cite{Becke, Perdew_jcp_1996}].

At $\lambda=0$, the exact value of $\mathcal{U}_{\rm XC}^{\lambda=0}$ is given by the HF exchange free energy (\textit{exact exchange}) [\cite{Janesko_pccp_2009, Mihaylov_prb_2020}]:
\begin{equation}\label{eq:HF_exchange}
\mathcal{U}_{\rm XC}^{\lambda=0}\equiv\mathcal{F}_{X}^{\rm HF}[\rho]=-\int \mathrm{d}\vec r_1 \int \mathrm{d}\vec r_2 \frac{\left|\rho\left(\vec r_1,\vec r_2\right)\right|^{2}}{\left| \vec r_1-\vec r_2\right|},
\end{equation}
where $\rho\left(\vec r_1,\vec r_2\right)$ is the exchange density,
\begin{equation}
    \rho\left(\vec r_1,\vec r_2\right)=\sum_j f_{\rm j}\phi_{\rm j}\left(\vec r_1\right)\phi_{\rm j}^{*}\left(\vec r_2\right),
\end{equation}
and  $f_{\rm i}$ is the occupation number defined by the Fermi–Dirac distribution.

At finite $\lambda$ values, $\mathcal{U}_{\rm XC}^{\lambda=1}$ can be approximated by using a GGA level XC density functional:
\begin{equation}\label{eq:U1_GGA}
    \mathcal{U}_{\rm XC}^{\lambda=1}\approx \mathcal{F}_{\rm XC}^{\rm GGA }[n]= \mathcal{F}_{\rm C}^{\rm GGA}[n]+ \mathcal{F}_{\rm X}^{\rm GGA}[n],
\end{equation}
where the XC functional is formally defined as the sum of the correlation and exchange parts [\cite{Pittalis_prl_2011, Perdew_jcp_1996}].

Following \cite{Perdew_jcp_1996}, the hybrid XC functionals constructed by using approximation Eq.(\ref{eq:U1_GGA}) and the following interpolation:
\begin{equation}\label{eq:V_l}
    \mathcal{U}_{\rm XC}^{\lambda}=\mathcal{F}_{\rm XC}^{\rm GGA}[n]+\left(\mathcal{F}_{X}^{\rm HF}[\rho]-\mathcal{F}_{\rm X}^{\rm GGA}[n]\right)\left(1-\lambda\right)^{m-1}.
\end{equation}

Using Eq. (\ref{eq:V_l}) in the adiabatic connection formula (\ref{eq:adi_con}), one arrives at:
\begin{equation}\label{eq:pbe0}
    \mathcal{F}_{\rm XC}[n]=\int_0^{1}\mathrm{d}\lambda\, \mathcal{U}_{\rm XC}^{\lambda}\approx \mathcal{F}_{\rm XC}^{\rm GGA}[n]+\frac{1}{m}\left(\mathcal{F}_{X}^{\rm HF}[\rho]- \mathcal{F}_{\rm X}^{\rm GGA}[n]\right),
\end{equation}
which is a finite temperature extension of the expression derived by \cite{Perdew_jcp_1996} for the ground state energy functional. 

In Eq. (\ref{eq:pbe0}), the Fermi-Dirac smearing of the occupation numbers in $F_{x}^{\rm HF}[\rho_{\sigma}]$  implicitly introduces the dependence on the electronic temperature.
Setting $m=4$  in Eq. (\ref{eq:pbe0}), and choosing the PBE functional [\cite{PerdewPBE}], one arrives at a finite temperature version of the PBE0 hybrid XC functional [\cite{Adamo1999}]. Setting $m=3$ in Eq. (\ref{eq:pbe0}), one deduces a finite temperature version of PBE0-1/3 [\cite{Cortona_jcp_2012}].
Similarly, defining the occupation numbers according to the Fermi–Dirac distribution allows one to integrate a part of the thermal exact-exchange contribution into other widely used hybrid XC functionals such as HSE06 [\cite{Heyd2006, Krukau2006}], HSE03 [\cite{Heyd2003}], and B3LYP [\cite{Stephens1994}]. 

In HSE06 and HSE03, to reduce the computational cost and convergence issues due to the long-range character of the Coulomb potential, a screened potential $1/r\to {\rm erfc}(\Omega r)/r$  in $\mathcal{F}_{\rm X}^{\rm HF}[\rho_{\sigma}]$ replaces the bare Coulomb potential  [\cite{Janesko_pccp_2009}]. In HSE06 and HSE03, the screening parameter is set to $\Omega=0.11~{\rm Bohr^{-1}}$ [\cite{Krukau2006}] and $\Omega=0.15/\sqrt{2}~{\rm Bohr^{-1}}$ [\cite{Heyd2003}].
For WDM applications, tHSE03 was shown to be valuable in describing the experimental X-ray spectra of warm dense aluminum [\cite{Witte_prl_2017}], and the reflectivity and dc electrical conductivity of liquid ammonia at high temperatures [\cite{Ravasio_prl_2021}]. 
Recently, \cite{Moldabekov_jcp_hybrid_2023} showed that the PBE0, PBE0-1/3, HSE06, HSE03 hybrid XC  functionals provide a significant improvement in the description of the electronic density response in the ground state as well as  WDM regime compared to the GGA level functionals.

Finally, an alternative promising route to construct fully non-local XC functional is the combination of the adiabatic connection formula with the fluctuation--dissipation theorem (AC-FDT)~[\cite{pribram}]. However, realistic simulations of warm dense matter using AC-FDT have not yet been achieved.

\begin{figure}[t!]
\centering
\includegraphics[width=1\textwidth]{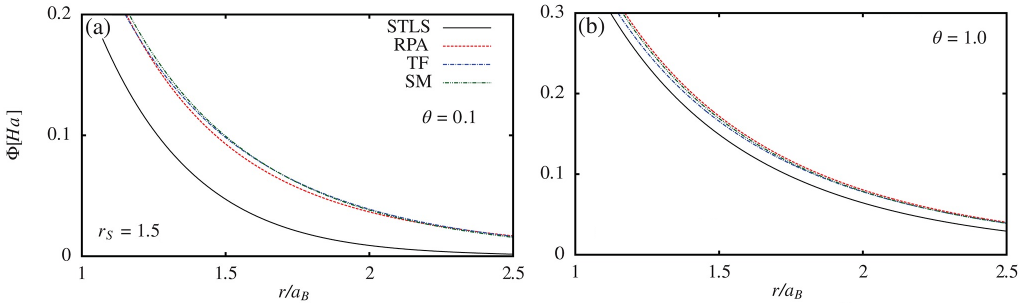}
\caption{Screened ion potential at $r_s=1.5$ and two temperatures. Left: $\theta =0.1$, right: $\theta=1.0$. 
The RPA curve corresponds to the potential of a test charge computed using random phase approximation for the dielectric function of the UEG. The Thomas-Fermi (TF) and Stanton-Murillo (SM) [\cite{Stanton_pre_2015, Moldabekov_pop_2015}] are the long wavelength approximations of the RPA result. The solid line corresponds to the the results computed taking into account electronic XC effect in the STLS approximation, which, at considered parameters, is in close agreement with the exact QMC results \cite{Moldabekov_pre_2018}. Adapted from  \cite{Moldabekov_cpp_2022}.}
\label{fig:potential}
\end{figure}

\begin{figure}
\centering
\includegraphics[width=0.5\textwidth]{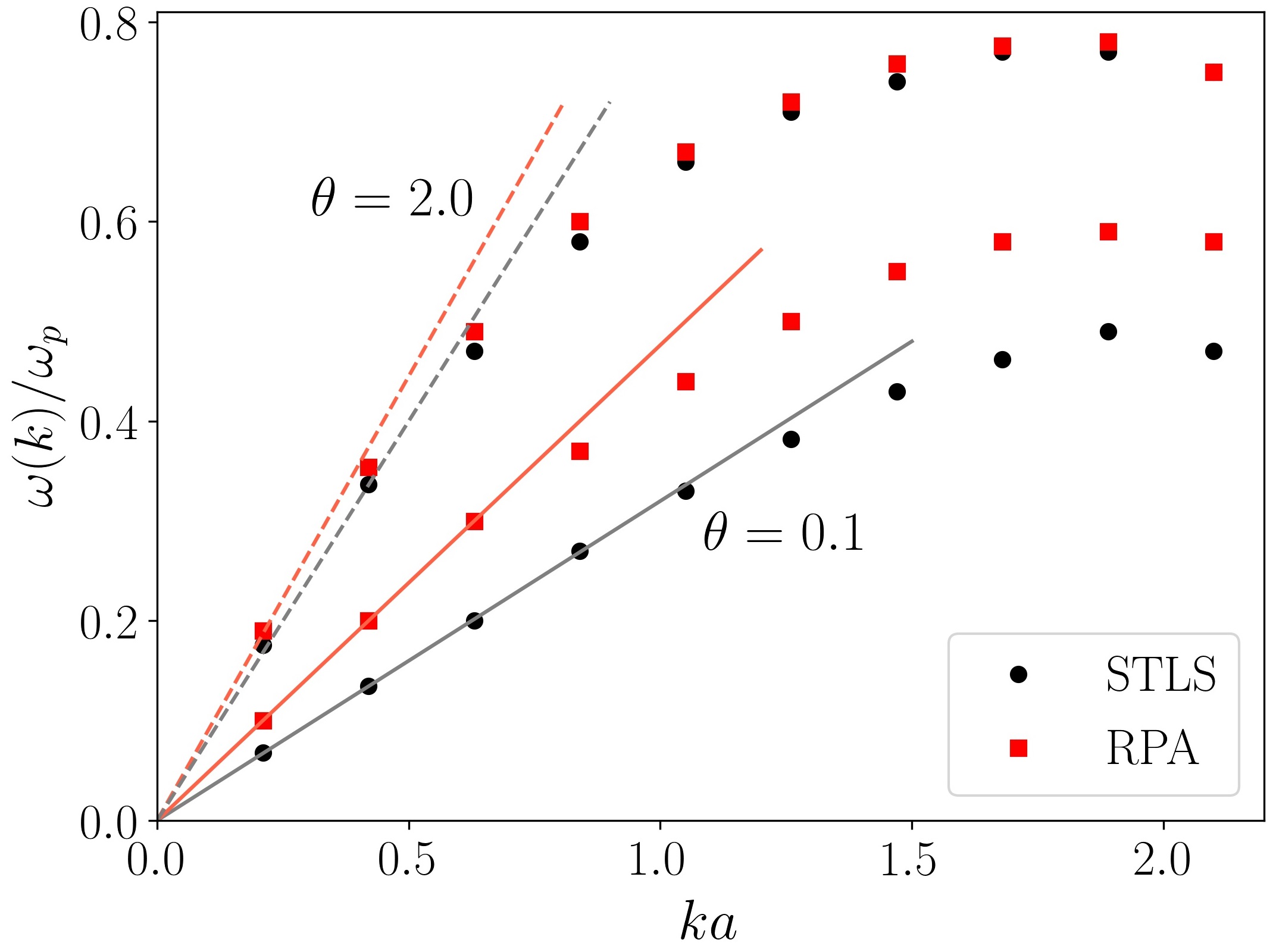}
\caption{Ion-acoustic peak positions at $r_s=1.5$ and different values of degeneracy parameter in warm dense hydrogen. Solid and dashed lines correspond to the linear dispersion at small wavenumbers. The frequency values are in the units of the plasma frequency of ions and the wavenumbers are in the units of the mean inter-ionic distance. Adapted from \cite{Moldabekov_pre_2019}.}
\label{fig:dispersion}
\end{figure}

\begin{figure}
\centering
\includegraphics[width=0.5\textwidth]{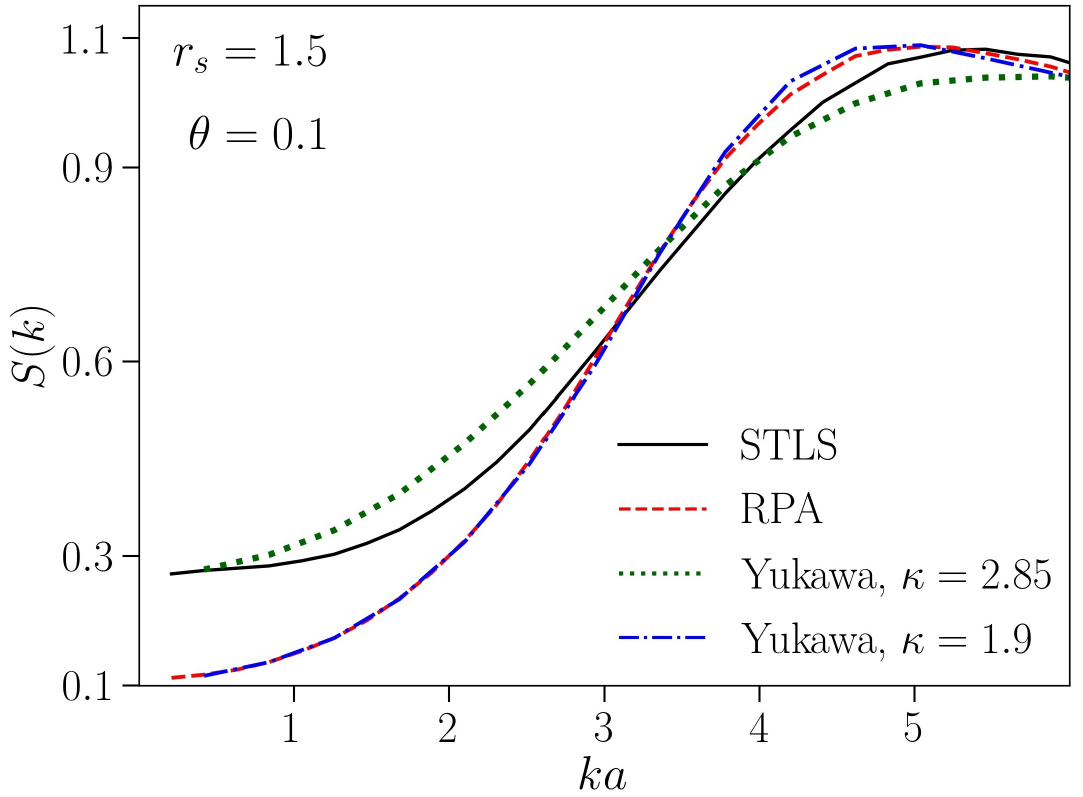}
\caption{Static structure factor of the ions in warm dense hydrogen. The STLS- and RPA-potential-based data are compared to the results of the  Yukawa model (the system of ions interacting via exponentially screened potential). For details, see \cite{Moldabekov_pre_2019}.}
\label{fig:Sii}
\end{figure}

\begin{figure}
\centering
\includegraphics[width=0.8\textwidth]{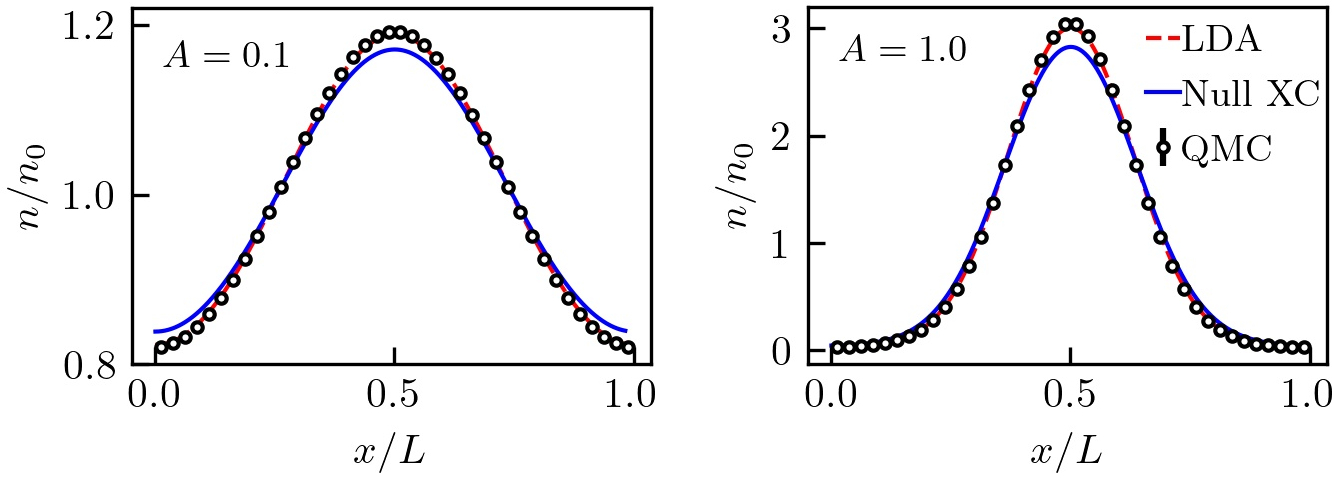}
\caption{\label{fig:NullXC}
The electron density along the perturbation direction of an external potential $2A\cos(\vec q\vec r)$ for two different amplitudes $A=0.1~[\rm Ha]$ (left) and $A=1.0~[\rm Ha]$ (right)  at wavenumber $q=0.843q_F$. The results are for strongly perturbed warm dense free electrons at $r_s=2$ and $\theta=1$. The comparison of the data computed using LDA XC functional and by setting XC functional to zero (Null XC) with the QMC data demonstrates the importance of the XC functional for the adequate calculation of the electronic density. The KS-DFT calculations were performed using GPAW [\cite{GPAW_2024, ASE_17}].  
}
\end{figure}





\subsection{Importance of the UEG limit for WDM simulations}\label{s:UEG_importance}

For WDM, XC functionals that incorporate the exact UEG limit as a constraint are particularly relevant.
First, the WDM regime is characterized by the partial or full ionization of atoms due to high temperatures and densities. The fraction of electrons that are weakly bound plays a decisive role in shaping structural characteristics [\cite{Gericke_pre_2010, Rudek_prl_2012, Moldabekov_pre_2018}], transport properties [\cite{White_PTRSA_2023, Moldabekov_pre_2019}], optical properties [\cite{Plagemann_2012}] and equation of state [\cite{Hu_pre_2014}]. This can be intuitively understood by considering the screening of a test charge and the electronic response to an external harmonic perturbation. 

First, let us illustrate the relevance of XC effects for the screening of the charge of ions and, as a consequence, their impact on transport properties and structural characteristics. To this end, we consider the screening of an ion charge (test charge) by a free electron gas. At large distances, the perturbation induced by the test charge is weak and one can describe the electronic screening cloud using linear response theory. The inclusion of electronic XC effects via a local field correction~[\cite{kugler1}] results in the stronger screening of the ion potential as illustrated in Fig. \ref{fig:potential} for $r_s=1.5$ at $\theta=0.1$ and $\theta=1$. 
Here, XC effects are taken into account based on the finite-temperature version of the
Singwi–Tosi–Land–Sj\"olander (STLS) approximation [\cite{tanaka_86}], which provides a close agreement with the exact QMC data at the considered parameters [\cite{Moldabekov_cpp_2022}].
In contrast, the random phase approximation (RPA) is exclusively based on the Lindhard density response function of the ideal electron gas and represents the screening without electronic correlations (i.e., with the XC kernel set to zero). In addition, we show results computed using long wavelength approximations to the dielectric function RPA according to the Thomas-Fermi model (TF) and the
model derived by Stanton and Murillo (SM)[\cite{Stanton_pre_2015}]. More details of the calculations are provided in Ref. [\cite{Moldabekov_cpp_2022}].
From Fig. \ref{fig:potential}, one can see that the increase in the temperature from $\theta=0.1$ to $\theta=1.0$ results in smaller corrections to the screening of the ion charge due to XC effects. To have a picture of the ramifications of the electronic XC-induced screening enhancement, we show results for the ion-acoustic mode in Fig. \ref{fig:dispersion}.
The ion-acoustic dispersion was calculated using the screened ion potential in molecular dynamics (MD). These calculations show that 
electronic XC effects lead to a $33~\%$ ($10 ~\%$) reduction of the ion-sound speed  $C_s$  at $\theta=0.1$ ($\theta=1.0$) [\cite{Moldabekov_pre_2019}]. In addition, we see that the importance of the electronic XC effects for $C_s$ is reduced with the increase in the temperature from $\theta=0.1$ to $\theta=2.0$. This observation and the data for the screened potential shown in Fig. \ref{fig:potential} demonstrate the general rule depicted in Fig. \ref{fig:diagram2}: The increase in the temperature of WDM results in the reduction of the relevance of the XC corrections.

The stronger screening of the potential of ions leads to an increase in the values of the static structure factor of ions at small wavenumbers $S(k\to0)$ as it is illustrated in Fig.\ref{fig:Sii} for $r_s=1.5$ and $\theta=0.1$. It increases about two times due to the inclusion of the electronic XC effects in the screening of the ion charge.  The $k\to 0$ limit of the static structure factor defines the compressibility $\mu$ of disordered systems  $ k_BT\mu=S(k\to0)$ [\cite{Jones_1973}] and, naturally, the equation of state. Furthermore, the failure to adequately describe screening might result in spurious effects such as ion-ion attraction in a dense plasma environment [\cite{Bonitz_pre_2013, Moldabekov_pop_2015}], as it was proposed by \cite{Shukla_prl_2012}.

The construction of XC functionals using the UEG limit is also crucial for the description of electronic density inhomogeneities induced by external fields.  This is illustrated in Fig. \ref{fig:NullXC}, where we show the density perturbation induced by an external cosinusoidal external field $2A\cos(\vec q\vec r)$ for two different amplitudes $A=0.1~[\rm Ha]$ (left panel) and $A=1.0~[\rm Ha]$ (right panel)  at the wavenumber $q=0.843q_F$ and compare the KS-DFT results to exact QMC data. 
In both cases, the inclusion of the XC functional (here using the ground state LDA) results in a substantial improvement in the accuracy of the density calculation. We stress that even in the case of a strong density inhomogeneity (compared to the uniform case) at $A=1.0$, the UEG limit-based LDA XC functional allows one to achieve highly accurate density calculations using KS-DFT. For applications at ambient conditions, the relevance of the UEG limit of the XC functional is discussed by, e.g.,  \cite{Tao_prb_2008},  \cite{PerdewPBE},  \cite{Perdew_prl_2008}, and \cite{SCAN}. 

\begin{figure}
\centering
\includegraphics[width=0.55\textwidth]{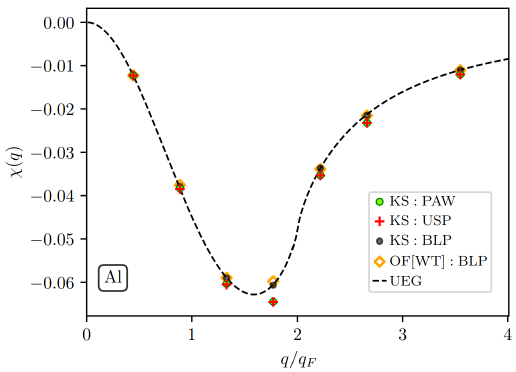}
\caption{Static density response function of the valence electrons in fcc Al at ambient conditions. 
KS-DFT and OF-DFT results are compared with UEG model calculations (dashed lines). 
The calculations were performed using pseudopotential from the projector augmented wave method (PAW), the ultrasoft pseudopotential (USP).
 and bulk-derived local pseudopotentials (BLP). For details, see Ref.[\cite{Moldabekov_Electronic_Structure_2024}].
The KS-DFT calculations were performed using Quantum ESPRESSO  [\cite{QE-2017, QE-2009}]. To compute the perturbed density and KS potential, we used the QEpy tool [\cite{qepy}] to introduce a static external harmonic perturbation. The OF-DFT results were calculated using the DFTpy code [\cite{DFTpy}].
Adapted from \cite{Moldabekov_Electronic_Structure_2024}. }
\label{fig:Al_chi}
\end{figure}

In Ref.~[\cite{Moldabekov_prb_2023}], it was shown recently that imposing the correct UEG limit of the non-interacting free energy functional $\mathcal{F}_{\rm S}[n(\vec r)]$ is critical for achieving high accuracy in modeling the density response to strong external perturbations in OF-DFT simulations. 
This is illustrated in Fig. \ref{fig:Al_chi} for the density response function of electrons in aluminum at ambient conditions.  
The electron density response function has been computed using the  Wang and Teter (WT) non-interacting (ideal) kinetic energy functional. The WT kinetic energy functional is derived using the Lindhard function describing the density response of the ideal Fermi gas [\cite{WT_paper}].
The OF-DFT results are compared with the KS-DFT data computed using different pseudopotentials (see Ref[\cite{Moldabekov_Electronic_Structure_2024}] for more details). From Fig.  \ref{fig:Al_chi}, it is evident that the UEG model WT functional allows to obtain good quality results for Al using the OF-DFT method.

To understand the connection of $\mathcal{F}_{\rm XC}[n(\vec r)]$ and 
$\mathcal{F}_{\rm S}[n(\vec r)]$ to linear density response properties, we provide a
derivation of the stiffness theorem in the following section \ref{ss:connection}. In the subsequent sections \ref{s:Test_electron_gas} and \ref{s:warm_dense_hydrogen}, we further expand the discussion of the relevance of the UEG limit for  $\mathcal{F}_{\rm XC}[n(\vec r)]$ and $\mathcal{F}_{\rm S}[n(\vec r)]$.

\section{Connection Between Energy Functionals and Density Response Functions}\label{ss:connection}

Let us next consider the change in the free energy of the system due to a weak perturbation of the equilibrium electron density. 
The connection between the energy response and the density response to a weak external perturbation is one of the key recipes for the construction of energy functionals. 
 In practice, the ionic dynamics in DFT are described using the Born-Oppenheimer approximation. This means that the ions are simulated using molecular dynamics (MD), while the electronic component is treated quantum mechanically using the DFT method.
In this approach, at every MD step, the equilibrium electronic density is computed in the field of fixed ions.
For uniform systems in equilibrium,  the ionic component of WDM has a disordered configuration for any given snapshot.
This results in a homogeneous charge distribution on average. 
For pedagogical reasons, we discuss the latter point in more detail after introducing the relevant concepts assuming density homogeneity.  

\subsection{Static density response of homogeneous systems}

The change in the KS potential (\ref{eq:v_tot_ks}) around the equilibrium value  due to the density perturbation  $\delta n$ has the following composition:

\begin{equation}\label{eq:delta_v_tot_ks}
    \Delta v_{\rm KS}[n](\vec r)= \Delta v_{\rm ext}[n](\vec r) + \Delta v_H[n](\vec r) + \Delta v_{\rm XC}[n](\vec r),
\end{equation}
or, using the Taylor expansion in terms of functional derivatives, we have in first order:
\begin{equation}\label{eq:delta_v_tot_ks2}
\begin{split}
    \int \left. \frac{\delta v_{\rm KS}[n](\vec r)}{\delta n(\vec r^{\prime})}\right\vert_{n=n_{\rm eq}} \delta n(\vec r^{\prime})~\mathrm{d}\vec r^{\prime}&= \int
    \left. \frac{\delta v_{\rm ext}[n](\vec r)}{\delta n(\vec r^{\prime})}\right\vert_{n=n_{\rm eq}} \delta n(\vec r^{\prime})~\mathrm{d}\vec r^{\prime} + \int
    \left. \frac{\delta v_H[n](\vec r)}{\delta n(\vec r^{\prime})}\right\vert_{n=n_{\rm eq}} \delta n(\vec r^{\prime})~\mathrm{d}\vec r^{\prime} \\
    &+ \int
    \left. \frac{\delta v_{\rm XC}[n](\vec r)}{\delta n(\vec r^{\prime})}\right\vert_{n=n_{\rm eq}} \delta n(\vec r^{\prime})~\mathrm{d}\vec r^{\prime},
    \end{split}
\end{equation}
where $n=n_{\rm eq}$ indicates that the functional derivative is computed for the equilibrium system.

From Eq. (\ref{eq:delta_v_tot_ks2}), it immediately follows that 
\begin{equation}\label{eq:delta_v_tot_ks3}
\begin{split}
     \left. \frac{\delta v_{\rm KS}[n](\vec r)}{\delta n(\vec r^{\prime})}\right\vert_{n=n_{\rm eq}} &= 
    \left. \frac{\delta v_{\rm ext}[n](\vec r)}{\delta n(\vec r^{\prime})}\right\vert_{n=n_{\rm eq}}  + 
    \left. \frac{\delta v_H[n](\vec r)}{\delta n(\vec r^{\prime})}\right\vert_{n=n_{\rm eq}}  + 
    \left. \frac{\delta v_{\rm XC}[n](\vec r)}{\delta n(\vec r^{\prime})}\right\vert_{n=n_{\rm eq}},
    \end{split}
\end{equation}
which is the functional derivative of the KS potential around the equilibrium density.

Next, we connect the variation in the KS potential with the density response function.
Within linear response theory, the static density response of a homogeneous system to the weak static external perturbation $\delta v_{\rm ext}=\delta v_{\rm ext}[n]$ reads
\begin{equation}\label{eq:chi_tot_r}
    \delta n(\vec r)=\int \chi (\vec r -\vec r^{\prime}) \delta v_{\rm ext}(\vec r^{\prime})~\mathrm{d}\vec r^{\prime},
\end{equation}
where $\chi (\vec r -\vec r^{\prime})$ is the static density response function. We note that Eq. (\ref{eq:chi_tot_r}) is the general definition of the density response function for homogeneous systems and is not limited to the KS model system. 

From Eq. (\ref{eq:chi_tot_r}), we find by taking the functional derivative with respect to density:
\begin{equation}\label{eq:chi_tot_r_new}
    \frac{\delta n(\vec r)}{\delta n(\vec r^{\prime \prime})}=\delta (\vec r - \vec r^{\prime \prime}) =
    \int \chi \left(\vec r -\vec r^{\prime }\right) \frac{\delta v_{\rm ext}(\vec r^{\prime })}{\delta n(\vec r^{\prime \prime})}~\mathrm{d}\vec r^{\prime }.
\end{equation}

Taking the Fourier transform of Eq. (\ref{eq:chi_tot_r_new}) and using the convolution theorem, we have 
\begin{equation} \label{eq:chi_tot_q_r}
  \widehat  F\left[ \frac{\delta v_{\rm ext}(\vec r^\prime)}{\delta n(\vec r^{\prime \prime})}\right]= \frac{1}{\chi (\vec q)} ,
\end{equation}
where  $\widehat  F$ is the operator of the Fourier transform.

Similarly, one can consider the density response of the non-interacting KS system to a weak perturbation in the KS potential $\delta v_{\rm KS}=\delta v_{\rm KS}[n]$ and introduce the KS response function [\cite{Kollmar}],
\begin{equation}\label{eq:chi_KS_r}
    \delta n(\vec r)=\int \chi_{\rm KS} (\vec r -\vec r^{\prime}) \delta v_{\rm KS}(\vec r^{\prime})~\mathrm{d}\vec r^{\prime},
\end{equation}
where $\chi_{\rm KS} (\vec r -\vec r^{\prime})$ is the static  KS density response function.

In the same way as we derived Eq. (\ref{eq:chi_tot_q_r}) from Eq. (\ref{eq:chi_tot_r}), we find from Eq. (\ref{eq:chi_KS_r}) the connection between the variation of the KS potential in real space and the inverse KS density response function in Fourier space:  
\begin{equation} \label{eq:chi_KS_q_r}
   \widehat  F\left[ \frac{\delta v_{\rm KS}[n](\vec r)}{\delta n(\vec r^{\prime})}\right]= \frac{1}{\chi_{\rm KS} (\vec q)} .
\end{equation}

For the functional derivative of the Hartree mean field energy density $\delta v_{H}=\delta v_{H}[n]$ with respect to the density, we use the well-known result:
\begin{equation} \label{eq:coulomb_q}
   \widehat  F\left[ \frac{\delta v_{\rm H}(\vec r)}{\delta n(\vec r^{\prime})}\right]= \frac{4\pi e^2}{q^2}.
\end{equation}

Taking the Fourier transform of Eq.~(\ref{eq:delta_v_tot_ks3}) and utilizing Eq.~(\ref{eq:chi_tot_q_r}), Eq.~(\ref{eq:chi_KS_q_r}), and Eq.~(\ref{eq:coulomb_q}) yields
\begin{equation}\label{eq:inverse_chi}
    \frac{1}{\chi_{\rm KS}(\vec q)}=\frac{1}{\chi (\vec q)}+\frac{4\pi e^2}{q^2}+\widehat  F\left[ \frac{\delta v_{\rm XC}(\vec r)}{\delta n(\vec r^{\prime})}\right].
\end{equation}

The last term on the right-hand side of Eq. (\ref{eq:inverse_chi}) is referred to as the XC kernel,
\begin{equation}\label{eq:Kxc_def}
    K_{\rm xc}(\vec q)=\widehat  F\left[ \frac{\delta v_{\rm XC}(\vec r)}{\delta n(\vec r^{\prime})}\right]=
    \widehat  F\left[ \frac{\delta^2 \mathcal{F}_{\rm XC}[n]}{\delta n(\vec r)\delta n(\vec r^{\prime})}\right],
\end{equation}
where Eq. (\ref{eq:v_ks}) is used for $v_{\rm XC}(\vec r)$.

Substituting Eq. (\ref{eq:inverse_chi}) into Eq. (\ref{eq:Kxc_def}) and re-arranging terms, we arrive at the important relation
\begin{equation}\label{eq:Kxc_chi_conn}
    K_{\rm XC}(\vec q)=\frac{1}{\chi_{\rm KS}(\vec q)}-\frac{1}{\chi (\vec q)}-\frac{4\pi e^2}{q^2},
\end{equation}
which we utilize to benchmark XC functionals and which allows one to use equilibrium KS-DFT for the calculation of the XC kernel in the cases when the explicit calculation of the functional derivative in Eq. (\ref{eq:Kxc_def}) is not possible.  
This is valuable e.g.~for linear response time-dependent DFT (LR-TDDFT) calculations of the dynamic density response function [\cite{Moldabekov_prr_2023, Moldabekov_jcp_2023_averaging}]. 

In the UEG limit, the static density response function is given by [\cite{quantum_theory}]:
\begin{equation}\label{eq:inv_chi_ueg}
     \frac{1}{\chi (\vec q)}=\frac{1}{\chi_{0}(\vec q)}-\frac{4\pi e^2}{q^2}-K_{\rm XC}(\vec q),
\end{equation}
where $\chi_0(\vec q)$ is the Lindhard function. Therefore, 
it holds $\chi_{\rm KS}(\vec q)=\chi_0(\vec q)$ in this case.

\subsection{Static density response of inhomogeneous disordered systems}

Let us next consider in more detail the homogeneity of the density of a disordered system after averaging over different ionic configurations. Without loss of generality, we consider the perturbation by a harmonic external field  $v_{\rm ext}=2A\cos(\vec q \vec r)$.
As an example, we consider the scenario where the dynamics of a certain number of ions (atoms) are simulated in a cubic box with a side length $L$ and with periodic boundary conditions, e.g., using the MD method. 
In such simulations, disordered systems exhibit various ionic configurations at different times, meaning that the electronic density distributions ${n_{A=0}^{i}(\mathbf{r})}$ for different ionic configurations are not equivalent (with $i$ being the label of a given ionic configuration).
The averaged value of ${n_{A=0}^{i}(\mathbf{r})}$ over large number of ionic configurations tends to a constant 
\begin{equation}\label{eq:n0}
    \braket{n_e(\mathbf{r})}_{A=0}=\lim_{N_s \to \infty}\frac{1}{N_s}\sum_{i=1}^{N_s} {n_{A=0}^{i}(\mathbf{r})} = n_0,
\end{equation}
where  $N_s$ is the number of ionic configurations (snapshots).

For a weak perturbation, we have, in the first order, the following  induced  change in the  density 
\begin{equation}\label{eq:expansion}
\Delta n_{\mathbf{q},A}^{i}(\mathbf{r})={n_{A}^{i}(\mathbf{r})}-{n_{A=0}^{i}(\mathbf{r})}= 2 \sum_{\vec G}
{\rho_{\vec G}^{i}(\vec q)}
 \textnormal{cos}\big(
 \left(\mathbf{q}+\mathbf{G}\right)\cdot\mathbf{r}
 \big)\ ,    
\end{equation}
where $\vec G$ is the reciprocal lattice vector with the length $2\pi/L$, $\rho_{\vec G}^{i}(\vec q)$ is the Fourier component of the density perturbation,  and the factor two is conventional. 

After averaging over ionic configurations, only the term with $\vec G=0$ survives for disordered systems,
\begin{equation}\label{eq:delta_n}
\begin{split}
     \Delta n(\mathbf{r})_{\mathbf{q}, A} &= \lim_{N_s \to \infty} \frac{1}{N_s}\sum_{i=1}^{N_s} \Delta n_{\mathbf{q},A}^{i}(\mathbf{r})=  2\left(\lim_{N_s \to \infty} \frac{1}{N_s} \sum_{i=1}^{N_s} {\rho_{\vec G=0}^{i}(\vec q)}\right)  \textnormal{cos}\left(
\mathbf{q}\cdot\mathbf{r}
 \right)=2\braket{\rho(\vec q)}_{\vec G=0}  \textnormal{cos}\left(
\mathbf{q}\cdot\mathbf{r}
 \right)\ ,
\end{split}
\end{equation}
as for all other terms with $\vec G\neq 0$,  we have after averaging,
\begin{equation}\label{eq:rho_to_0}
    \braket{\rho(\vec q)}_{\vec G\neq0}=\lim_{N_s \to \infty} \frac{1}{N_s}\sum_{i=1}^{N_s} {\rho_{\vec G\neq0}^{i}(\vec q)} =0\ ,
\end{equation}
where $\braket{\rho(\vec q)}_{\vec G\neq0}$ denotes the average value of the density perturbation components induced for reciprocal lattice vectors with $\vec G\neq0$. 

\begin{figure} 
\centering
\includegraphics[width=0.9\textwidth]{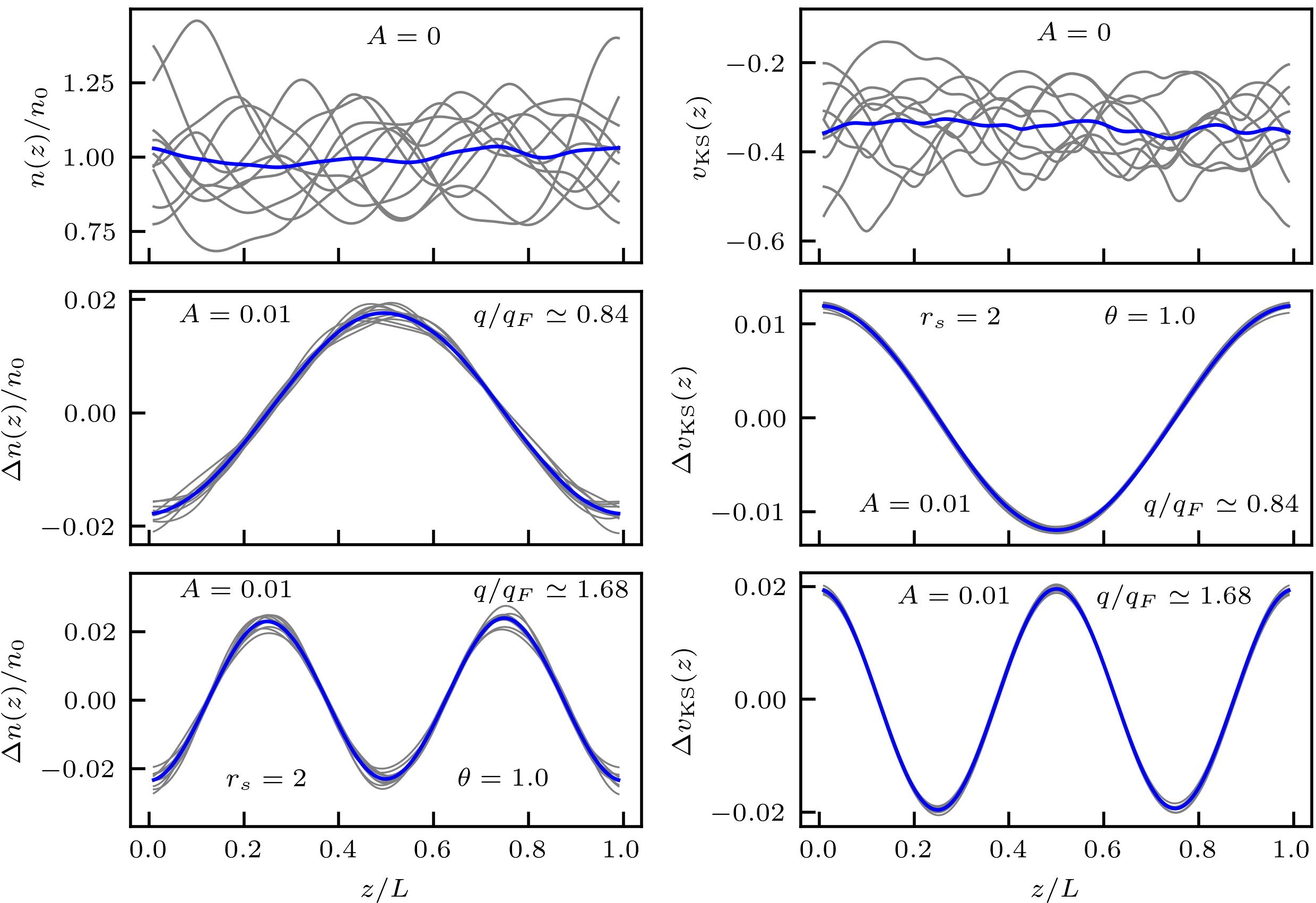}
\caption{ \label{fig:averaging_n_ks} Top panels: Density distribution (left) and KS potential distribution (right) along the z-axis of the unperturbed system. Middle panels:  density perturbation (left) and KS potential distribution (right) for the perturbation amplitude $A=0.01$ and wavenumber $q\simeq 0.84q_F$, where $q_F$ is the Fermi wavenumber computed using the mean density of electrons. Bottom panels: density perturbation (left) and KS potential distribution (right) for the perturbation amplitude $A=0.01$ and wavenumber $q\simeq 1.68q_F$. The results are computed for warm dense hydrogen with 14 protons in the main simulation cell at $r_s=2$ and $\theta=1$. From Ref. [\cite{Moldabekov_jcp_2023_averaging}]. The KS-DFT calculations were performed using Abinit [\cite{Abinit_2020}]. }
\end{figure}


Therefore, in the linear response regime, we have the static linear density response function $\chi(\mathbf{q})$ of a disordered system (that is homogeneous on average):
\begin{eqnarray}\label{eq:delta_n_LRT}
   \Delta n(\mathbf{r})_{\mathbf{q},A}  = 2 A \textnormal{cos}\left(\mathbf{q}\cdot\mathbf{r}\right)\chi(\mathbf{q})\ ,
\end{eqnarray}
where 
\begin{equation}\label{eq:my_chi}
    \chi(\mathbf{q})=\frac{1}{N_s}\sum_{i=1}^{N_s} \chi_{\vec G=0}^{i}(\mathbf{q})=\frac{\braket{\rho(\vec q)}_{\vec G=0}}{A}\ .
\end{equation}

In the same way as for the density distribution, we have for the averaged value of the KS potential:    
\begin{equation}\label{eq:v_ks_0}
    \braket{v_{\rm KS}(\mathbf{r})}_{A=0}=\lim_{N_s \to \infty}\frac{1}{N_s}\sum_{i=1}^{N_s} {v_{\rm KS,~A=0}^{i}(\mathbf{r})} = v_{\rm KS}^{0},
\end{equation}
where $v_{\rm KS,~A=0}^{i}(\mathbf{r})\neq {\rm const}$  is the KS potential of an individual ionic snapshot. 

The application of the cosinusoidal external perturbation results in the change of the KS potential, which can be represented  in a Fourier cosine series,
\begin{equation}\label{eq:expansion_vks}
\Delta v_{\rm KS, A}^{i}(\mathbf{r})={v_{\rm KS, A}^{i}(\mathbf{r})}-{v_{\rm KS, A=0}^{i}(\mathbf{r})}= 2 \sum_{\vec G}
{u_{\vec G}^{i}(\vec q)}
 \textnormal{cos}\big(
 \left(\mathbf{q}+\mathbf{G}\right)\cdot\mathbf{r}
 \big)\ .  
\end{equation}

Using the same reasoning as for the electron density, we find for the mean value of the perturbation of the KS potential
\begin{equation}
\begin{split}
      \Delta v_{\rm KS}(\mathbf{r})_{\mathbf{q},A} &= \lim_{N_s \to \infty} \frac{1}{N_s}\sum_{i=1}^{N_s} \Delta v_{\rm KS, A}^{i}(\mathbf{r})=2 \left(\lim_{N_s \to \infty} \frac{1}{N_s}\sum_{i=1}^{N_s} {u_{\vec G=0}^{i}(\vec q)}\right) \textnormal{cos}\left(\mathbf{q}\cdot\mathbf{r}\right)
      = 2 \braket{u(\vec q)}_{\vec G=0} \textnormal{cos}\left(\mathbf{q}\cdot\mathbf{r} \right)
\end{split} \label{eq:delta_vks}
\end{equation}
where $ {u_{\vec G=0}^{i}(\vec q)}$ denotes the Fourier component of the KS potential perturbation at $\vec G=0$.

Now, we introduce the KS response function describing the density response $\Delta n(\mathbf{r})_{\mathbf{q}, A}$ of the system to the perturbation in the KS potential $ \Delta v_{\rm KS}(\mathbf{r})_{\mathbf{q},A}$ on average as
\begin{eqnarray}\label{eq:delta_KS}
   \Delta n_{\mathbf{q},A}(\mathbf{r}) = \chi_{\rm KS}(\mathbf{q}) \Delta v_{\rm KS}(\mathbf{r})_{\mathbf{q},A}\ .
\end{eqnarray}

We stress that  the KS response function defined by Eq. (\ref{eq:delta_KS}) is not equivalent to a direct average over KS response functions computed for individual ionic configurations, i.e.,
\begin{equation}\label{eq:i_chi_KS}
     \chi_{\rm KS}(\mathbf{q})\neq  \braket{\chi_{\rm KS}(\mathbf{q})}=\frac{1}{N_s}\sum_{i=1}^{N_s} \chi_{\rm KS, \vec G=0}^{i}(\mathbf{q})\ .
\end{equation}

It was shown by \cite{Moldabekov_jcp_2023_averaging}  that the KS response function defined by Eq. (\ref{eq:delta_KS}) is consistent with the linear response theory formulation in the limit of the UEG. In contrast,  $\braket{\chi_{\rm KS}(\mathbf{q})}$ defined in Eq. (\ref{eq:i_chi_KS}) results in inconsistency with the standard linear response theory of the UEG. 

The validity of Eq. (\ref{eq:n0}), Eq. (\ref{eq:delta_n}), and Eq. (\ref{eq:rho_to_0}) is demonstrated numerically in Fig. \ref{fig:averaging_n_ks}, where the results from the KS-DFT based molecular dynamics simulations of warm dense hydrogen are presented at $\theta=1$ for $r_s=2$ [\cite{Moldabekov_jcp_2023_averaging}]. Specifically, we show the unperturbed electron density and KS potential (top panels) as well as the perturbations in the density and KS potential due to an external harmonic field (middle and bottom panels). The results were computed for 10 different snapshots (grey curves) with 14 protons in each.
The solid blue lines depict the corresponding averaged values. From the top panels of Fig. \ref{fig:averaging_n_ks}, one can clearly see that the densities and KS potentials of individual ionic configurations are inhomogeneous and that averaging over ionic configurations leads to a homogeneous density and KS potential profiles. In the middle and bottom panels, it is demonstrated that averaging results in a cancellation of the small deviations from the cosinusoidal shape of the density and KS potential perturbations. In other words, $\Delta n(\mathbf{r})_{\mathbf{q}, A}$ and $ \Delta v_{\rm KS}(\mathbf{r})_{\mathbf{q},A}$ follow the shape of the external perturbation. 

\section{Stiffness Theorem for Free Energy Functionals}\label{s:stiffness_section}

The first accurate QMC results for the static density response and the XC kernel of the UEG in the ground state were extracted from the change in the energy of electrons due to a weak harmonic external perturbation by \cite{Moroni_prl_1995}. The relation linking the change in the energy due to an external perturbation to the static density response function is known as the stiffness theorem [\cite{quantum_theory}].
 The same relation was used to compute the static density response function of 2D UEG and 3D $\ce{^{4}He}$ at $T=0$ [\cite{Moroni_prl_1992}]. The first ground state QMC results for the XC kernel of UEG  by \cite{Moroni_prl_1995} have been of paramount value for the design of GGA and meta-GGA functionals (e.g. see Ref. [\cite{SCAN, PBEsol}]). A detailed discussion of the stiffness theorem for $T=0$ can be found, e.g., in the textbook by  \cite{quantum_theory}. In contrast, the stiffness theorem for applications at finite temperatures has been sparsely discussed in the literature [\cite{Dornheim_jcp_2023, Moldabekov_pop_2018}].   Here, we provide a detailed discussion of the extension of the stiffness theorem for the free energy.
 Furthermore, to the best of our knowledge, we show the stiffness theorem for the intrinsic part of free energy for the first time. 

\subsection{Homogeneous systems}

Within the canonical ensemble, the minimization of the free energy in Eq. (\ref{eq:F}) leads to [\cite{Hansen_book}]
\begin{equation}\label{eq:Fmin}
    \frac{\delta F^{i}[n]}{\delta n^{i}(\vec r)}=0~\Longrightarrow \frac{\delta \mathcal{F}^{i}[n]}{\delta n(\vec r)} = v^{i}(\vec r),
\end{equation}
where $v^{i}(\vec r)$ is the sum of the Hartree mean field and an external field for a given configuration of ions.

Since the density is constant on average and the system is charge neutral in total, after averaging over snapshots, we write  
\begin{equation}\label{eq:Fmin_av}
    \left<\frac{\delta \mathcal{F}^{i}[n]}{\delta n(\vec r)}\right> = \frac{\delta \left<\mathcal{F}^{i}[n]\right>}{\delta n(\vec r)}= \lim_{N_s \to \infty}\frac{1}{N_s}\sum_{i=1}^{N_s} {v^{i}(\mathbf{r})} =const.
\end{equation}

Next, let us consider the change in the free energy on average due to some external perturbation, for which, taking into account total charge neutrality, we have:
\begin{eqnarray}\label{eq:F_tot_for_p}
     \Delta \left<F[n]\right>= \left<F[n]\right>-\left<F_0[n]\right>=\Delta \left<\mathcal{F}[n]\right> +\int  \left<\Delta n(\vec r)\right>  \left<\Delta v(\vec r)\right>~{\mathrm d}\vec r ,
\end{eqnarray}
where $\left<F[n]\right>$ and $\left<F_0[n]\right>$ are correspondingly  the free energy of the perturbed and unperturbed systems,  $\Delta \left<\mathcal{F}[n]\right>$ is the change in the intrinsic free energy, and 
\begin{equation}\label{eq:def_v_av}
    \left<\Delta v(\vec r)\right>=\frac{1}{2}\int  \frac{\left<\Delta n(\vec r^{\prime})\right>}{|\vec r-\vec r^{\prime}|} ~{\mathrm d}\vec r^{\prime} +\left<\Delta v_{\rm ext}(\vec r)\right>.
\end{equation}

In the second order,  the Taylor expansion in terms of the functional derivatives for the change in the intrinsic free energy yields
\begin{eqnarray}\label{eq:int_F_exp}
    \Delta \left<\mathcal{F}[n]\right>=\int \left. \frac{\delta \left<\mathcal{F}[n]\right>}{\delta n(\vec r)}\right\vert_{n=n_{\rm eq}} \delta n(\vec r)~\mathrm{d}\vec r+\frac{1}{2}
    \int \left. \frac{\delta^2 \left<\mathcal{F}[n]\right>}{\delta n(\vec r^{\prime})\delta n(\vec r)}\right\vert_{n=n_{\rm eq}} \delta n(\vec r^{\prime})\delta n(\vec r)~\mathrm{d}\vec r\mathrm{d}\vec r^{\prime}
\end{eqnarray}

Using Eq. (\ref{eq:Fmin_av}) and $\int \delta n(\vec r){\mathrm{d}}\vec r=0$,
from Eq. (\ref{eq:int_F_exp}) we find
\begin{eqnarray}\label{eq:int_F_exp2}
    \Delta \left<\mathcal{F}[n]\right>=\frac{1}{2}
    \int \left. \frac{\delta^2 \left<\mathcal{F}[n]\right>}{\delta n(\vec r^{\prime})\delta n(\vec r)}\right\vert_{n=n_{\rm eq}} \delta n(\vec r^{\prime})\delta n(\vec r)~\mathrm{d}\vec r\mathrm{d}\vec r^{\prime}.
\end{eqnarray}

Substituting Eq. (\ref{eq:int_F_exp2}) and  Eq. (\ref{eq:def_v_av}) into Eq. (\ref{eq:F_tot_for_p}), we get
\begin{eqnarray}\label{eq:F_tot_for_p2}
     \Delta \left<F[n]\right>= \frac{1}{2}
    \int \left. \frac{\delta^2 \left<\mathcal{F}[n]\right>}{\delta n(\vec r^{\prime})\delta n(\vec r)}\right\vert_{n=n_{\rm eq}} \delta n(\vec r^{\prime})\delta n(\vec r)~\mathrm{d}\vec r\mathrm{d}\vec r^{\prime} &+&\frac{1}{2}\int  \int    \frac{\left<\Delta n(\vec r)\right> \left<\Delta n(\vec r^{\prime})\right>}{|\vec r-\vec r^{\prime}|} ~{\mathrm d}\vec r^{\prime}~{\mathrm d}\vec r \nonumber\\
    &+&\int\left<\Delta n(\vec r)\right> \left<\Delta v_{\rm ext}(\vec r)\right>~{\mathrm d}\vec r .
\end{eqnarray}

Taking the functional derivative of Eq. (\ref{eq:F_tot_for_p2}) and using  the free energy minimization conditions for both unperturbed and perturbed systems give us
\begin{eqnarray}\label{eq:int_F_exp3}
    \left<\Delta v_{\rm ext}(\vec r)\right>+\int  \frac{\left<\Delta n(\vec r^{\prime})\right>}{|\vec r-\vec r^{\prime}|} ~{\mathrm d}\vec r^{\prime}=-
    \int \left. \frac{\delta^2 \left<\mathcal{F}[n]\right>}{\delta n(\vec r^{\prime})\delta n(\vec r)}\right\vert_{n=n_{\rm eq}} \delta n(\vec r^{\prime})~\mathrm{d}\vec r^{\prime}.
\end{eqnarray}

From Eq. (\ref{eq:int_F_exp3}), taking one more functional derivative, we write
\begin{eqnarray}\label{eq:dF_chi_rr}
    \frac{\delta \left<\Delta v_{\rm ext}(\vec r)\right>}{\delta n(\vec r^{\prime})}+\frac{1}{|\vec r-\vec r^{\prime}|}=-
     \left. \frac{\delta^2 \left<\mathcal{F}[n]\right>}{\delta n(\vec r^{\prime})\delta n(\vec r)}\right\vert_{n=n_{\rm eq}}.
\end{eqnarray}

We can express Eq.~(\ref{eq:dF_chi_rr}) in terms of the density response function. For that, we first take the inverse Fourier transform of  Eq.~(\ref{eq:chi_tot_q_r}) to express
\begin{eqnarray}\label{eq:chi_inv_rr}
    \left<\Delta v_{\rm ext}(\vec r)\right>=\int \left[\chi(\vec r-\vec r^{\prime})\right]^{-1} \left<\Delta n(\vec r^{\prime})\right>~{\mathrm d}\vec r^{\prime},
\end{eqnarray}
with $\left[\chi(\vec r-\vec r^{\prime})\right]^{-1}$ being defined as the inverse Fourier transform of $\chi^{-1}(\vec q)$,
\begin{equation}
    \left[\chi(\vec r-\vec r^{\prime})\right]^{-1}=\widehat  F^{-1}\left[ \frac{1}{\chi(\vec q)}\right].
\end{equation}

The functional derivative of Eq. (\ref{eq:chi_inv_rr}) yields
\begin{equation}\label{eq:d_chi_inv_rr}
    \frac{\delta \left<\Delta v_{\rm ext}(\vec r)\right>}{\delta n(\vec r^{\prime})}=\left[\chi(\vec r-\vec r^{\prime})\right]^{-1}.
\end{equation}

In this way, from Eq.(\ref{eq:d_chi_inv_rr}) and Eq.(\ref{eq:dF_chi_rr}), we deduce 
\begin{eqnarray}\label{eq:int_F_exp3_v22}
   \left[\chi(\vec r-\vec r^{\prime})\right]^{-1}+\frac{1}{|\vec r-\vec r^{\prime}|}=-
     \left. \frac{\delta^2 \left<\mathcal{F}[n]\right>}{\delta n(\vec r^{\prime})\delta n(\vec r)}\right\vert_{n=n_{\rm eq}},
\end{eqnarray}
which in the Fourier space reads
\begin{eqnarray}\label{eq:dF_chi_qr}
    \frac{1}{\chi(\vec q)}+\frac{4\pi}{q^2}=-
    \widehat  F\left[ \left. \frac{\delta^2 \left<\mathcal{F}[n]\right>}{\delta n(\vec r^{\prime})\delta n(\vec r)}\right\vert_{n=n_{\rm eq}}\right].
\end{eqnarray}

Before proceeding further, let us take a moment to consider Eq. (\ref{eq:dF_chi_qr}) from the KS-DFT perspective.
Using the partition of the intrinsic free energy into non-interacting and XC parts, from Eq. (\ref{eq:dF_chi_qr}), we 
find the connection of the non-interacting free energy $\mathcal{F}_{\rm S}[n]$ with the KS density response function:
\begin{eqnarray}\label{eq:dFs_chi_ks}
    -\widehat  F\left[ \left. \frac{\delta^2 \left<\mathcal{F}_{\rm S}[n]\right>}{\delta n(\vec r^{\prime})\delta n(\vec r)}\right\vert_{n=n_{\rm eq}}\right]=\frac{1}{\chi(\vec q)}+\frac{4\pi}{q^2}+K_{\rm XC}(\vec q)=\frac{1}{\chi_{\rm KS}(\vec q)},
\end{eqnarray}
where definitions (\ref{eq:Kxc_def}) and (\ref{eq:Kxc_chi_conn}) were used.

From  Eq.(\ref{eq:dFs_chi_ks}), it is evident that the non-interacting free energy of the KS system has an intrinsic connection to the XC functional.  The relation (\ref{eq:dFs_chi_ks}) is the key ingredient in constructing the non-interacting free energy functionals for the OF-DFT applications. 
In the limit of the UEG, due to Eq. (\ref{eq:inv_chi_ueg}), one finds that  $\chi_{\rm KS}(\vec q)=\chi_{0}(\vec q)$. Numerous non-interacting free energy functionals were constructed using the Lindhard function  $\chi_{0}(\vec q)$ in Eq. (\ref{eq:dFs_chi_ks}) [e.g., see \cite{Mi_ChemRev_2023, Huang_prb_2010, Moldabekov_prb_2023}]. This UEG limit-based non-interacting free functionals turned out to be accurate even for the systems with a strong density inhomogeneity. In addition, the connection (\ref{eq:dFs_chi_ks}) can be used to construct non-interacting free energy functionals using the results for the KS response function from KS-DFT.  Further discussions of this topic are provided in Sec. \ref{s:of_dft_test}. 
 
In practice, we note that the polarization function is often used instead of the density response function:
\begin{equation}
    \frac{1}{\Pi(\vec q)}=\frac{1}{\chi(q)}+\frac{4\pi}{q^2},
\end{equation}
which in real space reads
\begin{eqnarray}\label{eq:Pi_inv_qq}
    \left[\Pi(\vec r-\vec r^{\prime})\right]^{-1}= \left[\chi(\vec r-\vec r^{\prime})\right]^{-1}+\frac{1}{|\vec r-\vec r^{\prime}|},
\end{eqnarray}
with the inverse polarisation function being defined as 
\begin{equation}\label{eq:Pi_inv_rr}
     \left[\Pi(\vec r-\vec r^{\prime})\right]^{-1}=\widehat  F^{-1}\left[ \frac{1}{\Pi(\vec q)}\right].
\end{equation}

By substituting Eq. (\ref{eq:int_F_exp3_v22}) into Eq.(\ref{eq:int_F_exp2}) and using the definition of the polarization function (\ref{eq:Pi_inv_rr}), we derive the relation defining \textit{the stiffness theorem for the intrinsic part of the free energy density functional}:
\begin{equation}\label{eq:stiffness_Pi_rr}
      \Delta \left<\mathcal{F}[n]\right>=-\frac{1}{2}
    \int \left. \left[\Pi(\vec r-\vec r^{\prime})\right]^{-1}\right.\delta n(\vec r^{\prime})\delta n(\vec r)~\mathrm{d}\vec r\mathrm{d}\vec r^{\prime}.
\end{equation}

The relation (\ref{eq:stiffness_Pi_rr}) expresses the change in  $\mathcal{F}[n]$ in terms of the density variation and polarisation function (or equivalently density response function).     
For calculation purposes,  it is more convenient to express the integral in the right-hand side of Eq.(\ref{eq:stiffness_Pi_rr}) in Fourier space:
\begin{equation}\label{eq:stiffness_Pi_qq}
      \Delta \left<\mathcal{F}[n]\right>=-\frac{1}{2}
    \int \left.\frac{\left|\delta n(\vec q)\right|^{2}}{\Pi(\vec q)}\right. ~\mathrm{d}\vec q,
\end{equation}
where we used the convolution theorem, Plancherel theorem,  and the equation for complex conjugate $\delta n^{*}(\vec q)=\delta n(-\vec q)$ following from the fact that $\delta n(\vec r)$ is a real function.

For completeness, we note that one can express Eq. (\ref{eq:stiffness_Pi_qq}) in terms of $\delta v_{\rm ext}(\vec q)= \left<\Delta v_{\rm ext}(\vec q)\right>$ using the connection $\delta n(\vec q)=\chi(\vec q)\delta v_{\rm ext}(\vec q)$ (see Eq.(\ref{eq:chi_KS_r})):
\begin{equation}\label{eq:stiffness_Pi_qq_mix}
      \Delta \left<\mathcal{F}[n]\right>=-\frac{1}{2}
    \int \left.\frac{\chi^2(\vec q)}{\Pi(\vec q)}\left|\delta v_{\rm ext}(\vec q)\right|^{2}\right. ~\mathrm{d}\vec q,
\end{equation}
or equivalently 
\begin{equation}\label{eq:stiffness_Pi_qq2}
      \Delta \left<\mathcal{F}[n]\right>=-\frac{1}{2}
    \int \left.\Pi(\vec q)\left|\delta v(\vec q)\right|^{2}\right. ~\mathrm{d}\vec q,
\end{equation}
where $v(\vec q)=4\pi/q^2+v_{\rm ext}(\vec q)$.

To find a similar result for the change in the total free energy in terms of the density response function, we use Eq. (\ref{eq:int_F_exp3}) in Eq. (\ref{eq:F_tot_for_p2}) to identify
\begin{eqnarray}\label{eq:F_tot_for_p4}
     \Delta \left<F[n]\right>= \frac{1}{2}\int\left<\Delta n(\vec r)\right> \left<\Delta v_{\rm ext}(\vec r)\right>~{\mathrm d}\vec r .
\end{eqnarray}

Now, applying Eq. (\ref{eq:chi_inv_rr}) to Eq. (\ref{eq:F_tot_for_p4}), we establish \textit{the stiffness theorem for the total free energy}:
\begin{eqnarray}\label{eq:tot_F_stif_r}
     \Delta \left<F[n]\right>= \frac{1}{2}\int\int\left[\chi(\vec r-\vec r^{\prime})\right]^{-1} \delta n(\vec r)  \delta n(\vec r^{\prime})~{\mathrm d}\vec r^{\prime}{\mathrm d}\vec r ,
\end{eqnarray}
where we set $\delta n(\vec r)= \left<\Delta n(\vec r)\right>$.

In Fourier space, using the convolution theorem, we can rewrite Eq. (\ref{eq:tot_F_stif_r}) in two equivalent forms
\begin{eqnarray}\label{eq:tot_F_stif_q}
     \Delta \left<F[n]\right>= \frac{1}{2}
    \int \left.\frac{\left|\delta n(\vec q)\right|^{2}}{\chi(\vec q)}\right. ~\mathrm{d}\vec q, 
    \,~\mathrm{and}\,~
    \Delta \left<F[n]\right>= \frac{1}{2}
    \int \left.\chi(\vec q)\left|\delta v_{\rm ext}(\vec q)\right|^{2}\right. ~\mathrm{d}\vec q.
\end{eqnarray}

In the limit $T\to 0$, we have $F[n]\to E[n]$, and Eq. (\ref{eq:tot_F_stif_q}) reproduces the relation used by \cite{Moroni_prl_1992, Moroni_prl_1995} in their seminal works on the QMC results for $\chi(\vec q)$ of the UEG. For finite temperatures, \cite{Moldabekov_pop_2018} derived Eq. (\ref{eq:tot_F_stif_q}) for the UEG but with a minus sign on the right-hand side of Eq. (\ref{eq:tot_F_stif_q}), which is attributed to using a different definition of the polarization function as a response in terms of the charge density, namely  $\Pi(\vec q)=-\delta n(\vec q)/v(\vec q)$.

\subsection{Inhomogeneous systems within the Born–Oppenheimer approximation}

Let us now consider the free energy of inhomogeneous electrons in the external field of a given fixed configuration of ions. This is the standard situation in DFT-MD simulations within the Born–Oppenheimer approximation~[\cite{wdm_book}].  

The Hartree potential for equilibrium electrons reads
\begin{equation}\label{eq:H_n_eq}
    v_{\rm H}[n_{\rm eq}](\vec r)=\int \frac{n_{\rm eq}(\vec r^{\prime})}{\left|\vec r-\vec r^{\prime}\right|}{\mathrm{d}}\vec r^{\prime}+v_{\rm ion}(\vec r),
\end{equation}
where $v_{\rm ion}(\vec r)$ is the ionic potential.

An external perturbation results in the change of the Hartree potential
\begin{equation}\label{eq:deltav_H}
    \Delta v_{\rm H}[n](\vec r)=\int \frac{\Delta n(\vec r^{\prime})}{\left|\vec r-\vec r^{\prime}\right|}{\mathrm{d}}\vec r^{\prime},
\end{equation}
where we set $\Delta v_{\rm ion}(\vec r)=0$ since ions are in a fixed configuration.

The potential energy due to $v_{H}$ of the unperturbed electrons reads
\begin{equation}
    V_0[n]=\frac{1}{2}\int  \frac{n_{\rm eq}(\vec r^{\prime}) n_{\rm eq}(\vec r)}
    {\left|\vec r -\vec r^{\prime}\right|} ~{\mathrm d}\vec r{\mathrm d}\vec r^{\prime}
    +
    \int  n_{\rm eq}(\vec r) v_{\rm ion}(\vec r)~{\mathrm d}\vec r 
    \label{eq:V0}
\end{equation}
and for the mean-field potential energy of the electrons perturbed by $\Delta v_{\rm ext}(\vec r)$, we have:
\begin{equation}
    V_1[n]=\frac{1}{2}\int  \frac{(n_{\rm eq}+\Delta n(\vec r))(n_{\rm eq}(\vec r^{\prime})+\Delta n(\vec r^{\prime}))}
    {\left|\vec r -\vec r^{\prime}\right|} ~{\mathrm d}\vec r{\mathrm d}\vec r^{\prime}
    +
    \int  (n_{\rm eq}(\vec r)+\Delta n(\vec r)) v_{\rm ion}(\vec r)~{\mathrm d}\vec r 
    +
    \int  (n_{\rm eq}(\vec r)+\Delta n(\vec r))\Delta v_{\rm ext}(\vec r)~{\mathrm d}\vec r 
    \label{eq:V1}
\end{equation}

Using Eqs. (\ref{eq:V0}) and (\ref{eq:V1}), we deduce the change in the potential energy of electrons due to the net effect of the Hartree potential and the external perturbation:
\begin{eqnarray}
    \Delta V[n]&=&V_1[n]-V_0[n]\nonumber\\
    &=&
    \int  \frac{n_{\rm eq}(\vec r) \Delta n(\vec r^{\prime})}
    {\left|\vec r -\vec r^{\prime}\right|} ~{\mathrm d}\vec r{\mathrm d}\vec r^{\prime}
    +\frac{1}{2}\int  \frac{\Delta n(\vec r)\Delta n(\vec r^{\prime})}
    {\left|\vec r -\vec r^{\prime}\right|} ~{\mathrm d}\vec r{\mathrm d}\vec r^{\prime}
    +\int  \Delta n(\vec r) v_{\rm ion}(\vec r)~{\mathrm d}\vec r\nonumber\\
    &&
    +\int  (n_{\rm eq}(\vec r)+\Delta n(\vec r))\Delta v_{\rm ext}(\vec r)~{\mathrm d}\vec r. 
    \label{eq:DV_tot}
\end{eqnarray}

For inhomogeneous systems, the linear response formalism allows one to introduce the  linear density response function,
\begin{eqnarray}
    \Delta n(\vec r)=\int \chi(\vec r, \vec r^{\prime})\Delta v_{\rm ext}[n](\vec r^{\prime})~{\mathrm d}\vec r^{\prime}, 
\end{eqnarray}
and the inverse density response function,
\begin{eqnarray}\label{eq:inv_chi_inhom_rr}
    \Delta v_{\rm ext}[n](\vec r)=\int \left[\chi(\vec r, \vec r^{\prime})\right]^{-1} \Delta n(\vec r^{\prime}).
\end{eqnarray}

By virtue of Eq. (\ref{eq:inv_chi_inhom_rr}), $\Delta v_{\rm ext}[n](\vec r)$ can be considered as a functional of the density; writing the functional Taylor expansion in first order then leads to:
\begin{equation}
    \Delta v_{\rm ext}[n](\vec r)=\int \left .\frac{\delta \Delta v_{\rm ext}[n](\vec r)}{\delta n(\vec r^{\prime})}\right|_{n=n_{\rm eq}} \delta n(\vec r^{\prime})~{\mathrm d}\vec r^{\prime}=\int \left[\chi (\vec r, \vec r^{\prime})\right]^{-1} \delta n(\vec r^{\prime})~{\mathrm d}\vec r^{\prime}.
\end{equation}

The change in the total free energy is given by:
\begin{eqnarray}
     \Delta F[n]&=& \Delta \mathcal{F}[n]+\Delta V[n]\nonumber\\
     &=& \int \left. \frac{\delta \mathcal{F}[n]}{\delta n(\vec r)}\right\vert_{n=n_{\rm eq}} \delta n(\vec r)~\mathrm{d}\vec r+\frac{1}{2}
    \int \left. \frac{\delta^2 \mathcal{F}[n]}{\delta n(\vec r^{\prime})\delta n(\vec r)}\right\vert_{n=n_{\rm eq}} \delta n(\vec r^{\prime})\delta n(\vec r)~\mathrm{d}\vec r\mathrm{d}\vec r^{\prime} \nonumber\\
    &&
    +\int  \frac{n_{\rm eq}(\vec r) \Delta n(\vec r^{\prime})}
    {\left|\vec r -\vec r^{\prime}\right|} ~{\mathrm d}\vec r{\mathrm d}\vec r^{\prime}
    +\frac{1}{2}\int  \frac{\Delta n(\vec r)\Delta n(\vec r^{\prime})}
    {\left|\vec r -\vec r^{\prime}\right|} ~{\mathrm d}\vec r{\mathrm d}\vec r^{\prime}
    +\int  \Delta n(\vec r) v_{\rm ion}(\vec r)~{\mathrm d}\vec r \nonumber\\
    &&
    +\int  (n_{\rm eq}(\vec r)+\Delta n(\vec r))\Delta v_{\rm ext}(\vec r)~{\mathrm d}\vec r 
    \label{eq:DFtot_inhom}
\end{eqnarray}
where we used Eq. (\ref{eq:DV_tot}) for $\Delta V[n]$, and where we further used the second order functional Taylor expansion for $ \Delta \mathcal{F}[n]$.

Recall that the free energy minimization of the equilibrium system gives:
\begin{eqnarray}\label{eq:DF_dn_inhom_vH}
    \left. \frac{\delta \mathcal{F}[n]}{\delta n(\vec r)}\right\vert_{n=n_{\rm eq}}=-\left(\int \frac{n_{\rm eq}(\vec r^{\prime})}{\left|\vec r-\vec r^{\prime}\right|}{\mathrm{d}}\vec r^{\prime}+v_{\rm ion}(\vec r)\right);
\end{eqnarray}
substituting this relation into the first integral on the second line of Eq. (\ref{eq:DFtot_inhom}) to cancel the term   $\int \Delta n[n](\vec r)v_{\rm ion}(\vec r){\mathrm{d}}\vec r$ leads to:
\begin{eqnarray}\label{eq:DFtot_inhomv2}
     \Delta F[n]   &=& \frac{1}{2}
    \int \left. \frac{\delta^2 \mathcal{F}[n]}{\delta n(\vec r^{\prime})\delta n(\vec r)}\right\vert_{n=n_{\rm eq}} \delta n(\vec r^{\prime})\delta n(\vec r)~\mathrm{d}\vec r\mathrm{d}\vec r^{\prime}
    +\frac{1}{2}\int  \frac{\Delta n(\vec r)\Delta n(\vec r^{\prime})}
    {\left|\vec r -\vec r^{\prime}\right|} ~{\mathrm d}\vec r{\mathrm d}\vec r^{\prime}
    \nonumber\\
    &&
    +\int  \Delta n(\vec r)\Delta v_{\rm ext}(\vec r)~{\mathrm d}\vec r +\int n_{\rm eq}(\vec r)\Delta v_{\rm ext}(\vec r)~{\mathrm d}\vec r .
\end{eqnarray}

Comparing Eq. (\ref{eq:DFtot_inhomv2}) with Eq. (\ref{eq:F_tot_for_p}) for a homogeneous system, we see that, for a fixed ionic snapshot, an extra term $\int n_{\rm eq}(\vec r)\Delta v_{\rm ext}[n](\vec r){\mathrm{d}}\vec r$ appears. 

Applying the free energy minimization condition in Eq. (\ref{eq:DFtot_inhomv2}) allows one to derive
\begin{eqnarray}\label{eq:D2Ftot_inhomv}
     -\int \left. \frac{\delta^2 \mathcal{F}[n]}{\delta n(\vec r^{\prime})\delta n(\vec r)}\right\vert_{n=n_{\rm eq}} \delta n(\vec r^{\prime})~\mathrm{d}\vec r^{\prime}  &=& \int \left[\chi (\vec r, \vec r^{\prime})\right]^{-1} \delta n(\vec r^{\prime})~{\mathrm d}\vec r^{\prime}+\int  \frac{\Delta n(\vec r^{\prime})}{|\vec r-\vec r^{\prime}|} ~{\mathrm d}\vec r^{\prime}+\sigma [n_{\rm eq}](\vec r),
\end{eqnarray}
where we denoted
\begin{equation}
    \sigma [n_{\rm eq}](\vec r)=\frac{\delta}{\delta n(\vec r)}\left(\int n_{\rm eq}(\vec r)\Delta v_{\rm ext}[n](\vec r){\mathrm{d}}\vec r\right)=\int n_{\rm eq}(\vec r)\left[\chi(\vec r, \vec r^{\prime})\right]^{-1}~{\mathrm d}\vec r^{\prime}, 
\end{equation}
and used
\begin{equation}
    \frac{\delta}{\delta n(\vec r)}\left(\int  \Delta n(\vec r) \Delta v_{\rm ext}[n](\vec r)~{\mathrm d}\vec r \right)=\int \left[\chi (\vec r, \vec r^{\prime})\right]^{-1} \delta n(\vec r^{\prime})~{\mathrm d}\vec r^{\prime}
\end{equation}
following from Eq. (\ref{eq:inv_chi_inhom_rr}).

Taking one more functional derivative of Eq. (\ref{eq:D2Ftot_inhomv}) yields
\begin{eqnarray}\label{eq:D2Ftot_inhomv_v2}
     -\left. \frac{\delta^2 \mathcal{F}[n]}{\delta n(\vec r^{\prime})\delta n(\vec r)}\right\vert_{n=n_{\rm eq}}  &=&  \left[\chi (\vec r, \vec r^{\prime})\right]^{-1}+\frac{1}{\left|\vec r-\vec r^{\prime}\right|}=\left[ \Pi (\vec r, \vec r^{\prime})\right]^{-1} ,
\end{eqnarray}
where it is taken into account that $\frac{\delta \sigma [n_{\rm eq}](\vec r)}{\delta n(\vec r^{\prime})}=0$, since $\sigma [n_{\rm eq}](\vec r)$ is a potential field independent from density variation (like $\left[\chi (\vec r, \vec r^{\prime})\right]^{-1}$ and $\chi (\vec r, \vec r^{\prime})$).
In Eq. (\ref{eq:D2Ftot_inhomv_v2}), we introduced the inverse polarization function  $\left[\Pi (\vec r, \vec r^{\prime})\right]^{-1}$ connecting $\Delta v[n](\vec r)= \Delta v_{\rm H}[n](\vec r)+\Delta v_{\rm ext}[n](\vec r)$ and the density perturbation:
\begin{equation}
     \Delta v_{\rm H}[n](\vec r)+\Delta v_{\rm ext}[n](\vec r)=\int \left[\Pi (\vec r, \vec r^{\prime})\right]^{-1} \delta n(\vec r^{\prime})~{\mathrm d}\vec r^{\prime}.
\end{equation}

Substituting Eq. (\ref{eq:DF_dn_inhom_vH}) and  Eq. (\ref{eq:D2Ftot_inhomv_v2}) in the second-order functional Taylor expansion of the change in the intrinsic free energy yields:
\begin{eqnarray}\label{eq:Inhom_stiff_intr_F}
    \Delta \mathcal{F}[n]
    &=& 
    -\frac{1}{2}
    \int \left[\Pi (\vec r, \vec r^{\prime})\right]^{-1}  \delta n(\vec r^{\prime})\delta n(\vec r)~\mathrm{d}\vec r\mathrm{d}\vec r^{\prime} 
    -\int v_{H}[n_{\rm eq}](\vec r) \delta n(\vec r)~\mathrm{d}\vec r.
\end{eqnarray}

Next, recalling Eq. (\ref{eq:D2Ftot_inhomv}) it follows from Eq. (\ref{eq:DFtot_inhomv2}) that
\begin{equation}\label{eq:Inhom_stiff_tot_F}
    \Delta F[n]  = \frac{1}{2}
    \int \left[\chi (\vec r, \vec r^{\prime})\right]^{-1} \delta n(\vec r^{\prime})\delta n(\vec r)~\mathrm{d}\vec r\mathrm{d}\vec r^{\prime}
    +\frac{1}{2}\int n_{\rm eq}(\vec r)\delta v_{\rm ext}(\vec r)~\mathrm{d}\vec r,
\end{equation}
where $\left[\chi (\vec r, \vec r^{\prime})\right]^{-1}$
is defined by Eq. (\ref{eq:inv_chi_inhom_rr}).

Eq. (\ref{eq:Inhom_stiff_tot_F}) and 
Eq. (\ref{eq:Inhom_stiff_intr_F}) are the versions of the stiffness theorem for the electrons in the field of fixed ions corresponding to 
the Born–Oppenheimer approximation in KS-DFT.
The last two terms in Eq. (\ref{eq:Inhom_stiff_tot_F}) and 
Eq. (\ref{eq:Inhom_stiff_intr_F}) are due to the field of ions and appear because the initial unperturbed density is inhomogeneous. 
 For a homogeneous case $n(\vec r)={\rm const}$ and $v_{H}(\vec r)={\rm const}$, the last term on the right-hand side of Eq.~(\ref{eq:Inhom_stiff_intr_F}) vanishes since $\int \delta n(\vec r){\mathrm{d}}\vec r=0$ and Eq.~(\ref{eq:Inhom_stiff_intr_F}) reduces to  Eq.~(\ref{eq:stiffness_Pi_rr}).  In this limit, the last term on the right-hand side of Eq.~(\ref{eq:Inhom_stiff_tot_F}) does not necessarily vanish. To reproduce the result (\ref{eq:tot_F_stif_r}), one needs to add the neutralizing background charge, which has been taken into account in the derivation of Eq. (\ref{eq:tot_F_stif_r}). The further extension of the presented results requires a fully self-consistent formula for the total free energy of a multicomponent system, taking into account the polarization of the ions.

A potentially useful relation for the construction of the 
non-interacting free energy functionals
is deduced from Eq. (\ref{eq:D2Ftot_inhomv_v2}) by using the partition of the intrinsic free energy into non-interacting and XC parts:
\begin{eqnarray}\label{eq:d2F_chi_rr}
     -\left. \frac{\delta^2 \mathcal{F}_S[n]}{\delta n(\vec r^{\prime})\delta n(\vec r)}\right\vert_{n=n_{\rm eq}}  &=&  \left[\chi_{\rm KS} (\vec r, \vec r^{\prime})\right]^{-1},
\end{eqnarray}
where we used the connection:
\begin{equation}
    \left[\chi (\vec r, \vec r^{\prime})\right]^{-1}+\frac{1}{\left|\vec r-\vec r^{\prime}\right|}+K_{\rm XC}(\vec r, \vec r^{\prime})=\left[\chi_{\rm KS} (\vec r, \vec r^{\prime})\right]^{-1},
\end{equation}
with the XC kernel of the inhomogeneous system being defined as:
\begin{equation}
    K_{\rm XC}(\vec r, \vec r^{\prime})=
     \frac{\delta^2 \mathcal{F}_{\rm XC}[n]}{\delta n(\vec r)\delta n(\vec r^{\prime})}.
\end{equation}

In Eq.(\ref{eq:d2F_chi_rr}), $\left[\chi_{\rm KS} (\vec r, \vec r^{\prime})\right]^{-1}$ can be computed in Fourier space by inverting the matrix for the KS response function $\chi^{\rm KS}_{\vec G,\vec G^\prime}(\vec k)$, where $\vec G$ and $\vec G^{\prime}$ are reciprocal lattice vectors, and $\vec k= \vec q - \vec G$ (with $\vec q$ being the perturbation wavenumber).
The Fourier coefficients $\chi^{\rm KS}_{\vec G,\vec G^\prime}(\vec k)$ were derived by Adler [\cite{Adler_pr_1962}] and Wiser [\cite{Wiser_pr_1963}]. The calculation of 
$\chi^{\rm KS}_{\vec G,\vec G^\prime}(\vec k)$  is implemented in a number of openly available KS-DFT codes, such as GPAW [\cite{LRT_GPAW2}].
One can use relation (\ref{eq:d2F_chi_rr}) to design advanced material-specific non-interacting free energy functionals for OF-DFT applications, e.g., to scale up simulations to a large number of particles. Moreover, the stiffness theorem and related relations demonstrate a direct connection between the density response and energy functionals. Hence, it allows one to assess the quality of different XC functionals on various length scales by analyzing the density response for different $q$. This is further discussed in Sec. \ref{s:Test_electron_gas}.

\section{Quantum Monte Carlo methods for WDM}\label{s:QMC}

Arguably, the gold standard for the estimation of equilibrium properties of WDM systems is given by the path integral quantum Monte Carlo method~\cite{cep}. Having originally been introduced for the simulation of ultracold $^4$He~[\cite{Fosdick_PR_1966,Jordan_PR_1968}], the PI-QMC method is capable of giving results that are exact within the given statistical uncertainty. For quantum degenerate fermions such as the electrons in WDM, PI-QMC is afflicted with the notorious fermion sign problem~[\cite{dornheim_sign_problem}] that manifests as an exponential increase in the compute time that is required to reach a certain level of accuracy with important parameters such as the system size $N$ or decreasing the temperature $T$.
Nevertheless, highly accurate simulations are possible down to the electronic Fermi temperature $\Theta\sim1$ for the UEG~[\cite{dornheim_ML,dornheim_sign_problem,dornheim_dynamic,dynamic_folgepaper}] and hydrogen~[\cite{Dornheim_JCP_2024,dornheim2024ab}].

For completeness, we note that, due to the pressing need to describe extreme states of matter, a gamut of approximate and semi-empirical approaches to deal with the sign problem have been suggested in the literature, e.g.~\cite{Brown_PRL_2013,Schoof_PRL_2015,Dornheim_NJP_2015,Malone_PRL_2016,Hirshberg_JCP_2020,Dornheim_Bogoliubov,Joonho_JCP_2021,Xiong_JCP_2022,Xiong_PRE_2023,Filinov_PRE_2023}. A particularly important approach is given by restricted PI-QMC~[\cite{Ceperley1991,Driver_PRL_2012,Militzer_PRL_2015}]. It formally avoids the sign problem, but this comes at the cost of the de-facto uncontrolled fixed-node approximation~[\cite{Anderson_fixed_nodes}].
Second, we mention the $\xi$-extrapolation method~[\cite{Xiong_JCP_2022,Dornheim_jcp_2023}] that is based on the PIMC simulation of fictitious identical particles, and which extends direct PI-QMC to much larger systems, e.g.~\cite{Dornheim_JPCL_2024,xiong2024gpu}. A key advantage of this method is that, in contrast to restricted PI-QMC, it retains its access to the full imaginary time structure, which is important to estimate a variety of linear~[\cite{dornheim_ML,Dornheim_prl_esa_2020,Dornheim_pop_2023,Tolias_JCP_2024,dornheim2024dynamic}] and nonlinear~[\cite{JCP21_nonlin,Dornheim_cpp_2022_non_lr,Tolias_2023}] response properties, as a starting point for the \emph{analytic continuation} of dynamic properties~[\cite{dornheim_dynamic,Hamann_PRB_2020}], and for the interpretation of X-ray Thomson scattering experiments~[\cite{Dornheim_T_2022,Dornheim_insight_2022,dornheim2023xray}]. Very recently, highly accurate PI-QMC results based on the $\xi$-extrapolation method have been presented for warm dense  hydrogen~[\cite{dornheim2024ab}] and strongly compressed beryllium~[\cite{Dornheim_Science_2024}] as it has been realized experimentally at the National Ignition Facility~[\cite{Tilo_Nature_2023}].

\begin{figure}
\centering
\includegraphics[width=0.48\textwidth]{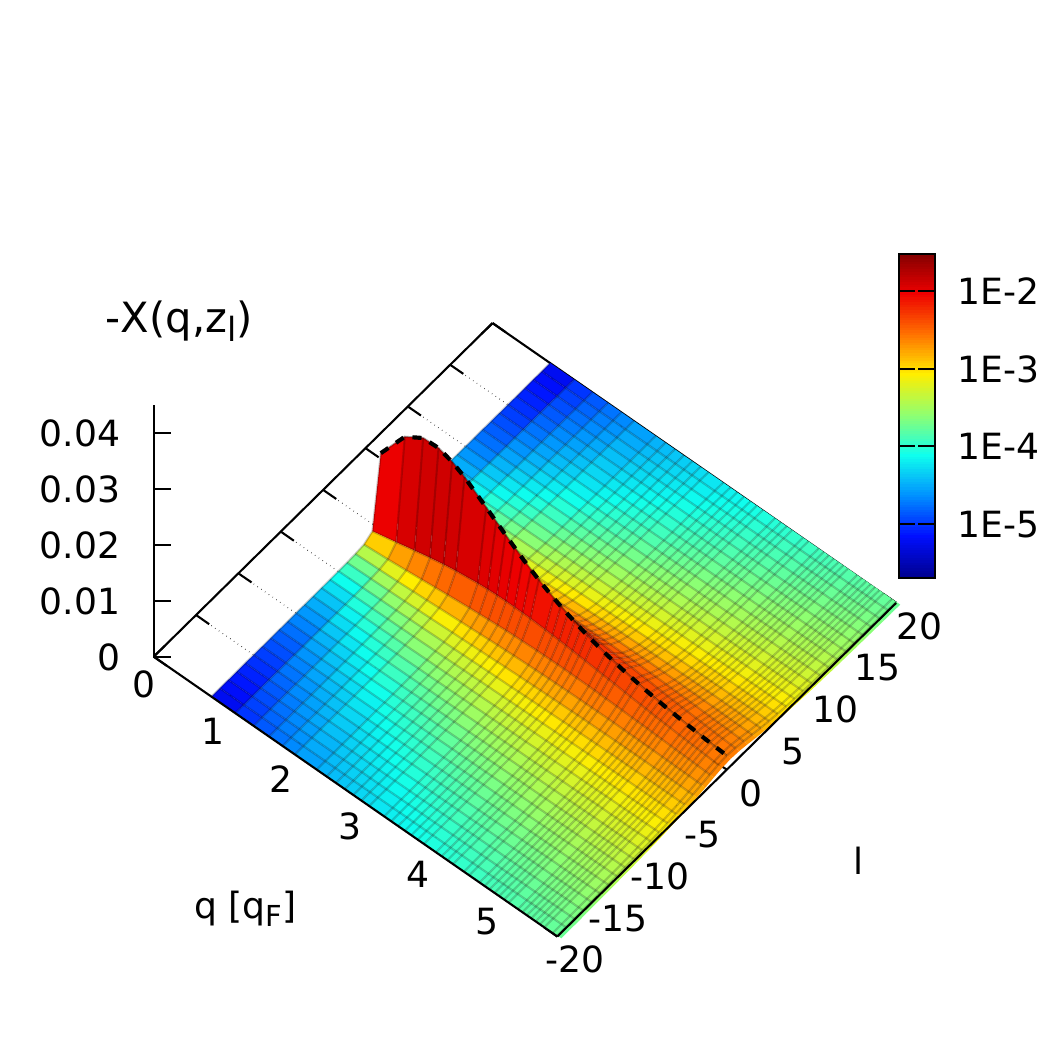}
\includegraphics[width=0.48\textwidth]{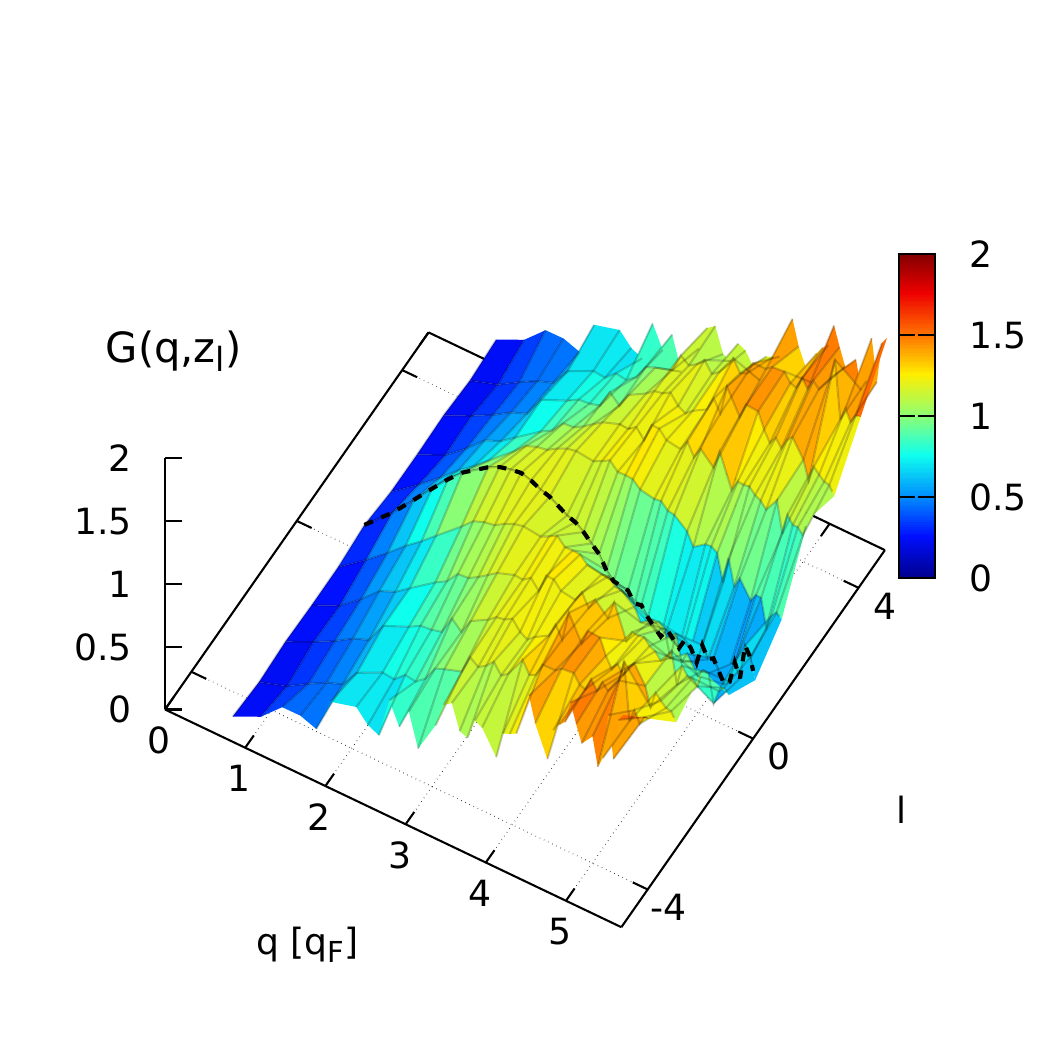}
\caption{\label{fig:imagination}
Exact \emph{ab initio} PI-QMC results for the dynamic Matsubara density response function $\widetilde{\chi}(q,z_l)$ [left] and dynamic Matsubara local field correction $\widetilde{G}(q,z_l)$ [right] of the warm dense UEG at $r_s=3.23$ and $\Theta=1$.
}
\end{figure}

In the context of the present work, a key strength of PI-QMC simulations is their capability of estimating the electronic density response of high energy density matter; see the recent review article by~\cite{Dornheim_pop_2023} for a comprehensive overview.
The most obvious route towards this goal is to harmonically perturb the system, and subsequently measure its response~[\cite{Moroni_prl_1992}]. This idea was first explored in the ground state to explore the static linear response function and static local field correction of the UEG~[\cite{Moroni_prl_1995,Bowen_PRB_1994}]; the latter was subsequently parametrized by \cite{cdop} and used for a variety of applications.
This idea is straightforwardly extended to finite temperatures, e.g.,~\cite{dornheim_pre,groth_jcp}, and has been used to obtain first highly accurate results for the static XC-kernel of warm dense hydrogen in Ref.~[\cite{Boehme_PRL_2022,Bohme_PRE_2023}]. Moreover, it allows one to study nonlinear effects~[\cite{Dornheim_PRL_2020,PhysRevResearch.3.033231}], which can be important for the description of effective potentials and forces~[\cite{Gravel,Dornheim_Force_2022}] and energy loss properties such as the stopping power~[\cite{Nagy_PRA_1989,PhysRevB.37.9268}].
While being formally exact, however, the direct perturbation approach requires one to perform a set of independent QMC simulations for different perturbation amplitudes for each individual wave number $q$; this can be prohibitive for the characterization of the density response over a broad range of densities, temperatures, and wavenumbers.
An elegant and more efficient alternative is given by the estimation of imaginary-time correlation functions (ITCF)~[\cite{JCP21_nonlin}].
Arguably the most important example is given by the imaginary-time density--density correlation function $F_{ee}(\mathbf{q},\tau)=\braket{\hat{n}_e(\mathbf{q},\tau)\hat{n}_e(-\mathbf{q},0)}_0$, where $n_e(\mathbf{q},t)$ denotes the single-electron density evaluated in Fourier space and the expectation value $\braket{\dots}_0$ is evaluated with respect to the original, i.e., unperturbed Hamiltonian. We note that $F_{ee}(\mathbf{q},\tau)$ corresponds to the usual intermediate scattering function $F_{ee}(\mathbf{q},t)$~[\cite{siegfried_review}], but evaluated for an imaginary-time argument $t=-i\hbar\tau$, where $\tau\in[0,\beta]$.
In this context, an important relation is given by the imaginary-time version of the fluctuation--dissipation theorem~[\cite{Dornheim_insight_2022}]
\begin{eqnarray}\label{eq:static_chi}
    \chi(\mathbf{q},0) = -2n \int_0^{\beta/2} \textnormal{d}\tau\ F_{ee}(\mathbf{q},\tau)\ ,
\end{eqnarray}
which allows one to obtain the full wavenumber dependence of the static density response function from a single simulation of the unperturbed system.
Indeed, Eq.~(\ref{eq:static_chi}) has been used as the basis for extensive PI-QMC results for both $\chi(\mathbf{q},0)$ and $G(\mathbf{q},0)$, see Refs.~[\cite{dornheim_ML,Dornheim_prl_esa_2020,Dornheim_prb_2021}].
We further note that it has been generalized very recently~[\cite{Tolias_JCP_2024,dornheim2024dynamic}] for the estimation of the dynamic density response $\widetilde{\chi}(q,z_l)$ and dynamic local field correction $\widetilde{G}(q,z_l)$ in the imaginary Matsubara frequency domain,
\begin{eqnarray}\label{eq:MDR}
    \widetilde{\chi}(\mathbf{q},z_l) = -2n\int_0^{\beta/2}\textnormal{d}\tau\ F(\mathbf{q},\tau)\ \textnormal{cos}\left(i z_l \tau\right)\,,
\end{eqnarray}
where $z_l=i2\pi l/\beta$ are the discrete imaginary bosonic Matsubara frequencies with $l=\dots,-1,0,1,\dots$.
In Fig.~\ref{fig:imagination}, we show the first PIMC results for these properties for the warm dense UEG at $r_s=3.23$ and $\Theta=1$. These conditions are highly relevant for contemporary WDM research and can be realized in experiments e.g.~using hydrogen jets~[\cite{Zastrau_2021,Fletcher_Frontiers_2022}].
The left panels shows $\widetilde{\chi}(\mathbf{q},z_l)$, which is dominated by its static limit of $\widetilde{\chi}(\mathbf{q},0)=\chi(\mathbf{q},0)$ for small $q$, see the dashed black line at $l=0$. Indeed, all contributions for $l\neq0$ are a consequence of quantum delocalization of the electrons and would be entirely absent in a classical system~[\cite{dornheim2024dynamic,dornheim2024quantum}];
such quantum delocalization plays an increasingly important role on decreasing length scales, which explains the increasing importance of dynamic terms for large wavenumbers.
In this context, a revealing relation is given by the Matsubara series of the static structure factor [\cite{dornheim2024dynamic}]:
\begin{eqnarray}\label{eq:Matsubara_Series}
    S(\mathbf{q}) = -\frac{1}{n\beta}
\sum_{l=-\infty}^\infty \widetilde{\chi}(\mathbf{q},z_l)\ ,
\end{eqnarray}
which reduces to the $l=0$ contribution in the classical limit.
Eq.~(\ref{eq:Matsubara_Series}) directly implies that it is, in principle, possible to compute the electronic static structure factor (which is the Fourier transform of the electron--electron pair correlation function) from KS-DFT results for the Matsubara density response $\widetilde{\chi}_\textnormal{KS}(\mathbf{q},z_l)$
and a corresponding Matsubara XC-kernel $\widetilde{K}_\textnormal{XC}(\mathbf{q},z_l)=-4\pi/q^2 \widetilde{G}(\mathbf{q},z_l)$.
In practice, it might be promising to approximate the latter based on the Matsubara local field correction of the UEG, which is shown in the right panel of Fig.~\ref{fig:imagination}. Indeed, it appears feasible to construct a parametrization $\widetilde{G}(\mathbf{q},z_l;r_s,\Theta)$ based on PIMC results and other constraints~[\cite{Hou_PRB_2022,LeBlanc_PRL_2022,PhysRevB.101.245135,PhysRevB.34.4989,holas_limit}] that covers the entire WDM parameter space.
In addition to being interesting in its own right, such a tool would open up the intriguing possibility to construct non-local and fully thermal XC-functionals based on the adiabatic connection formula and the fluctuation--dissipation theorem~[\cite{pribram}].

A further important use case of $F_{ee}(\mathbf{q},\tau)$ is given by its relation to the dynamic structure factor $S_{ee}(\mathbf{q},\omega)$,
\begin{eqnarray}\label{eq:Laplace}
    F_{ee}(\mathbf{q},\tau) = \int_{-\infty}^\infty \textnormal{d}\omega\ S_{ee}(\mathbf{q},\omega)\ e^{-\hbar\omega\tau}\ .
\end{eqnarray}
Eq.~(\ref{eq:Laplace}) constitutes the basis for a so-called \emph{analytic continuation}~[\cite{JARRELL1996133}], i.e., its numerical inversion to solve for $S_{ee}(\mathbf{q},\omega)$. On the one hand, accurate PI-QMC based results for such a dynamic property would be of tremendous value either as an input or as a benchmark for time-dependent DFT applications [\cite{Moldabekov_prr_2023}].
On the other hand, it is well known that the \emph{analytic continuation} constitutes a notoriously difficult and, in fact, ill-posed problem~\cite{Goulko_PRB_2017}, and no general solution exists at the present time.
For the special case of the UEG, this problem has been solved on the basis of the constrained sampling of the dynamic LFC~[\cite{dornheim_dynamic,dynamic_folgepaper,Hamann_PRB_2020}], which has given important insights about dynamic XC-effects, such as the emergence of a roton-type feature in $S_{ee}(\mathbf{q},\omega)$~[\cite{Dornheim_Nature_2022,koskelo2023shortrange,Takada_PRB_2016}] that can potentially be detected in experiments with hydrogen jets~[\cite{Hamann_PRR_2023}].

Finally, we note that Eq.~(\ref{eq:static_chi}) has been generalized by~\cite{JCP21_nonlin} for the estimation of nonlinear density response functions based on PI-QMC results for higher-order imaginary-time correlation functions~[\cite{Dornheim_JPSJ_2021}].

\section{Benchmarking Free Energy Functionals for Electron Gas at WDM conditions}\label{s:Test_electron_gas}

To this date, most of the existing KS-DFT simulations of WDM have been performed using ground-state XC functionals. This is due to multiple reasons. First, the electrons remain sufficiently quantum degenerate with $\theta\sim 0.1$ even at a few electronvolts [\cite{Lee_prl_2009, Descamps2020, Cho_SciRep_2016, moldabekov_cu_2024}] at high densities. Second, for WDM and dense plasmas applications at very high temperatures with $\theta\gg1$, the inaccuracies in the XC functional are not critical since for quasi-free electrons $\mathbf{\mathcal{F}_{\rm XC} /\mathcal{F}_{\rm S} \ll 1}$ (cf. Fig. \ref{fig:diagram2}). Third,  highly accurate QMC data for the free energy of the UEG over a wide range of temperatures and densities relevant to the WDM applications have become available only recently [\cite{Groth_prl_2017,Dornheim_prl_2016}] and were implemented into the Libxc library of XC functionals even later.
Fourth, experimental data for WDM often have substantial uncertainties both with respect to temperature and density due to diagnostic challenges related to the extreme conditions and the short time scales. Therefore, the inaccuracies in the XC functional alone usually do not lead to the results deviating from the experimental data beyond the experimental uncertainty range [\cite{Witte_prl_2017, Rueter_prl_2014, Vinko_prl_2010, Kritcher_prl_2009}]. Finally, ground-state XC functionals on the meta-GGA level, as well as hybrid XC functionals, are designed to use the occupation numbers of orbitals. Therefore, these functionals automatically incorporate certain information about thermal effects when used in combination with the Fermi-Dirac smearing of the occupation numbers. This was demonstrated for the meta-GGA level XC functional SCAN  on the example of the warm dense hydrogen [\cite{Moldabekov_jctc_2024}]. In the following sections, we discuss this aspect of the meta-GGA and hybrid XC functionals in more detail. 

Despite the aforementioned justifications,  certain natural concerns and questions remain regarding the use of the ground-state XC functionals in the WDM regime. For example, one can ask whether the ground state approximation for the XC functional provides corrections in the "right direction" and what reasons may account for it. To answer this question from an \textit{ab initio} perspective, several often used ground state XC functionals and the thermal LDA functional by \cite{Groth_prl_2017} were tested by comparing to the exact QMC data for a weakly and strongly perturbed electron gas, and for warm dense hydrogen [\cite{Moldabekov_jcp_2021, Moldabekov_prb_2022, Moldabekov_jctc_2023, Moldabekov_jcp_hybrid_2023, Moldabekov_jpcl_2023,bonitz2024principles}].  

\subsection{Harmonically Perturbed Electron Gas}\label{s:harm_per_EG}

The analysis of the accuracy of the KS-DFT simulations using several commonly used XC functionals has been performed by benchmarking against exact QMC data for the harmonically perturbed electron gas by \cite{Moldabekov_jcp_2021, Moldabekov_prb_2022, Moldabekov_jctc_2023}. In these works, the LDA [\cite{Perdew_Zunger_PRB_1981}], thermal-LDA [\cite{Groth_prl_2017}], PBE [\cite{PerdewPBE}], PBEsol [\cite{PBEsol}], SCAN [\cite{Sun_prl_2015}], and AM05 [\cite{AM05}] XC functionals were tested for both moderate ($r_s=2$) and strong ($r_s=6$) coupling regimes. PBE and PBEsol are GGA-level functionals that are arguably most often used in solid-state and  WDM applications.  SCAN and AM05 are meta-GGA level functionals, which are also popular among KS-DFT practitioners studying matter under extreme conditions. For example, SCAN was used to study phase transition in warm dense hydrogen [\cite{Hinz_prr_2020}] and for simulation of carbon ionization at gigabar pressures [\cite{Bethkenhagen_prr_2020}]\footnote{We note that in Ref. [\cite{Bethkenhagen_prr_2020}], the SCAN functional was used for testing purposes, 
while the bulk of the work was done with the PBE functional.}.
AM05 was used for the calculation of the equation of state of silicon dioxide [\cite{Sjostrom_prb_2015}] and argon [\cite{osti_1055894, Sun_jcp_2016}] under WDM conditions. 
The analysis of the XC functionals using the harmonically perturbed electron gas has been extended to the hybrid functionals at $r_s=2$ in Ref. [\cite{Moldabekov_jcp_hybrid_2023}], where PBE0 [\cite{Adamo1999}], PBE0-1/3 [\cite{Cortona_jcp_2012}], HSE06 [\cite{Heyd2006, Krukau2006}], HSE03 [\cite{Heyd2003}], and B3LYP [\cite{Stephens1994}] XC functionals were tested. 
These assessments of the quality of the XC functionals at WDM conditions have demonstrated conditions where the considered XC functionals fail to correctly describe the electronic structure and where the KS-DFT method can be used to obtain nigh QMC-level accuracy.

Let us first discuss local and semi-local XC functionals. The main findings for LDA, PBE, PBEsol, SCAN, and AM05 are summarized as follows:

\begin{itemize}
    \item The KS-DFT results are accurate for small wave numbers of the density perturbation, $q<q_F$. In particular, the SCAN functional yields an excellent agreement with the reference QMC data and, thus, is gauged to be a highly reliable choice.

    \item In a wider range of wave numbers with $q<3q_F$, the thermal-LDA, LDA, PBE, PBEsol, and AM05 functionals provide results with a relative error not exceeding a few percent if $\delta n/n_0<1$. 

    \item  At $q<2q_F$, the thermal-LDA-based density response of the warm dense electron gas has a similar quality to the results computed using the ground state LDA, PBE, PBEsol, and AM05 functionals. At $q\gtrsim2q_F$, thermal-LDA is less accurate for the density response description than these ground state XC functionals.
    
    \item  As a general trend, the performance of all considered XC functionals deteriorates with the increase in the wavenumber at $q\lesssim 2.5q_F$. Among the considered XC functionals, the SCAN-based results at $q\simeq 2q_F$ are less accurate than the data computed using LDA, PBE, PBEsol, and AM05. 

    \item For a  metallic density with $r_s=2.0$, at large wavenumbers $q\gtrsim 5~q_F$, all considered XC functionals yield errors of less than $6~\%$ if the density perturbation is weak $\delta n/n_0\ll 1$. In contrast, in the regime of strong perturbations, $\delta n/n_0>1$, and large wavenumbers, all considered XC functionals fail to provide accurate results with errors in the density about $10~\%$. In this regime, the considered XC functionals are considered to be unreliable.

    \item In general, for a given $\theta$, the accuracy of the KS-DFT results deteriorates with the increase in $r_s$, i.e, with the decrease in the density. 

    \item There is a strong correlation between the performance of the XC functionals in the linear response regime and the quality of the results when the perturbation is strong, i.e., beyond the linear response regime. The XC functionals that provide a more accurate density description of the weakly perturbed electron gas also work better in the regime of strong perturbation.  

\end{itemize}

\begin{figure}
\centering
\includegraphics[width=0.5\textwidth]{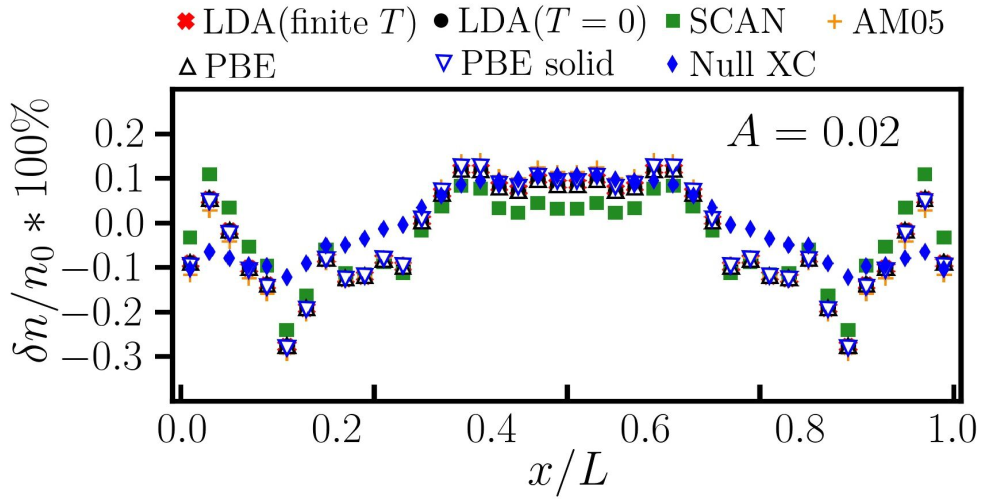}
\caption{\label{fig:density_A001_qmin}
Relative deviation in the density between the KS-DFT data and the reference QMC data for the perturbed UEG  at $r_s=2$ with $A=0.01~[{\rm Ha}]$ and $q_1=0.843q_F$. Adapted from \cite{Moldabekov_scipost_2022}. The KS-DFT calculations were performed using GPAW [\cite{GPAW_2024, ASE_17}].}
\end{figure}

\begin{figure}
\centering
\includegraphics[width=0.6\textwidth]{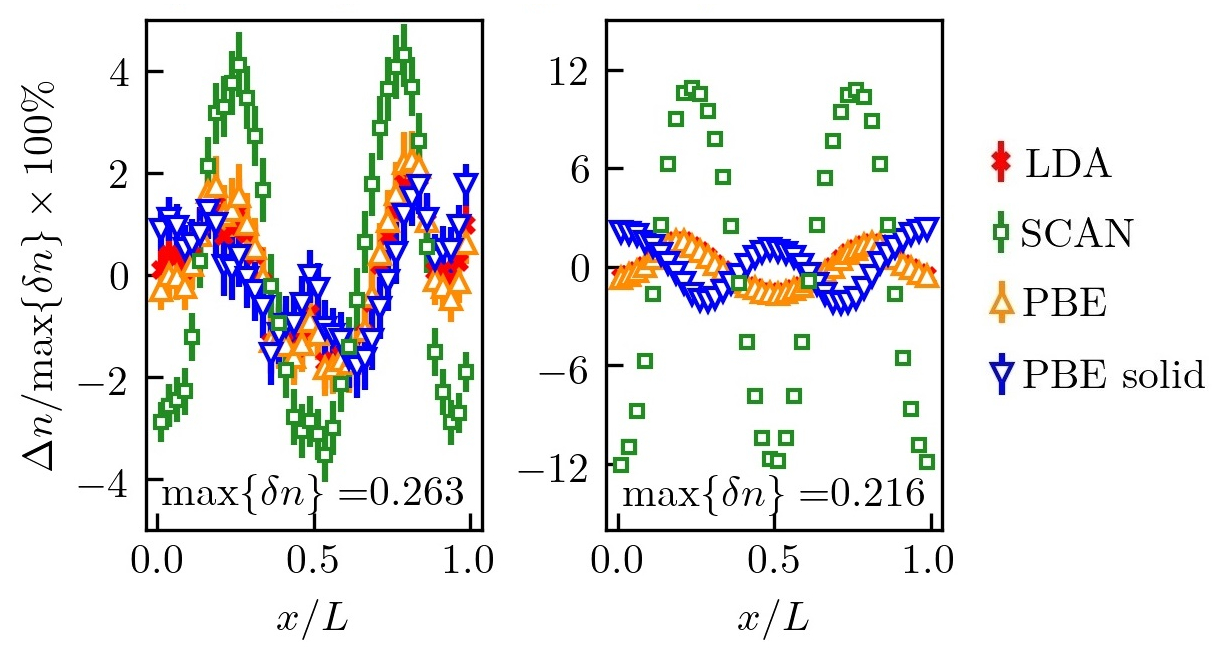}
\caption{\label{fig:density_err_two_A}
Relative deviation in the density $\Delta n/{\rm max }~\delta n \times 100~\%$ between the KS-DFT data and the reference QMC data for UEG at $r_s=2$ with $A=0.1$ (left) and at $r_s=6$ with $A=0.01$ (right). Adapted from Ref. [\cite{Moldabekov_jcp_2021}]. The KS-DFT calculations were performed using GPAW [\cite{GPAW_2024, ASE_17}].}
\end{figure}

As an illustration, we show the difference in the densities between the exact QMC data and the KS-DFT results in Fig. \ref{fig:density_A001_qmin} for various local and semi-local XC functionals at $q=0.843q_F$ and $A=0.02~{\rm Ha}$ (including the finite temperature version of the LDA by \cite{Groth_prl_2017}). We find that KS-DFT is capable of providing excellent results for the weakly perturbed warm dense electron gas at relatively small wavenumbers  $q\lesssim q_F$. We note a certain level of generality of the conclusions achieved using a single harmonic perturbation. This is shown to be the case considering perturbations that can be represented as a linear combination of harmonic perturbations with different wavenumbers [\cite{Moldabekov_jcp_2021, Moldabekov_prb_2022}]. As an illustration, the results computed for $r_s=2$ and $r_s=6$ at $\theta=1$ are shown  in Fig. \ref{fig:density_err_two_A}. The inhomogeneity is induced by a double harmonic perturbation $2A\left[\cos(\vec q_1\vec r)+\cos(\vec q_2\vec r)\right]$ with $q_1=0.843~q_{F}$ and $q_2=1.685~q_{F}$. The amplitude of the perturbations is set to $A=0.1$ for $r_s=2$  and to $A=0.01$ for $r_s=6$. These perturbation amplitudes lead to the density deviation of the order of $0.1~n_0$ from the mean value.
From Fig. \ref{fig:density_err_two_A}, one can see that the LDA, PBE, and PBEsol-based results show a good agreement with the QMC data at $r_s=2$ and $r_s=6$. The SCAN functional deviates from the QMC data significantly due to the perturbation component with $q_2=1.685~q_{F}$. As mentioned,  the quality of the SCAN-based results around $2q_F$ is worse than the quality of the data computed using LDA, PBE, and PSEsol. 

For the hybrid PBE0, PBE0-1/3, HSE06, HSE03, and B3LYP  XC functionals, the main findings are summarized as the following: 
 \begin{itemize}
     \item PBE0, PBE0-1/3, HSE06, and HSE03 are more accurate for the density response in a wide range of wavenumbers ($q\lesssim 3q_F$) compared with LDA (both ground-state version and thermal version), PBE, PBEsol, SCAN, and AM05.
     \item B3LYP is significantly less accurate for the perturbed electron gas compared with the considered hybrid, local, and semilocal XC functionals.
     \item Among the considered XC functionals, it was revealed that PBE0 provides the closest agreement with the exact QMC data for both weakly perturbed cases with $\delta n\lesssim 0.1n_0$ and strongly perturbed electron gas with $\delta n\sim n_0$  in a wide range of wavenumbers $q\lesssim3q_F$.
 \end{itemize}
 
These conclusions are valid for both the ground state and the partially degenerate case.
We note that B3LYP was designed for atoms and molecules without enforcing a correct UEG limit. This explains its failure to adequately describe an inhomogeneous electron gas. 

At $q\lesssim3_F$, the high accuracy of the KS-DFT results at small wavenumbers and the increasingly large errors at large wavenumbers of the density perturbation can be understood by considering the static linear density response function:
\begin{equation}
    \frac{1}{\chi(q)}=\frac{1}{\chi_{\rm KS}}-\left[\frac{4\pi}{q^2} + K_{\rm xc}(q)\right],
\end{equation}
which follows from Eq.(\ref{eq:Kxc_chi_conn}).

The UEG model-based local and semi-local XC functionals reproduce the correct long wavelength limit of the XC kernel due to the compressibility sum rule~[\cite{Sjostrom_prb_2013,Tolias_Vashishta_JCP_2024}]. 
As the result of this fact and the connection (\ref{eq:chi_tot_r}) between density perturbation and the density response function, the KS-DFT calculations using the UEG-based local and semi-local XC functionals provide density perturbation values with high accuracy at small wavenumbers.
With the increase in the perturbation wavenumber, the quality of the long wavelength approximation of $K_{\rm xc}(q)$ declines for the description of the density response.  
As we show in the next section \ref{ss:hybrid_Kxc}, the hybrid XC functionals provide much more accurate results for the XC kernel of the UEG than the aforementioned local and semi-local XC functionals. 
In the limit of large wavenumbers, the importance of $K_{\rm xc}(q)$ diminishes as the quantum kinetic energy $\sim \hbar q^2/(2m)$ of electrons becomes dominant over the XC energy. 

The correlation between the accuracy of the KS-DFT results in the regime of weak perturbation and in the case of strong inhomogeneities can be understood by considering the connection between the linear density response function and non-linear density response functions. Within non-linear response theory, the static density perturbation of the uniform electron gas due to an external harmonic perturbation   $v(\vec r)=2A\cos{\left(\vec q\cdot \vec r\right)}$ can be expressed in the form of the Fourier expansion [\cite{PhysRevResearch.3.033231}]:
\begin{align}\label{eq:rho_tot}
n(\mathbf{r}) &=n_0(\mathbf{r}) + 2 \sum_{\eta=1}^\infty \braket{\hat\rho_{\eta\mathbf{q}}}_{q,A} \textnormal{cos}\left(\eta\mathbf{q}\cdot\mathbf{r} \right),
\end{align}
where $\braket{\hat\rho_\mathbf{k}}_{q,A}$ are the density perturbation components in the Fourier space. 

In Eq.~(\ref{eq:rho_tot}), $\braket{\hat\rho_\mathbf{k}}_{q,A}$ 
has non-zero components at multiples of the perturbing field wavenumber, i.e. at $k=\eta \vec q$ with $\eta$ being a positive integer number. The components $\braket{\hat\rho_{\eta \mathbf{q}}}_{q,A}$ at $\eta=1$, $\eta=2$, and $\eta=3$ are referred to as density perturbations at the first, second and third harmonics, correspondingly.  These density perturbation components $\braket{\hat\rho_\mathbf{k}}_{q,A}$ are used to define the non-linear density response functions [\cite{PhysRevResearch.3.033231, Mikhailov_Annalen, Moroni_prl_1995}]: 
\begin{eqnarray}\label{eq:rho1}
\braket{\hat\rho_\mathbf{q}}_{q,A} &=& \chi^{(1)}(q) A + \chi^{(1,\textnormal{cubic})}(q) A^3  {+\dots}\ ,\\
\label{eq:rho2}
\braket{\hat\rho_\mathbf{2q}}_{q,A} &=& \chi^{(2)}(q) A^2 {+\dots} \ , \\
\label{eq:rho3}
\braket{\hat\rho_\mathbf{3q}}_{q,A} &=& \chi^{(3)}(q) A^3 {+\dots} \ ,
\end{eqnarray}
where $\chi^{(1,\textnormal{cubic})}(q)$ is the \textit{cubic response function at the first harmonic}, $\chi^{(2)}(q)$ is the \textit{quadratic response function}, and  $\chi^{(3)}(q)$ is the \textit{cubic response function at the third harmonic}.

For the UEG, it was shown by \cite{PhysRevResearch.3.033231} that  $\chi^{(2)}(q)$ and $\chi^{(3)}(q)$ can be computed with high precision using the following relations:
 \begin{eqnarray}
     \chi^{(2)}_{\rm LFC}( q) =  \frac{\chi^{(2)}_{0}( q)} { \left[1-v(q)\left[1-G(q)\right]\chi^{(1)}_{0}(q)\right]^{2} \left[1-v(2q)\left[1-G(2q)\right]\chi^{(1)}_{0}( 2q)\right]}, \label{eq:chi2_LFC}
 \end{eqnarray}
and 
 \begin{eqnarray}
     \chi^{(3)}_{\rm LFC}( q) =  \frac{\chi^{(3)}_{0}( q)} {\left[1-v(q)\left[1-G(q)\right]\chi^{(1)}_{0}(q)\right]^{3}\times \left[1-v(3q)\left[1-G(3q)\right]\chi^{(1)}_{0}( 3q)\right]}.\label{eq:chi3_LFC},
 \end{eqnarray}
where $G(q)=-v(q)K_{\rm XC}(q)$ is referred to as local field correction, $\chi^{(2)}_{0}( q)$ and $\chi^{(3)}_{0}( q)$ are the corresponding quadratic and cubic response functions of the ideal electron gas.

 In works by \cite{Mikhailov_Annalen,Mikhailov_PRL}, it was proven that the ideal electron gas response functions $\chi^{(2)}_{0}( q)$ and $\chi^{(3)}_{0}( q)$ are expressed recursively in terms of the finite-temperature Lindhard function $\chi^{(1)}_{0}( q)$ [in atomic units]:
\begin{eqnarray}\label{eq:Mikhailov2}
\chi^{(2)}_0(q) = \frac{2}{q^2}\left( \chi^{(1)}_0(2q)-\chi^{(1)}_0(q)\right),
\end{eqnarray}
\begin{eqnarray}\label{eq:Mikhailov3}
\chi_0^{(3)}(q)=\frac{3\chi_0^{(1)}(3q)-8\chi^{(1)}_0(2q)+5\chi_0^{(1)}(q)}{3q^4},
\end{eqnarray} 
which was subsequently generalized by \cite{Tolias_2023} to higher orders.

From Eqs. (\ref{eq:chi2_LFC}), (\ref{eq:chi3_LFC}), (\ref{eq:Mikhailov2}), and (\ref{eq:Mikhailov3}) it is evident that the quality of the quadratic density response at the second harmonic and cubic density response at the third harmonic depends on the accuracy of the calculations in the linear response regime. These equations remain valid in KS-DFT as it was shown in Ref. [\cite{Moldabekov_jctc_2022}] by considering the jellium model. Expressions (\ref{eq:chi2_LFC}), (\ref{eq:chi3_LFC}), (\ref{eq:Mikhailov2}) and (\ref{eq:Mikhailov3}) are generalized to the disordered systems (e.g., WDM and liquid metals) by substituting the KS response function $\chi_{\rm KS}(q)$ instead of $\chi_0(q)$, where the KS response function is defined by Eq. (\ref{eq:chi_KS_r}). The general dynamic case follows by introducing the frequency dependence in the density response functions and XC kernel. Within LR-TDDFT, the calculation of the dynamic density response function and KS response function are discussed in detail in Refs. [\cite{Moldabekov_prr_2023, Moldabekov_jcp_2023_averaging}]. For more details about various aspects of the non-linear response theory of  UEG, we refer the reader to Refs. [\cite{PhysRevResearch.3.033231, Dornheim_PRL_2020, Tolias_2023, CPP_nonlin21, JCP21_nonlin, Dornheim_JPSJ_2021, Dornheim_cpp_2020_non_lr, Dornheim_cpp_2022_non_lr}].

To date, there is no theoretical solution for the cubic response function at the first harmonic $\chi_0^{(1,\textnormal{cubic})}(q)$ of the ideal electron gas (introduced in Eq. (\ref{eq:rho1})) which provides agreement with the exact QMC calculations. Nevertheless, there is a formal relation between  $\chi^{(1,\textnormal{cubic})}_0$, and  the cubic response function of correlated UEG  [\cite{PhysRevResearch.3.033231}]:
\begin{equation}\label{eq:cubic_first_LFC}
    \chi^{(1,\textnormal{cubic})}( q)= \frac{\chi^{(1,\textnormal{cubic})}_{0}( q)}{\left[1-v(q)\left[1-G(q)\right]\chi^{(1)}_{0}(q)\right]^{4}}.
\end{equation}

\begin{figure*}[!t]
\minipage{0.333\textwidth}
  \includegraphics[width=\linewidth]{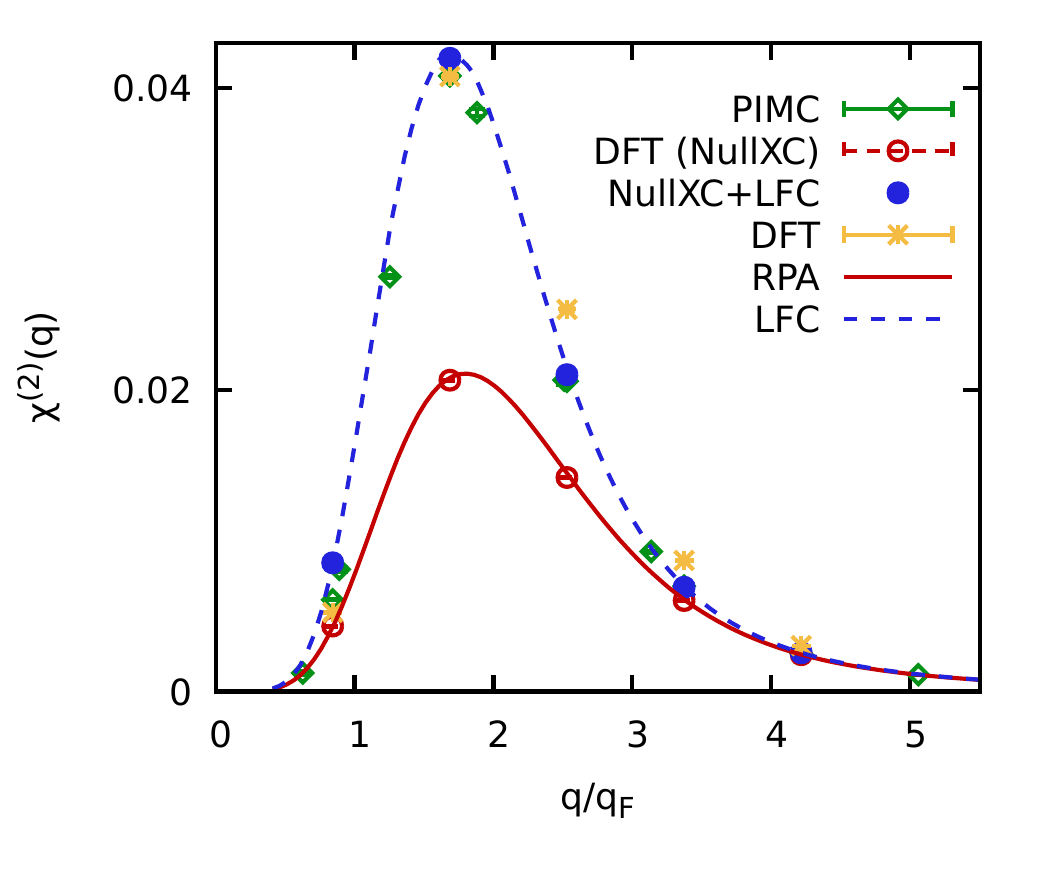}
\endminipage
\minipage{0.333\textwidth}
  \includegraphics[width=\linewidth]{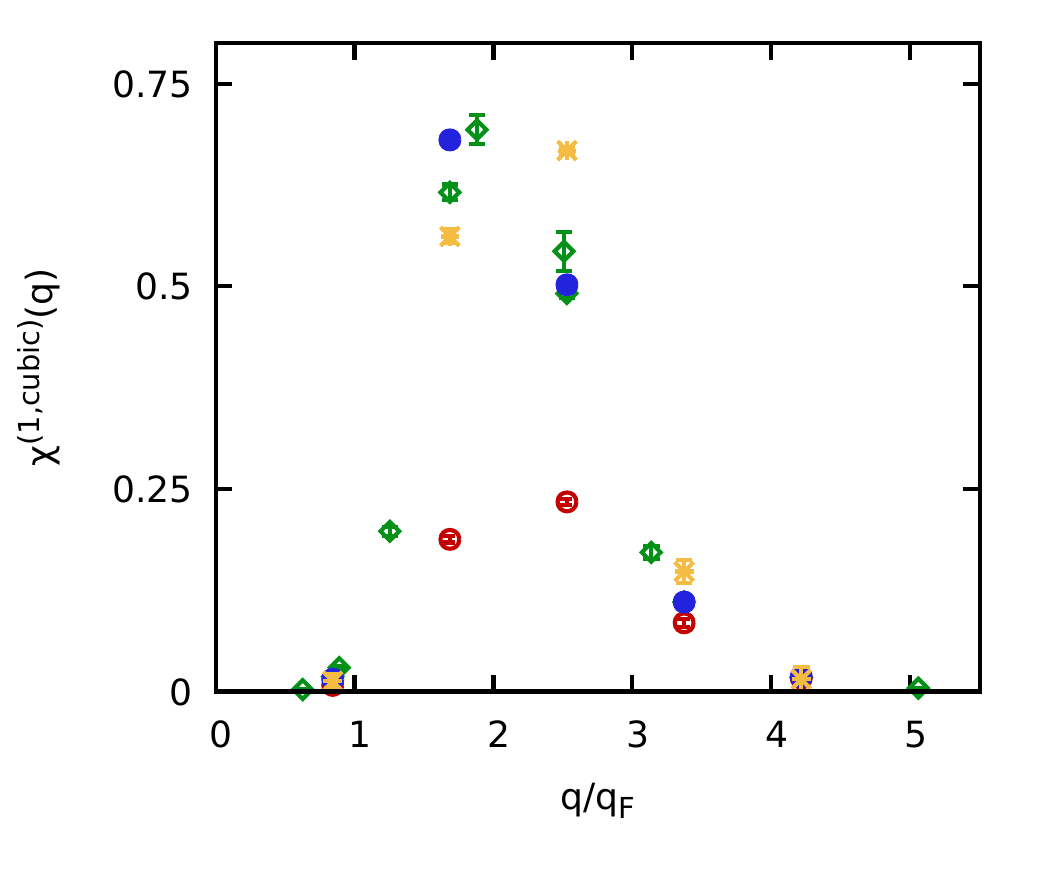}
\endminipage
\minipage{0.333\textwidth}%
  \includegraphics[width=\linewidth]{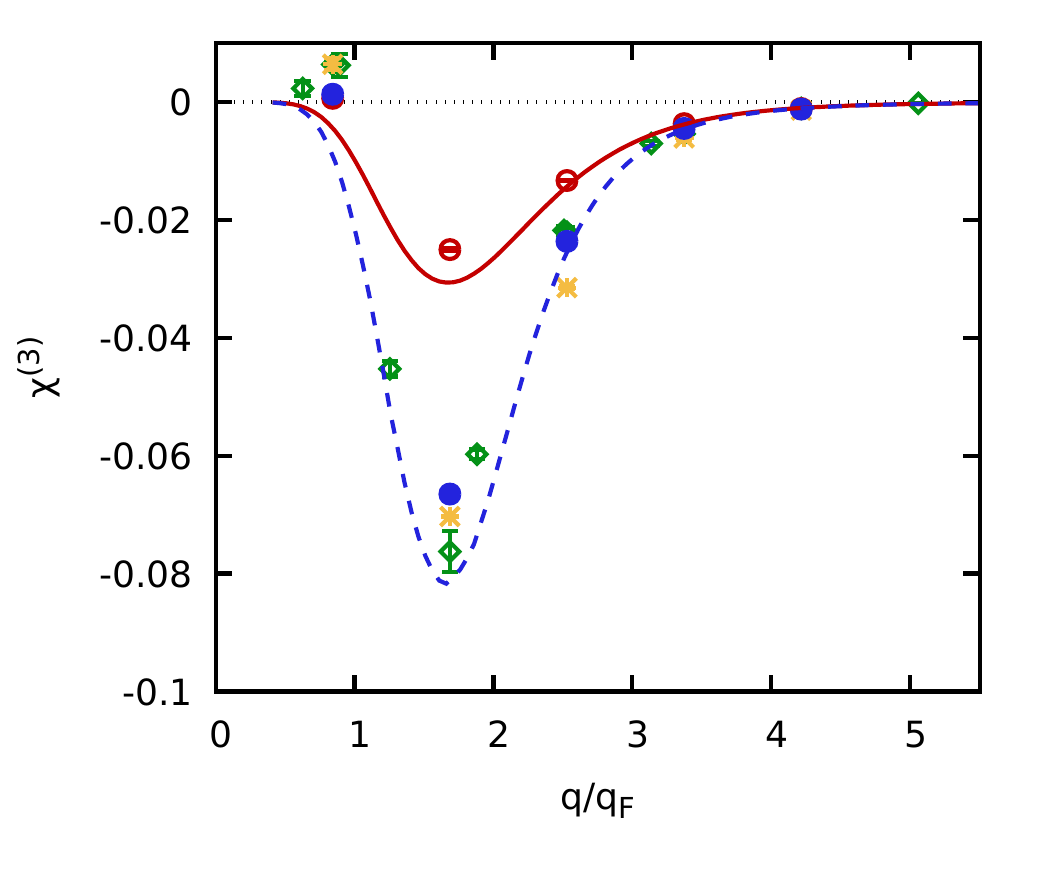}
\endminipage
  \caption{
  The results for non-linear static density response functions of UEG as defined in Eq. (\ref{eq:rho1}), Eq. (\ref{eq:rho2}), and Eq. (\ref{eq:rho3}) for $r_s=6$ and $\theta=1.0$.
  Left:  The quadratic response function at the second harmonic. 
  Middle: The cubic response function at the first harmonic.
  Right:  The cubic response function at the third harmonic. From \cite{Moldabekov_jctc_2022}. The KS-DFT calculations were performed using GPAW [\cite{GPAW_2024, ASE_17}]. }
  \label{fig:NLR_rs6_theta1}
  \end{figure*}

This relation has been shown empirically to hold for the UEG by QMC simulations. In the case of KS-DFT, $\chi^{(1,\textnormal{cubic})}_{0}( q)$  and $\chi^{(1,\textnormal{cubic})}( q)$ can be computed using Eq.(\ref{eq:rho1}), with $\braket{\hat\rho_\mathbf{q}}_{q,A}$ being generated by applying an external harmonic perturbation with different amplitudes $A$ [\cite{Moldabekov_jctc_2022}]. It was shown by \cite{Moldabekov_jctc_2022} that KS-DFT with a ground-state LDA XC functional allows one to achieve good agreement with exact QMC data for $\chi^{(2)}(q)$, $\chi^{(1,\textnormal{cubic})}(q)$, and  $\chi^{(3)}(q)$. This is illustrated for $r_s=6$ and $\theta=1$ in Fig. \ref{fig:NLR_rs6_theta1}, where the exact QMC results are compared with the LDA-based KS-DFT results (labeled as DFT), the KS-DFT results with zero XC functional (labeled as NullXC), the analytical results given by Eqs. (\ref{eq:chi2_LFC}), (\ref{eq:chi2_LFC}), (\ref{eq:cubic_first_LFC}) (labeled as LFC) with the machine learning representation of the LFC  $G(q)$  by \cite{dornheim_ML}, and the results computed using the same analytical solutions  with $G(q)=0$ (labeled as RPA)) [see \cite{Moldabekov_jctc_2022} for more details]. Clearly, one can see good agreement between the QMC data and the KS-DFT results.

To summarize, it is evident that the correct UEG limit of the XC functional is critical for the description of the strongly perturbed electron gas.  Next, let us discuss the XC kernels generated using hybrid XC functionals in WDM regime. 

\subsection{Hybrid XC Kernels for WDM based on the UEG}\label{ss:hybrid_Kxc}

In the following, we will consider hybrid XC functionals constructed by mixing HF exchange with a GGA level XC functional.
According to definition (\ref{eq:Kxc_def}), taking the second order functional derivative of Eq. (\ref{eq:pbe0}), we find:
\begin{equation}\label{eq:Kxc_mixing}
    K^{\rm hyb}_{\rm XC}[n]=K_{\rm XC}^{\rm GGA}[n]+a\left(K_{X}^{\rm HF}[\rho_{\sigma}]- K_{\rm X}^{\rm GGA}[n]\right),
\end{equation}
where  $a=\frac{1}{m}$ is the mixing coefficient defining the fraction of HF exchange introduced into the exchange part of the GGA XC functional.

For the calculation of $K_{\rm XC}$, the standard approach in KS-DFT is to use analytical formulas for the second-order functional derivatives. This is not problematic for the LDA and GGA level functionals due to their relatively simple forms. However, for extended systems, such an approach is not feasible for the meta-GGA and hybrid functionals explicitly using orbitals (e.g., see Hartree-Fock exchange (\ref{eq:HF_exchange})) [\cite{TIMROV2015460, GE20142080}].
This used to limit the application of XC kernels in LR-TDDFT to the levels of LDA and GGA. This problem was resolved in Refs. [\cite{Moldabekov_jctc_2024, Moldabekov_prr_2023, Moldabekov_jcp_2023_averaging}] by using the harmonic perturbation method for the calculation of the static XC kernel from KS-DFT without the need for any additional input despite the usual XC-functional, and without the need to explicitly evaluate any functional derivatives. 

\begin{figure}[!t]
\centering
\includegraphics[width=1\textwidth]{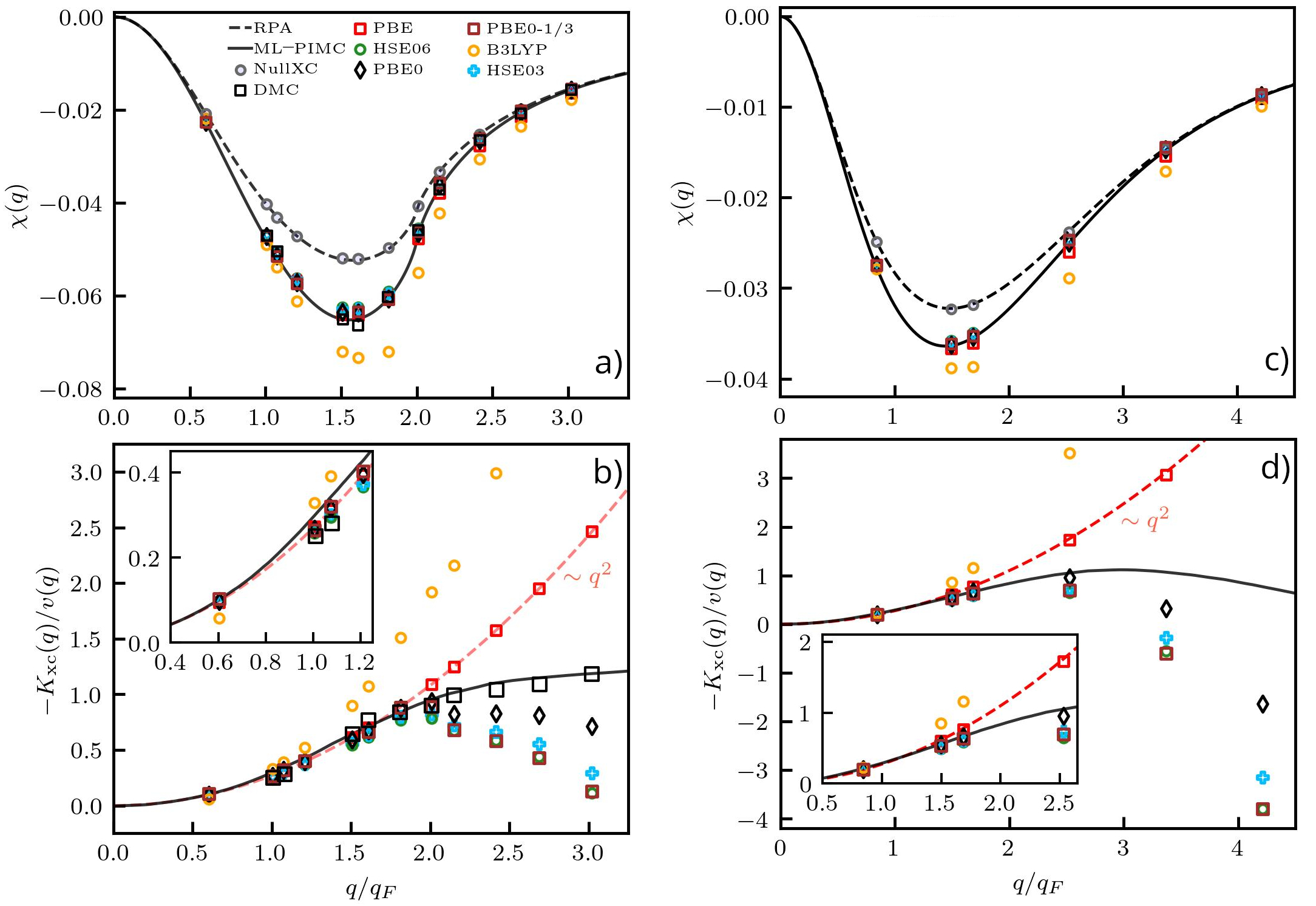}
\caption{\label{fig:Kxc_hybrid_jcp}
The linear static electron density response function of UEG for $\theta=0.01$ (top left) and $\theta=1$ (top right) at$r_s=2$.  The corresponding static XC kernels are shown in the bottom panels. Solid black line: exact UEG results based on the neural-net representation of Ref.~[\cite{dornheim_ML}]. Black squares: exact QMC results by ~\cite{Moroni_prl_1995} for the ground state. The RPA result for the density response function is given by the dashed line. The grey circles in the top panel are the KS-DFT results for the density response function with the XC functional set to zero (denoted as NullXC). The other symbols distinguish the KS-DFT results computed using different XC-functionals. Adapted from \cite{Moldabekov_jcp_hybrid_2023}. The KS-DFT calculations were performed using Abinit [\cite{Abinit_2020}].}
\end{figure}

The harmonic perturbation method for KS-DFT is similar to the approach by \cite{Moroni_prl_1995} for the calculation of the XC kernel of the UEG.
First, one needs to perform a KS-DFT simulation of an equilibrium system with a given configuration of ions to compute the electronic equilibrium density distribution $n^{i}(\vec r)$ (with $i$ indicating the considered ionic snapshot). Second, for the same material (and for the same snapshot), the KS-DFT calculation is repeated but with an additional weak external perturbation $v_{\rm ext}=2A\cos{(\vec q \vec r)}$, where the wavenumber must be commensurate with the simulation box length. This provides the density distribution of the perturbed system $n^{i}_{\vec q}( \vec r)$. Then,  the density perturbation values $\delta n^i_{\vec q}(\vec r)$ are extracted by taking the difference between the perturbed and unperturbed density distributions. 
Using the expansion (\ref{eq:expansion}) for $\delta n^{i}_{\vec q}(\vec r)$ allows one to find the Fourier coefficients of the density perturbation $\rho^{i}_{\vec G}(\vec q)$. The corresponding static density response for a given $\vec G$ is given by $\chi^{i}_{\vec G}(\vec q)=\rho^{i}_{\vec G}(\vec q)/A$.
The static KS response $\chi_{\rm KS, \vec G}^{i}(\mathbf{q})$ follows from the definition (\ref{eq:chi_KS_q_r}). Using $\chi^{i}_{\vec G}(\vec q)$ and $\chi_{\rm KS, \vec G}^{i}(\mathbf{q})$, from Eq. (\ref{eq:Kxc_chi_conn}) one finds the XC kernel for a given fixed ionic configuration.

We reiterate that, for homogeneous systems, all components of the density response function and KS response function with $\vec G\neq 0$ vanish after averaging [\cite{Moldabekov_jcp_2023_averaging}]. Therefore, we consider only $\chi(\vec q)=\left<\chi^{i}_{\vec G=0}(\vec q)\right>$. The corresponding averaged value of the KS response function $\chi_{\rm KS}(\vec q)$ is found by inverting  Eq. (\ref{eq:delta_KS}).  The mean values $\chi(\vec q)$ and  $\chi_{\rm KS}(\vec q)$ are used in Eq. (\ref{eq:Kxc_chi_conn}) to compute the XC kernel.

\begin{figure}[!t]
\centering
\includegraphics[width=0.9\textwidth]{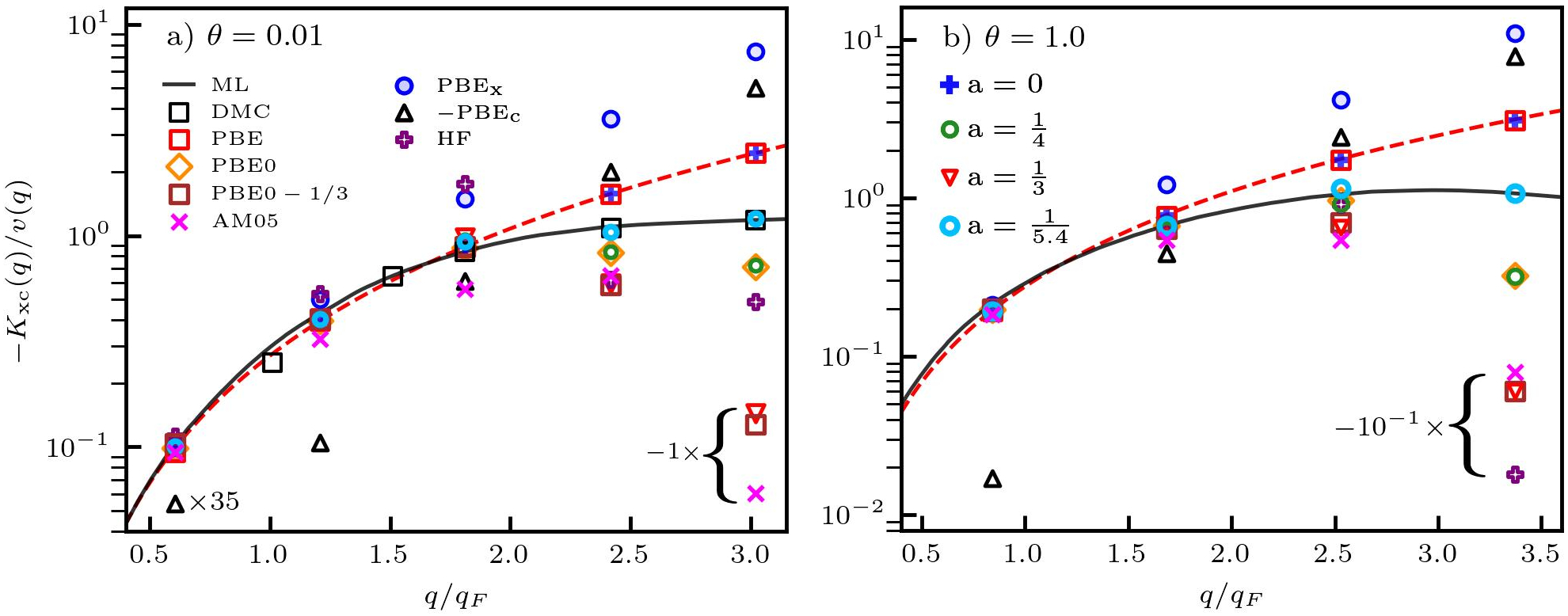}
\caption{\label{fig:kernel} Static XC-kernel of the UEG at a) $\theta=0.01$  and b) $\theta=1$ for $r_s=2$. 
The diffusion quantum Monte Carlo (DMC) results by \cite{Moroni_prl_1995} are presented by black squares for $\theta=0.01$. The solid line is the ML representation of the quantum Monte Carlo results by \cite{dornheim_ML}.
The dashed line illustrates the quadratic dependence of the PBE data on $q$.
The mixing parameter $a$ indicates the data obtained by combining the results for the kernels from separate HF, the exchange-only PBE (labeled as ${\rm PBE_x}$), and the correlation-only PBE (labeled as ${\rm PBE_c}$) calculations according to Eq.~(\ref{eq:Kxc_mixing}). Other symbols correspond to the KS-DFT data computed using PBE, PBE0, PBE0-1/3, and AM05 as described in the main text. 
Some data points are multiplied by a constant factor for a better illustration of the results. The corresponding multiplications are explicitly shown next to the curly brackets in the graphs. From \cite{Moldabekov_jpcl_2023}.
The KS-DFT calculations were performed using Abinit [\cite{Abinit_2020}].}
\end{figure} 

Applying the harmonic perturbation method, the static density response function and hybrid XC kernels based on PBE0, PBE0-1/3, HSE03, and HSE06 were computed for the UEG and then analyzed by comparing with exact QMC reference data in Ref. [\cite{Moldabekov_jcp_hybrid_2023}]. For the static density response, the results of this investigation clearly showed the superiority of the considered hybrid XC functionals compared to the local and semilocal level XC functionals. For $r_s=2$, this is illustrated in Fig. \ref{fig:Kxc_hybrid_jcp} for the ground state and for partially degenerate electrons with $\theta=1$. Specifically, we show the static density response functions (top panels) and local field corrections (LFC) $G(q)=-v(q)K_{\rm XC}(q)$ (bottom panels). 
The results computed using hybrid XC functionals are compared with the diffusion QMC data by \cite{Moroni_prl_1995} (DMC) for $T=0$ and with the PI-QMC data-based machine learning model by \cite{dornheim_ML}. In addition, the XC kernels computed using the GGA-level PBE functional and the hybrid B3LYP are provided. The B3LYP is one of the commonly used XC functionals in KS-DFT simulations at low temperatures. The main difference of B3LYP from other considered hybrid XC functionals is that B3LYP does not use the correct UEG limit as a constraint. As a result, B3LYP is not able to adequately describe the XC kernel of the UEG and, as a consequence, any system with a substantial amount of electronic delocalization (atomic ionization). The PBE is designed to respect the UEG limit at $q<2q_F$. This is also confirmed by the KS-DFT simulations using PBE. The hybrid  PBE0 and PBE0-1/3 [HSE03 and HSE06] are constructed using PBE [screened PBE] in Eq.~(\ref{eq:pbe0}). As a result, the XC kernels (or LFCs) based on these functionals are in close agreement with the PBE-based data at $q<2q_F$. In contrast to the PBE-based results, $G(\vec q)$ computed using PBE0, PBE0-1/3, HSE03, and HSE06 do not show a quadratic increase with the wavenumber as PBE-based data at at $q>2q_F$. Instead,  $G(\vec q)$ computed using  PBE0, PBE0-1/3, HSE03, and HSE06
decrease with the increase in the wavenumber at $q>2q_F$. Consequently, at $\theta=1$, $G(\vec q)$ values computed using these hybrid XC functionals are in good agreement with the QMC data also for larger wavenumbers. 
At $q>2.5q_F$, $G(\vec q)$ values of PBE0, PBE0-1/3, HSE03, and HSE06 start to deviate considerably from the exact QMC data. 

\begin{figure}\centering
\includegraphics[width=0.5\textwidth]{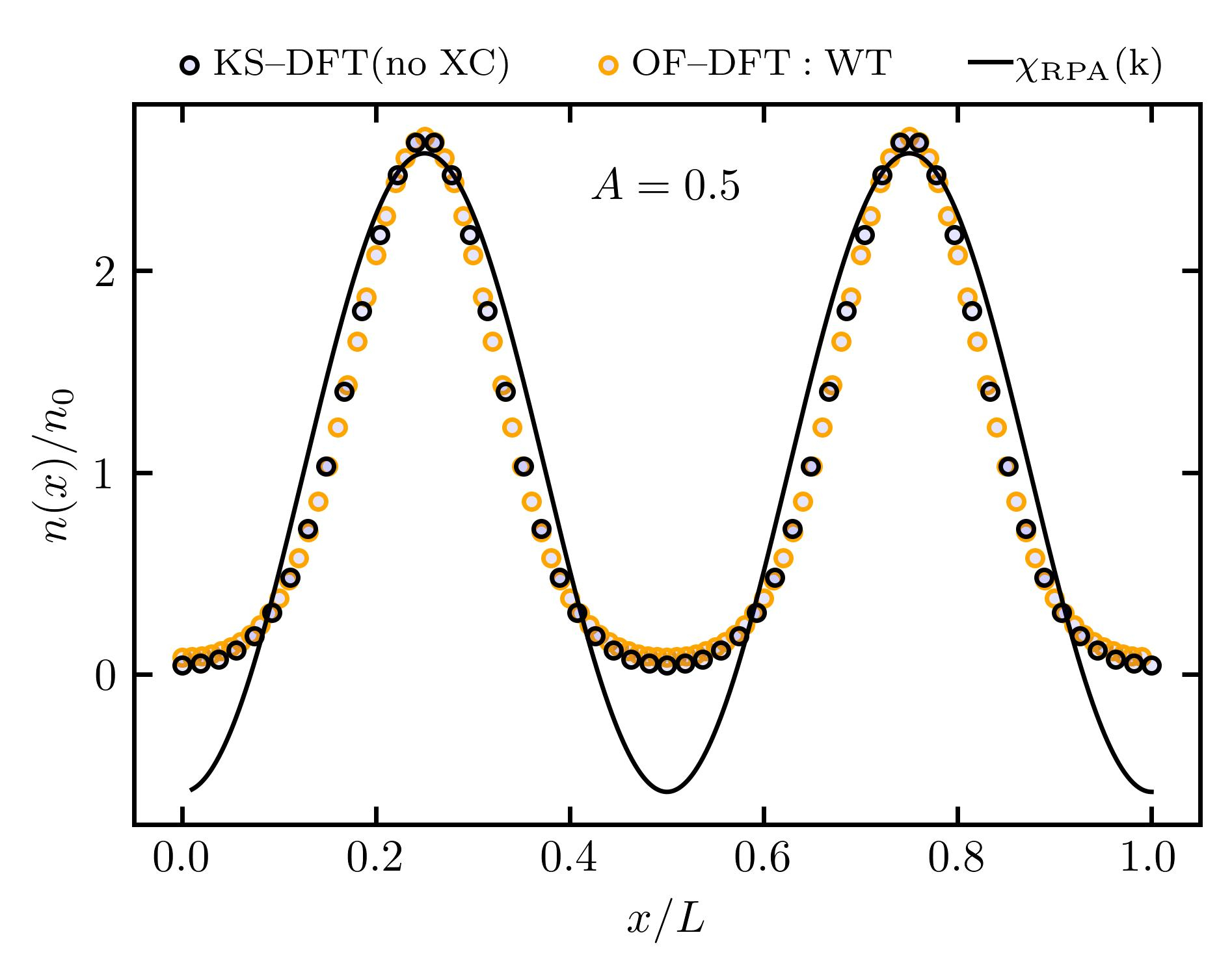}
\caption{\label{fig:den} Density profile along the perturbation direction at $A=0.5~{\rm Ha}$ and $q\simeq 1.208 q_F$. The results are computed by perturbing the UEG. The calculations are performed using OF-DFT with WT functional and KS-DFT and by setting the XC functional to zero for both. The solid line is the density value $n=n_0+\delta n$ with $\delta n$ computed using the RPA density response function of the UEG.  he KS-DFT calculations were performed using Abinit [\cite{Abinit_2020}]. The OF-DFT results were calculated using the DFTpy
code [\cite{DFTpy}].
}
\end{figure}

To understand the behavior of $G(\vec q)$ computed using PBE0, PBE0-1/3, HSE03, and HSE06, \cite{Moldabekov_jpcl_2023} have suggested computing different terms in Eq. (\ref{eq:Kxc_mixing}) separately. In Fig. \ref{fig:den}, we show the results for $G(\vec q)$ computed using exchange and correlation parts of PBE separately (denoted as $PBE_x$ and $PBE_c$, respectively) for $r_s=2$ at $\theta=0.01$ and $\theta=1$. Additionally, we provide the data computed by using the HF exchange with the correlation part set to zero. These results are compared with the QMC data and with the results computed using PBE, PBE0, PBE0-1/3, and AM05. 
The HF-based result for $G(\vec q)$ is in agreement with the $PBE_x$  data at $q<2q_F$ and decreases with the increase in the wavenumber at $q>2q_F$. The latter explains the behavior of PBE0 and PBE0-1/3 results for $G(\vec q)$ at $q>2q_F$. The $PBE_c$-based $G(\vec q)$ has a negative sign. Its contribution to the total PBE result is negligible at $q<q_F$. In total, the $PBE_c$ and $PBE_x$ results compensate each other in such a way that $G(\vec q)\sim q^2$. This balance between exchange and correlation parts of PBE is broken in the hybrid functionals replacing some part of $PBE_x$ by the HF exchange according to Eq. (\ref{eq:Kxc_mixing}). As a consequence, at large wavenumbers, $G(\vec q)$ computed using PBE0 and PBE0-1/3 decreases with the increase in $\vec q$. 

From this analysis, we see that the rate of the decrease of $G(\vec q)$ at $q>2q_F$ is controlled by the mixing coefficient $a$ in Eq. (\ref{eq:Kxc_mixing}). Therefore, one can choose the mixing coefficient to achieve better agreement between the KS-DFT and QMC results. This leads to the choice  $a=1/5.4$, which works for both $\theta=0.01$ and $\theta=1$. For comparison,  $a=1/4$ and $a=1/3$ are used in PBE0 and PBE0-1/3, respectively. The rationale for the choice of the mixing coefficient $a=1/4$ is based on the analysis of atomization errors of  molecules using the M{\o}ller-Plesset perturbation expansion 
[\cite{Perdew_jcp_1996}]. The choice of the mixing degree  $a=1/3$ for solids is justified empirically by computing material properties and comparing them with experimental measurements such as lattice constants, bulk moduli, the vacancy formation energy, and atomic data [\cite{Cortona_jcp_2012}].
However, it is a-priori unclear why these coefficients are a good choice for matter under extreme conditions, e.g., at high temperatures and densities.  The existing QMC data for the XC kernel of the UEG in a wide range of densities and temperatures [\cite{Dornheim_prb_2021, dornheim_ML, Dornheim_prl_esa_2020}] allow one to define the mixing coefficient in a non-empirical way [\cite{Moldabekov_jpcl_2023}] as we have demonstrated for $r_s=2$ and $\theta=1$.

\begin{figure}\centering
\includegraphics[width=1\textwidth]{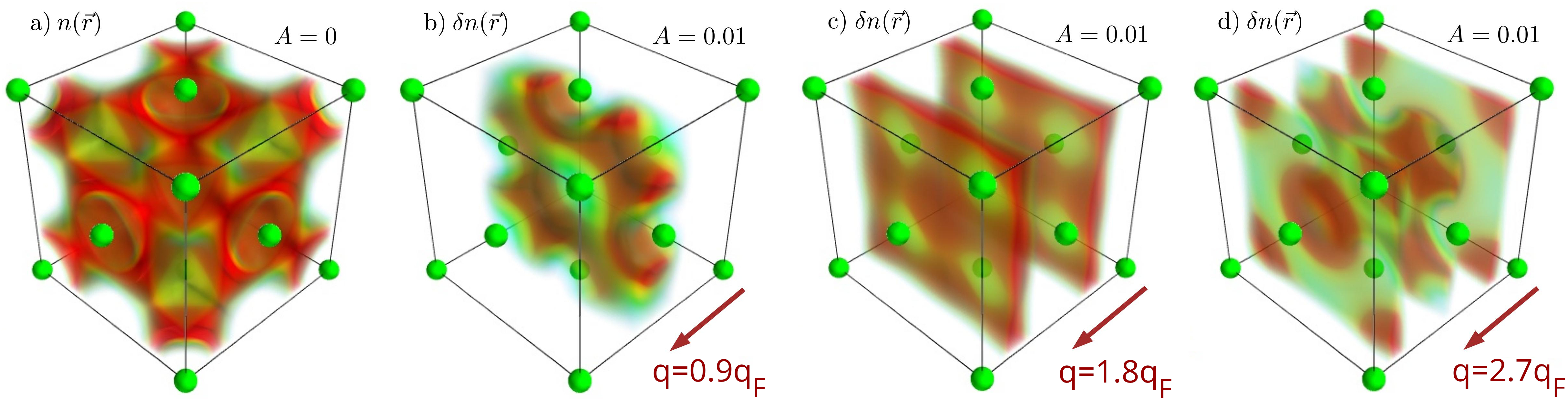}
\caption{\label{fig:3D_Al} Bulk electron density of Al with an fcc structure.
 The illustration of a) the electron density for the unperturbed case ($A=0$), and the density perturbation due to an external harmonic perturbation with the amplitude $A=0.01~{\rm Ha}$ and wavenumbers b) $q= 0.9q_F$, c) $q= 1.8q_F$, and d)  $q= 2.7q_F$. Al atoms are shown as green spheres. Adapted from \cite{Moldabekov_prb_2023}.
The KS-DFT calculations were performed using Abinit [\cite{Abinit_2020}].}
\end{figure} 

\subsection{Testing Non-Interacting Free Energies for Consistency with the UEG limit}\label{s:of_dft_test}

Let us now consider the non-interacting free energy functional $\mathcal{F}_{\rm S}[n]$, which is the main ingredient of the OF-DFT method. As mentioned,
the relation (\ref{eq:dFs_chi_ks}) connects the second-order functional derivative of $\mathcal{F}_{\rm S}[n]$ with the KS response function. This connection is the key ingredient for the construction of  $\mathcal{F}_{\rm S}[n]$. In the case of the UEG, Eq. (\ref{eq:dFs_chi_ks}) allows one to find
\begin{eqnarray}\label{eq:UEG_dF_chi0}
    \widehat  F\left[ \left. \frac{\delta \left<\mathcal{F}_{\rm S}[n]\right>}{\delta n(\vec r^{\prime})\delta n(\vec r)}\right\vert_{n=n_{0}}\right]=-\frac{1}{\chi_{0}(\vec q)},
\end{eqnarray}
where $\chi_{0}(\vec q)$ is the Lindhard function, and $n_0$ is the UEG density.

For example, the well-known Thomas-Fermi model with the first order gradient correction to $\mathcal{F}_{\rm S}[n]$ can be derived from Eq. (\ref{eq:UEG_dF_chi0}) using the lowest order term of the long wavelength ($q\to0$) expansion of $\chi^{-1}_{0}(\vec q)$ [\cite{Moldabekov_pop_2018}]. More advanced semi-local functionals use higher-order terms of the long wavelength expansion of $\chi^{-1}_{0}(\vec q)$.
For instance, a Laplacian-meta-GGA level KE functional, PGSL (Pauli-Gaussian second order and Laplacian), developed by~\cite{PGSL} has been designed to reproduce the Lindhard density response function of the UEG at $q<2q_F$. 
It was demonstrated that OF-DFT calculations using the PGSL functional provide an accurate description of the bulk properties of metals and semiconductors without relying on system-dependent parameters.
Another example of such a semi-local functional for the ground state is the GGA level kinetic energy functional by \cite{ground_LKT}, which was also tested successfully on simple metals and semiconductors. The extension of the LKT functional to finite temperatures has been provided by \cite{Luo_prb_2020}.

Fully based on Eq. (\ref{eq:UEG_dF_chi0}), a highly successful and often used non-local kinetic energy functional is the Wang-Teter (WT) functional [\cite{WT_paper, Sjostrom_Daligault_prb_2013}]:
\begin{equation}\label{eq:KE_WT}
\begin{split}
   \mathcal{F}_{\rm WT}[n(\vec r)]&=\mathcal{F}_{\rm TF}[n(\vec r)]+\mathcal{F}_{vW}[n(\vec r)]+\int \int {\mathrm{d}}\vec r {\mathrm{d}}\vec r^{\prime}~ n(\vec r)^{5/6} \mathcal{K}(\vec r-\vec r^{\prime};n_0)n(\vec r^{\prime})^{5/6},
   \end{split}
\end{equation}
where  $\mathcal{F}_{\rm TF}[n(\vec r)]$ is the Thomas-Fermi free energy, $\mathcal{F}_{vW}[n(\vec r)]=\int {\mathrm{d}}\vec r~  \left|\nabla n(\vec r) \right|^2/\left(8n(\vec r)\right)$ is the von Weizs\"acker (vW) gradient correction, and the kernel $\mathcal{K}(\vec r-{\vec r}^{\prime};n_0)$ is computed in Fourier space by using $n(\vec r)$ as the density of UEG in Eq. (\ref{eq:UEG_dF_chi0}). 
Originally, \cite{WT_paper} developed the non-interacting free energy functional (\ref{eq:KE_WT}) for $T=0$.
Later, \cite{ Sjostrom_Daligault_prb_2013} have extended this model to finite temperatures by using the finite-temperature versions of the Thomas-Fermi model and Lindhard function.

As mentioned in Sec. \ref{s:KS_model}, the OF-DFT method is an approximation to the KS-DFT model. 
Therefore, to analyze the quality of  $\mathcal{F}_{\rm S}[n]$, it is a standard practice to benchmark the OF-DFT results by comparing them with the data from KS-DFT simulations (e.g., see \cite{LMGP}). 
Recently, it was suggested to use the harmonic perturbation approach to compute the static density response function from OF-DFT to test whether Eq. (\ref{eq:UEG_dF_chi0}) is satisfied in the UEG limit [\cite{Moldabekov_prb_2023}].
It was found that the non-interacting free energy functionals designed using the Lindhard function in the UEG limit might not be traceable back numerically to Eq. (\ref{eq:UEG_dF_chi0}), leading to an inconsistency.
Furthermore, considering bulk Aluminum (Al), Silicon (Si), and semiconducting crystal diamond, it was shown that the correct UEG limit (\ref{eq:UEG_dF_chi0}) is key for accurate calculations of the density response by the OF-DFT method. A corresponding illustration of the density change due to a harmonic perturbation in bulk Al is provided in Fig. \ref{fig:3D_Al}.
The 3D distribution of the equilibrium density of the valence electrons without perturbation is shown in Fig. \ref{fig:3D_Al}a). In the other panels,  we demonstrate the change in the density induced by a cosinusoidal perturbation with different wavenumbers. In general, the electronic density
in bulk Al is inhomogeneous. However, this density inhomogeneity does not exceed $10\%$ of the mean density of the valence electrons $n_0\simeq 1.81\times 10^{23}~{\rm cm^{-3}}$. This can be gauged to be a relatively weak density inhomogeneity. As discussed in Sec. \ref{s:harm_per_EG} in the context of non-linear density response theory, for such weak density inhomogeneities, the correct description of the UEG limit is vital for accurate density perturbation calculations. This explains the success of the UEG limit-based non-interacting free energy models for the description of the structure of valence electrons in metals.

Even when the density perturbation is strong, imposing the correct UEG limit is crucial for non-interacting free energy functionals. This is demonstrated in Fig. \ref{fig:den} by considering strongly inhomogeneous electronic density formed in free electron gas under the harmonic perturbation. The magnitude of the density inhomogeneity exceeds  $2n_0$ and is generated with a periodicity corresponding to the wavenumber $1.21 q_F$. We show the results computed using KS-DFT, OF-DFT with WT functional, and the density response of the UEG (\ref{eq:inv_chi_ueg}). We set the XC functional (kernel) to zero in all calculations to unambiguously gauge the quality of  $\mathcal{F}_{\rm S}[n]$ in OF-DFT. 

First of all, we see that the density perturbation computed using  $\chi_{\rm RPA}$ of the UEG in Eq. (\ref{eq:chi_tot_r}) significantly overestimates the density depletion around the minima of $n(\vec r)$ compared to the KS-DFT data. This is unsurprising since the strong perturbation is well beyond linear response theory. Second, we conclude that OF-DFT provides excellent agreement with the KS-DFT results. As was shown by \cite{Moldabekov_prb_2023}, this high quality of the OF-DFT results is due to the enforcement of the correct UEG limit in the WT functional (\ref{eq:KE_WT}). Besides that, in the numerical solution of the OF-DFT equations, it is naturally required that $n(\vec r)\geq 0$, which allows one to avoid unphysical negative density values. Additionally, it is known that $\mathcal{F}_{vW}[n]$ is an exact solution for the single orbital limit ($n\to0$), and $\mathcal{F}_{TF}[n]$ is an exact solution for the high density limit ($n\to \infty$). These constraints are also crucial for the adequate behavior of $\mathcal{F}_{\rm S}[n]$. We note that  $\mathcal{F}_{TF}[n]$ and $\mathcal{F}_{vW}[n]$ can be derived by using the long-wavelength ($\vec q\ll \vec q_F$) and short-wavelength limits ($\vec q\gg \vec q_F$) of the inverse Lindhard function in Eq. (\ref{eq:UEG_dF_chi0}) [\cite{Moldabekov_pop_2018, Moldabekov_cpp_2D_2017}].

\begin{figure}[t!]
\center
\includegraphics[width=1\textwidth]{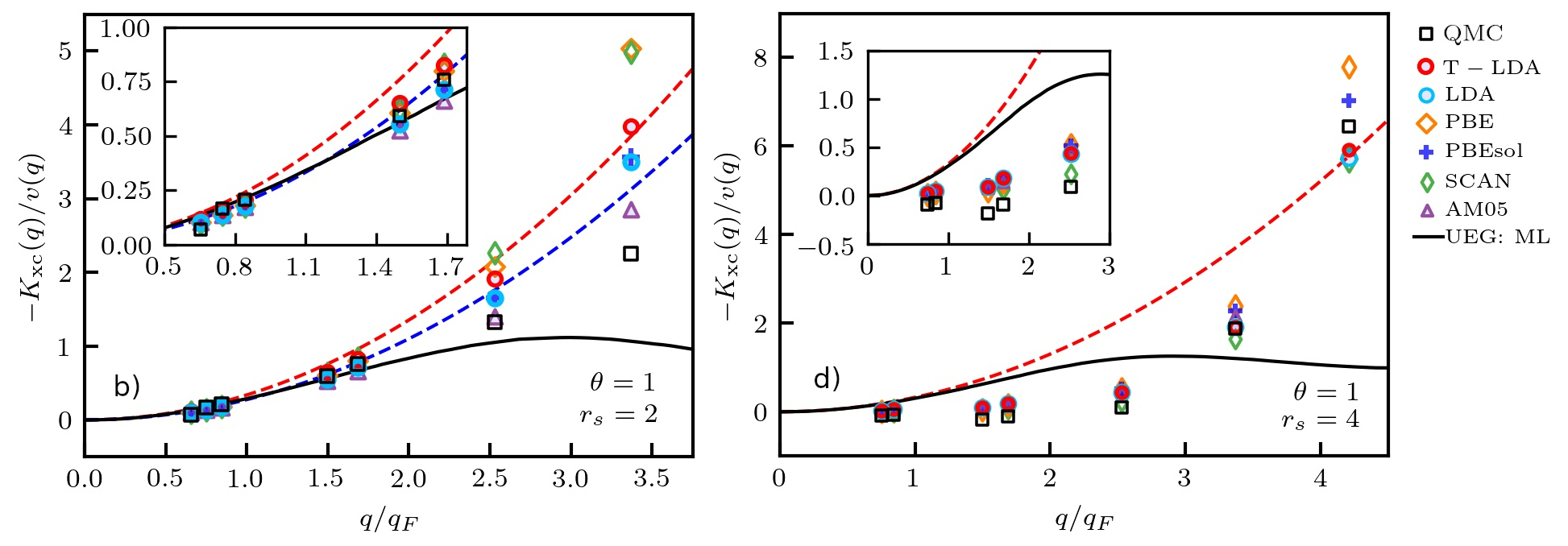}
\caption{ \label{fig:H_rs2_4} 
 XC-kernel $K_\textnormal{xc}(\mathbf{q})$ of hydrogen at WDM conditions ($\Theta=1$) for $r_s=2$ [left] and $r_s=4$ [right] computed using the static density response function  $\chi(\mathbf{q})$ from KS-DFT and QMC in combination with $\chi_0(\mathbf{q})$ in Eq. (\ref{eq:Kxc_chi_conn}). Black squares: exact QMC results for hydrogen by \cite{Boehme_PRL_2022}. Solid (dashed) black line: exact results for the UEG model at the same conditions based on the neural-net representation by \cite{dornheim_ML} (analytical RPA). 
 The other symbols correspond to KS-DFT calculations with different XC functionals. Adapted from \cite{Moldabekov_jctc_2023}.
 The KS-DFT calculations were performed using GPAW [\cite{GPAW_2024, ASE_17}]. 
}
\end{figure} 

\section{Testing XC Functionals for Warm Dense Hydrogen}\label{s:warm_dense_hydrogen}

The understanding of hydrogen at elevated pressures and temperatures is important for both fundamental science and applied technologies. Hydrogen constitutes the predominant material in stars and giant planet interiors~[\cite{Benuzzi_Mounaix_2014,fortov_review}] and its isotopes are used as fuel for inertial confinement fusion (ICF) experiments~[\cite{hu_ICF,Betti2023, Abu-Shawareb_prl_2022}]. 
Accurate simulations of warm dense hydrogen are of relevance for astrophysical models, 
such as heat transport in the Jovian atmosphere~[\cite{Bell_2018_L20, Tan_2019_26}], and for the exploration of the high pressure and high-temperature superconductivity~[\cite{Wang2022, Drozdov2015Nature}]. 
The exploratory research of warm dense hydrogen is continuously being pursued at large 
research facilities such as the National Ignition Facility (NIF)~\cite{Moses_NIF}, the Omega Laser Facility~[\cite{OMEGA_hydrogen}], and the Linac coherent light source (LCLS)~[\cite{Kraus_prr_2023}].
Yet, the diagnostics of these experiments is notoriously difficult~\cite{Fletcher_Frontiers_2022,Davis2016}.
Hence, our understanding of the physical processes in these experiments heavily relies on simulations~[\cite{bonitz2024principles}]. 

Very recently, \cite{Moldabekov_jctc_2023, Moldabekov_prr_2023, Moldabekov_jcp_2023_averaging} have analyzed the quality of different local and semi-local XC functionals for the description of the static density response and XC kernel of warm dense hydrogen against quasi-exact PI-QMC simulations~[\cite{Boehme_PRL_2022,Bohme_PRE_2023,Dornheim_PRE_2023}]. The list of tested XC functionals includes thermal-LDA (T-LDA) by \cite{Groth_prl_2017}, ground state LDA by \cite{Perdew_Zunger_PRB_1981}, GGA level PBE by \cite{PerdewPBE}, PBEsol by \cite{PBEsol}, meta-GGA level SCAN by  \cite{Sun_prl_2015}, TPSS by \cite{TPSS}, revTPSS by \cite{revTPSS}, AM05 by \cite{AM05},  and meta-GGA XC functionals by \cite{Tran} and by \cite{Rasanen}. At partial electronic degeneracy with $\theta=1$, it was found that the KS-DFT calculations provide good agreement with the PI-QMC results for the static density response at $r_s\leq4$. In addition, it was shown that T-LDA, LDA, PBE, PBEsol, SCAN, and AM05 provide good quality results for the XC kernel at $q\lesssim2q_F$.  In Fig. \ref{fig:H_rs2_4}, we show results for the XC kernel of warm dense hydrogen, providing a comparison between KS-DFT, PI-QMC data [\cite{Boehme_PRL_2022}], and for the UEG model at the same conditions. We find that the UEG model is in closer agreement with the QMC and KS-DFT results at $q\lesssim1.5q_F$ and strongly deviates for $q\gtrsim2q_F$.
Wavenumbers $q\lesssim1.5q_F$ correspond to wavelengths  $\lambda \gtrsim 2\pi/(1.5q_F)\simeq 4.4~{\rm Bohr}$. This can be compared with the mean distance between protons at $r_s=2$, which is about $4~{\rm Bohr}$. 
This implies that the electrons in warm dense hydrogen behave like a free electron gas at $r_s=2$ and $\theta=1$ in processes that involve length scales larger than the mean distance between particles.
At $r_s=4$ and $\theta=1$, the UEG model does not adequately describe the XC kernel of warm dense hydrogen indicating that electron-proton coupling is strong and a simple partially ionized plasma picture is not accurate. 

Naturally, it is preferable that the XC functional provides not only an accurate density response function but also an equilibrium electronic density structure.  \cite{Moldabekov_jctc_2024} investigated the performance of T-LDA, LDA, PBE, PBEsol, SCAN, AM05, TPSS, and revTPSS for the description of the electronic structure in warm dense hydrogen at $r_s=4$, $r_s=2$, and $r_s=1$ with $\theta=1$. Considering the electronic density and density gradients, it was established that T-LDA and SCAN provide the most accurate results among considered XC functionals. Moreover, the PBE XC functional provides accuracy close to that of T-LDA.

We note that the region where the density $n(\vec r)$ is small and the density-gradient $\nabla n(\vec r)$ is large (defined as $s[n]=\left| n(\vec r) \right|/\left(24\pi^2 n^4\right)^{1/3}\gg 1$)  has a crucial role for the accurate description of  the  weak binding  occurring in van der Waals systems\footnote{Van der Waals systems are weakly bound molecular complexes formed by attractive intermolecular interactions between closed-shell atoms or molecules.} [\cite{Zhang_jcp_1997, TPSS}]. The TPSS and revTPSS functionals are designed to provide an adequate description of such weak binding. In Ref.[\cite{Moldabekov_jctc_2024}], warm dense hydrogen at parameters relevant for ICF experiments with $s[n]\lesssim 2.5$ has been considered, which means that the conditions characteristic for van der Waals systems were not reached.  To the best of our  knowledge, a systematic benchmark of XC functionals for warm dense hydrogen at conditions with $s[n]\ll1$ against exact QMC data has not been reported yet.

Let us summarize the analysis of various XC functionals for warm dense hydrogen.
SCAN and T-LDA provide a similar quality for both the equilibrium density and density response properties. Although slightly less accurate for equilibrium electronic structure, PBE generally performs similarly well as T-LDA.
We note that meta-GGA functionals are computationally much more expensive and are often problematic to converge \footnote{For ground state applications, the meta-GGA-level SCAN functional was regularized yielding improved computational performance and enhanced numerical stability [\cite{r2SCAN, rSCAN}]. The computational performance of these regularized versions at WDM conditions compared to the original version has not been performed. }. Therefore, from a numerical point of view, it is more advantageous to use local T-LDA or semilocal PBE functionals instead of the meta-GGA level SCAN functional.

\section{Conclusions and Outlook}\label{s:end}

We have presented an extensive discussion of the connection between the free energy functional of electrons and the density response function for both homogeneous and inhomogeneous systems, with a particular focus on warm dense matter. One of the important relations in DFT applications is the stiffness theorem, which connects the free energy functional with the density response function. This connection was extensively discussed in prior works for the energy functionals in the ground state [\cite{quantum_theory}].
In contrast, the stiffness theorem for the free energy appropriate for finite temperature applications has not been given much attention previously. 
Remedying this situations has been one of the main aims of the present work; as a new result, we have shown the stiffness theorem for the intrinsic part of free energy, providing the connection between the intrinsic free energy functional and the electronic polarization function.
In addition to their practical usefulness for constructing XC and non-interacting free energy functionals, the relations discussed above also highlight the importance of the static density response and the KS response in assessing the accuracy of XC functionals and non-interacting free energy functionals utilized in DFT methods.

The analysis of a whole gamut of XC functionals and non-interacting free energy functionals unambiguously demonstrates that the UEG limit is the key constraint for an adequate description of the electronic structure of warm dense matter using DFT methods.  This is a strong motivation to further develop the theory of the UEG at WDM conditions.
In this regard, we note that the static XC kernel [\cite{dornheim_ML,Dornheim_prl_esa_2020}] and free-energy density [\cite{Dornheim_PhysRep_2018}] of the warm dense UEG are solved problems. 
In our opinion, the next major frontier is given by its dynamic XC-effects, which can be inferred from PIMC results via two related, though substantially distinct routes: i) via the analytic continuation of $F(\mathbf{q},\tau)$ [cf.~Eq.~(\ref{eq:Laplace})], which, while being notoriously difficult, would give one access to the real-frequency dynamics, and ii) in terms of the imaginary Matsubara frequency XC-kernel $\widetilde{K}_\textnormal{XC}(\mathbf{q},z_l)$, which can be directly extracted from PIMC simulations~[Eq.~(\ref{eq:MDR})], and which is fully sufficient to compute electron--electron correlation functions from KS-DFT, and for the construction of advanced, fully non-local XC-functionals.

The conclusions drawn for the warm dense electron gas are expected to be largely valid for WDM generated by driving metal or semiconductor samples to extreme conditions [e.g., see Refs. \cite{Descamps_sciadv, Grolleau_prl_2021, Tilo_Nature_2023}]. Collecting recommendations from Sec. \ref{s:Test_electron_gas} for the perturbed electron gas and from the findings for warm dense hydrogen discussed in Sec. \ref{s:warm_dense_hydrogen}, we conclude that, among the considered ground-state XC functionals, PBE provides a good balance between generality, accuracy, and computational cost for electronic structure simulations of warm dense matter. The overall quality of the T-LDA functional by \cite{Groth_prl_2017}  and the ground-state PBE functional are similar for the density response of the warm dense electron gas and hydrogen, with the former being slightly more accurate for the density gradients in warm dense hydrogen [\cite{Moldabekov_jctc_2024}].  
It is important to note that these conclusions are based on the currently available results and may be revised if new experimental observations (and upcoming PI-QMC simulations of other elements such as warm dense Be~[\cite{Dornheim_Science_2024}]) prove them inaccurate.
We emphasize that one cannot determine the quality of the  XC functionals for warm dense matter simulations using only KS-DFT, i.e., without additional data from more accurate theories or experiments~[\cite{bonitz2024principles}].

In the standard WDM regime with $r_s\sim 2$ and $\theta\sim 1$, easy-to-compute PBE and T-LDA functionals already provide good agreement with the QMC results for the UEG, for the harmonically perturbed warm dense electron gas, and for warm dense hydrogen. 
Nevertheless, further testing of XC functionals should be systematically extended in future works as the development of thermal XC functionals remains an active field. In a recent review paper by \cite{bonitz2024principles}, various ground state and thermal LDA and GGA level functionals were tested by analyzing equation of state (EOS) data and comparing with the results from QMC simulations [\cite{Filinov_PRE_2023, Militzer_pre_2021}]. An improvement in the quality of DFT results was reported when thermal XC effects were consistently included. 
The comparative analysis in Ref. [\cite{bonitz2024principles}]  showed that the pressure from the KS-DFT-based molecular dynamics simulations using the PBE functional exhibits deviations from the QMC data of up to $7\%$ at $\theta\lesssim 1$. The quality of the PBE-based results for the EOS of hydrogen improves with the increase in temperature due to the reduction of the importance of the XC contribution.    
The finite-temperature GGA functional by Karasiev \textit{et al.}~[\cite{Karasiev_prl_2018}] achieves significantly better agreement with the PIMC data [with an error up to a few percent in the WDM regime]. We note that the EOS quality depends not only on the accuracy of the electronic density $n(\vec r)$, but also on the free energy functional. 
As a way to get a better understanding of the error source in the EOS calculations, one might use $n(\vec r)$ computed with, e.g., PBE in  a finite temperature version of $F_{\rm xc}[n]$ to compute EOS and compare the result with the EOS calculated self-consistently using that $F_{\rm xc}[n]$, i.e., analyzing the error of the calculations in terms of contributions due to the approximate functional and due to the approximate density [\cite{Kim_prl_2013}]. It might happen that a better agreement between KS-DFT and QMC data originates from using the finite temperature form of the XC free energy rather than the more accurate density itself.
Therefore, generally, a good agreement in the EOS obtained by using certain XC functional does not automatically mean that other properties are also well described by this XC functional. Additionally, we note that there is still some debate on the best way of implementing the finite temperature XC functional on the GGA level [\cite{kozlowski2023generalized}].  In addition, when used at finite tempertures, it was shown that the ground-state meta-GGA level XC functionals inherit information about thermal effects from the kinetic energy denisty used in it [\cite{Moldabekov_jctc_2024}], which means a careful analysis is needed in the construction of the meta-GGA level functionals to avoid   `` double counting '' of the thermal effects in $F_{\rm xc}[n]$.
In this regard, we further mention the exciting possibility of rigorously benchmarking the available zoo of XC-functionals against upcoming PI-QMC simulations of heavier elements such as warm dense beryllium~[\cite{Dornheim_Science_2024,Dornheim_JCP_2024}], which is facilitated by improved methodologies such as the simulation of fictitious identical particles~[\cite{Xiong_JCP_2022,Xiong_PRE_2023,Dornheim_jcp_2023,Dornheim_JPCL_2024}], or even novel machine learning approaches~[\cite{Xie_PRL_2023}].

Summarizing the analysis of the electronic structure properties from this work and the analysis of the EOS and pair distribution function from Ref. [\cite{bonitz2024principles}], we can conclude that role of thermal effects in the XC functional is important in the WDM regime with partially degenerate electrons $0.1\lesssim \theta\lesssim 3$.



Finally, we note that the ideas and methodologies of the DFT approach for non-relativistic quantum systems are transferable to other fields such as nuclear physics [e.g., see \cite{Marino_prc_2021,  Naito_prc_2022, Duguet2023, Zurek_prc_2024}] and quark-gluon matter [e.g., see \cite{Blaschke_inbook, Ivanytskyi_prd_2022, Blacker_prd_2020}]. For instance, \cite{Marino_prc_2023} demonstrated the utility of the harmonic perturbation method within DFT  to study the static density response of nuclear matter. In this regard, similarly to the KS-DFT, the key problem of nuclear energy density functionals is to improve the predictive power for assisting experimental efforts. For WDM, this is achieved to a large degree by building XC functionals in a non-empirical way based on exact QMC results for the UEG model.


\section*{Acknowledgments}
This work was funded by the Center for Advanced Systems Understanding (CASUS) which is financed by Germany’s Federal Ministry of Education and Research (BMBF) and by the Saxon state government out of the State budget approved by the Saxon State Parliament. This work has received funding from the European Research Council (ERC) under the European Union’s Horizon 2022 research and innovation programme
(Grant agreement No. 101076233, "PREXTREME"). 
Views and opinions expressed are however those of the authors only and do not necessarily reflect those of the European Union or the European Research Council Executive Agency. Neither the European Union nor the granting authority can be held responsible for them.
We gratefully acknowledge computation time at the Norddeutscher Verbund f\"ur Hoch- und H\"ochstleistungsrechnen (HLRN) under grant shp00026 and mvp00024, and on the Bull Cluster at the Center for Information Services and High Performance Computing (ZIH) at Technische Universit\"at Dresden. 
\bibliographystyle{cas-model2-names}

\bibliography{cas-refs}



\end{document}